\documentstyle[12pt]{article}
\begin{document}

\title{The doublet of Dirac fermions in the field of the non-Abelian
       monopole, isotopic chiral symmetry, and parity selection rules}
\date {}

\author {V.M.Red'kov \\
     Institute of Physics, Belarus Academy of Sciences\\
F Skoryna  Avenue 68, Minsk 72, Republic ob Belarus\\
e-mail: redkov@dragon.bas-net.by}

\maketitle

\begin{abstract}

The paper concerns  a~problem  of  Dirac  fermion  doublet  in
the~external monopole potential obtained by embedding the~Abelian
$SU(2)$ monopole
solution in the~non-Abelian scheme. In this particular case,
the~doublet-monopole  Hamiltonian
is invariant under some symmetry  operations  consisting  of
a~(complex and one parametric) Abelian
subgroup in the complex rotational group
$SO(3.C) : [\hat{H} , \hat{F}(A)]_{-} = 0 , \hat{F}(A) \in  SO(3.C)$.
This symmetry results in a~certain  freedom  in  choosing  a~discrete
operator $\hat{N}_{A}$ ($A$ is a~complex number) entering the~complete
set of quantum variables.
The~same  complex
number  $A$   represents an~additional parameter at the~basis wave functions
$\Psi ^{A}_{\epsilon jm\delta \mu }(t,r,\theta ,\phi )$.
The  {\em generalized}  inversion-like  operator
$\hat{N}_{A}$  implies its own ($A$-dependent) definition for scalar and
pseudoscalar, and further affords certain generalized $N_{A}$-parity
selection rules. All the~different sets of basis functions
$\Psi ^{A}_{\epsilon jm\delta \mu }(x)$ determine globally the~same Hilbert
space. The~functions $\Psi ^{A}_{\epsilon jm\delta \mu }(x)$
decompose into linear combinations of
$\Psi ^{A=0}_{\epsilon jm\delta \mu }(x)$. However, the bases considered
turn out to be nonorthogonal ones when $A^{*}\neq A$; the latter correlates
with the non-self-conjugacy property of the~operator  $\hat{N}_{A}$
at  $A^{*}\neq  A$. The meaning of possibility to violate
the~known quantum-mechanical regulation on self-conjugacy as regards
the~inversion-like operator $\hat{N}_{A}$ is discussed.
The question  of possible physical understanding the~complex expectation
values for  $\hat{N}_{A}$  (at  $A^{*}\neq A$)  is examined. Also,
the~problem of possible physical status for the~matrix  $\hat{F}(A)$
at $A^{*} = A$  is considered in full detail: since the~matrix
belongs formally to the~gauge group  $SU(2)^{gauge.}_{loc.}$,  but  in
the~same time, being a~symmetry operation for the~Hamiltonian under
consideration, this operator generates linear transformations on basis
wave functions. It is emphasized that interpretation of the~$A$-freedom
as exclusively a~gauge one is not justified since this will leads to
a~logical collision with the~quantum superposition principle, and besides,
there  will arise  the~conclusion that two sorts of basis states
$\Psi ^{A=0}_{\epsilon jm,+1,\mu}(x)$
and $\Psi^{A=0}_{\epsilon jm,-1,\mu}(x)$  are to be  physically identical.
The~latter could be interpreted only as a~return to the
Abelian scheme again.


\end{abstract}

\newpage
\subsection*{1. Introduction\footnote{A more short version of the paper may be seen
in [101]}.}

While  there  not  exists  at  present  definitive  succeeded
experiments concerning  monopoles, it is nevertheless true  that
there exists a~veritable jungle  of  literature  on  the~monopole
theories. Moreover, properties of more  general
monopoles, associated with large gauge groups now thought to be
relevant in  physics.
As evidenced  even  by   a~cursory examination of some popular surveys
(see, for example, [1,2]),  the~whole monopole area covers and touches
quite a~variety of fundamental problems. The~most outstanding of them are:
the~electric charge quantization [3-10], $P$-violation in purely  electromagnetic
processes [11-16], scattering on the~Dirac string [17-19], spin from monopole
and spin from isospin [20-23], bound states in fermion-monopole system
and violation of the Hermiticity property [24-38],
fermion-number breaking in the~presence of a~magnetic monopole and
monopole catalysis of baryon decay [39-41].
The~tremendous volume of publications on monopole topics
(and there is no hint  that  its  raise  will  stop)  attests
the~interest which  they  enjoy among theoretical  physicists,  but
the~same token, clearly  indicates  the~unsettled  and  problematical
nature of those objects:  the~puzzle of monopole seems to be
one of  the~still yet unsolved problems of particle  physics\footnote{Very
physicists have contributed to investigation of the~monopole-based theories.
The~wide scope of the~field and the~prodigious number of investigators
associated with various of its developments make it all but hopleless to list
even the~principal contributors. The present study does not pretend to be
a~survey in this matter, so I give but a few of the~most important references
which may be useful to the~readers who wish some supplementary material or
are interested in more techical developments beyonds the~scope of the~present
treatment.}

In the same time,
the~study  of monopoles has now reached a~point where further progress
depends on a~clearer understanding of this object that had been available
so far.  Apparently  what is needed is neither the~search of decided
experiments, which are unlikely to be successful,  nor a~new solution of some nonlinear systems of equations,
but  rather  the analysis and careful criticism of already considered results.
In refwerence to this, leaving aside a~major part of various monopole
problems, much more comprehensive in themselves, just  some aspects
of the~$SU(2)$-model will be a~subject of the~present work. That of course
seriously restricts the~generality of consideration, but it should be
emphasized at once that though much more involved monopole-like
configurations are consistently (and somewhat routinely) invented and
reported in the~literature; in the~same time we should recognize
that certain purely Abelian or, the~most contiguous with it, $SU(2)$-model's
aspects came to light  when considering those generalized systems.
In view of that, the~particular $SU(2)$-model's features, being
considered here,  might be of reasonable interest for a~more large
number of non-Abelian models\footnote{Some more discussion on possible
extension to any different gauge theories are given in the~conclusive
Sec.~11 ;  and that possibility of generalization lends interest
to the~present study.}.

Once the~non-Abelian monopole had been brought
by 't~Hooft and Polyakov [42-44] into scientific usage, its main properties
had been noted and examined. The background of thinking of the~whole
(non-Abelian) monopole problem in that time can be easily traced:
it was obviously tied up with the~most outstanding points of its
Abelian counterpart, Dirac monopole; namely singularity properties,
quantization conditions, and some other contiguous to them.
This  reflected the~impress of old attitude towards the~monopole, which
had been imbibed by physicists from early Dirac's investigation on this matter
and consists in drawing special attention to the~known singularity aspects.
Evidently, the~most significant and noticeable achievement of that new
theory was
the~elimination  of the~singularity aspect from the~theory. It should be noted
that, from  the~very beginning, the~main emphasis  had been drawn to just
a~spatially radial non-singular structure of that new monopole-like system.
Thus,  the~required absence   of singularity had been achieved and therefore
the~clouds over this part of the~subject had been dispersed.

Much less attention has been given to a~number of other sides of that
non-Abelian  construct. For instance:
To the~'t Hooft-Polyakov construction has been assigned a~monopole-like
status; what is a~distinctive physical feature of it that provides
the~principal grounds for such an~assignment?
Or, which part of this system represents {\it a~trace} of Abelian monopole
and therefore  bears   this same  old quality, and which one is referred
to its own and purely non-Abelian nature?
So, the~question in issue is the role and status of the~Abelian monopole
in the non-Abelian theory. Although a~number of general and rigorous
relationships connecting these
aspects of these two models have been established to
date \footnote{Sometimes, it is considered to be solely
a~subsidiary construct, being appropriate to mimic the 't~Hooft-Polyakov
potential: at least, it can exactly simulate the latter far away
from the region $r=0$ (at spatial infinity).}
little work has been
done so far in linking them at the~level of a~specifically contrasting
examination  where all detailed  calculations have been carried out. This
article will endeavor to supply this work\footnote{Though evidently,ultimate
answers have not been found by this work as well, it might hoped that a~certain
exploration into and clearing up this matter have been achieved.}.

In general, there are several ways of approaching  the monopole problems.
As known, together with geometrically topological  way  of  exploration into
them,  another  approach   to   studying  such
configurations is possible, namely, that  concerns  any physical
manifestations  of monopoles when they are considered as external  potentials.
Moreover,
from the~physical standpoint, this latter method can  thought  of  as
a~more visualizable one in comparison with  less  obvious  and  more
direct  topological  language.  So,  the~basic   frame   of   our
further investigation  is  the~study of a~particle multiplet
in the~external monopole potentials. Much work
in studying quantum mechanical  particles in the~monopole potentials,
in both the~Abelian and non-Abelian cases, has been done in the~literature;
see, respectively, in [45-48] and [49-54].
For  definiteness, we restrict ourselves to the~simplest doublet case;
taking special attention  to any manifestations
 of just the~Abelian monopole on the non-Abelian background.
Many properties of this system turn out to
be of interest in themselves, producing in their totality an example of
the~theory of what may be called `Abelian monopole
in non-Abelian embodiment'. Generally speaking, this theory is
the~most important aspect of the~present study.

Now, for convenience of the readers, some remarks about the~approach and
technique used in the~work are to be given.
The~primary technical `novelty' is that, in the paper, the~tetrad (generally
relativistic) method [55-63] of Tetrode-Weyl-Fock-Ivanenko (TWFI) for
describing
a~spinor particle will be exploited. So, the~matter equation for an~isotopic
doublet of  spinor particles in the field of the~non-Abelian monopole is taken
in the~form
$$
[i\; \gamma ^{\alpha }(x) \; ( \partial _{\alpha } \; +
\; \Gamma _{\alpha }(x) \; - \;
i\; e\; t^{a} \; W^{(a)}_{\alpha } \; ) \; - \;
 (m \; + \; \kappa \;  \Phi ^{(a)} t^{a}) \; ]\; \Psi (x) = 0  \; .
\eqno(1.1)
$$

\noindent where $\gamma^{\alpha}(x)$ are the~generalized Dirac matrices,
$\Gamma_{\alpha}(x)$ stands for the~bispinor connection; $e$ and $\kappa$ are
certain constants. The choice of the~formalism to deal with
the~monopole-doublet  problem has turned out to be of great  fruitfulness
for examining this system.

Taking of just this method is  not an~accidental step. It is matter that,
as known (but seemingly not very vastly), the~use of a~special spherical tetrad
in the~theory of a~spin $1/2$ particle had led Schr\"odinger and Pauli [64, 65]  to a~basis
of remarkable features. In particular, the~following
explicit expression for (spin $1/2$ particle's) momentum operator components
had been calculated
$$
J_{1}= l_{1} + {{i \sigma ^{12} \cos \phi }\over{ \sin\theta}} ,\qquad
J_{2}= l_{2} + {{i \sigma ^{12} \sin \phi }\over{ \sin\theta}} ,\qquad
J_{3} = l_{3} \eqno(1.2)
$$

\noindent   just that kind of structure for $J_{i}$ typifies this frame in bispinor
space. This Schr\"{o}dinger's basis had been used with great efficiency by Pauli
in his  investigation [65] on   the~problem of allowed spherically
symmetric wave  functions in quantum mechanics.  For our purposes,
just several simple
rules extracted from the~much more comprehensive Pauli's analysis will be
quite   sufficient (those are almost mnemonic working regulations).

They can be explained on the~base of $S=1/2$ particle case.
To this end, using any representation of $\gamma$ matrices where
$\sigma^{12} = {1 \over 2}\;(\sigma_{3} \oplus \sigma_{3})$  (throughout the work,
the Weyl's spinor  frame is used) and taking into account the~explicit form for
$\vec{J}^{2}, J_{3}$ according to (1.2), it is readily verified that the~most
general bispinor functions  with fixed quantum numbers $j,m$ are to be
$$
\Phi _{jm}(t,r,\theta ,\phi ) =
                              \left ( \begin{array}{l}
          f_{1}(t,r) \; D^{j}_{-m,-1/2}(\phi ,\theta ,0) \\
          f_{2}(t,r) \; D^{j}_{-m,+1/2}(\phi ,\theta ,0) \\
          f_{3}(t,r) \; D^{j}_{-m,-1/2}(\phi ,\theta ,0) \\
          f_{4}(t,r) \; D^{j}_{-m,+1/2}(\phi ,\theta ,0)
                               \end{array} \right )
\eqno(1.3)
$$

\noindent where $D^{j}_{mm'}$ designates the Wigner's $D$-functions
(the~notation
and subsequently required formulas, according to [66], are adopted).
One should take notice of the low right indices
$-1/2$ and $+1/2$ of $D$-functions  in (1.3),  which correlate with the~explicit diagonal structure of
the~matrix $\sigma^{12} = {1 \over 2}\; ( \sigma_{3} \oplus \sigma_{3})$.
The~Pauli criterion allows only half integer values for $j$.

So, one may
remember some very primary facts of $D$-functions theory and then produce,
almost automatically, proper wave functions. It  seems rather likely,
that there may exist a~generalized analog of such a~representation
for $J_{i}$-operators, that  might be successfully used whenever
in a~linear problem there exists a~spherical symmetry, irrespective of
the~concrete embodiment of such a~symmetry.  In particular, the~case of
electron in the~external Abelian monopole field, together with
the~ problem of selecting the~allowed wave functions as well as
the~Dirac charge quantization condition,  completely come under
that Shr\"{o}dinger-Pauli method. In particular,  components of the generalized
conserved momentum can be expressed as follows (for more detail, see [67])
$$
j^{eg}_{1} = l_{1} + {{(i\sigma ^{12} - eg) \cos \phi}
\over { \sin \theta }} ,  \qquad
j^{eg}_{2} = l_{2} + {{(i\sigma ^{12} - eg) \sin\phi}
\over {\sin \theta }} ,   \qquad
j^{eg}_{3} = l_{3}
\eqno(1.4)
$$

\noindent where $e$ and $g$  are an~electrical and magnetic charge,
respectively. In accordance with the~above regulations,
the~corresponding wave functions are to be built up as
(also see (4.3a,b))
$$
\Phi^{eg} _{jm}(t,r,\theta ,\phi ) =
                              \left ( \begin{array}{l}
          f_{1}(t,r) \; D^{j}_{-m,eg-1/2}(\phi ,\theta ,0) \\
          f_{2}(t,r) \; D^{j}_{-m,eg+1/2}(\phi ,\theta ,0) \\
          f_{3}(t,r) \; D^{j}_{-m,eg-1/2}(\phi ,\theta ,0) \\
          f_{4}(t,r) \; D^{j}_{-m,eg+1/2}(\phi ,\theta ,0)
                               \end{array} \right )    \; .
\eqno(1.5)
$$

\noindent The~Pauli criterion produces two results: first,
$\mid eg \mid = 0, 1/2, 1, 3/2,\ldots $ (what is called the~Dirac charge
quantization condition; second, the quantum number $j$
in (1.5) may take the values $\mid eg \mid -1/2 , \mid eg \mid +1/2,
\mid eg \mid +3/2, \ldots$ that  selects the~proper spinor particle-monopole
functions.

So, it seems rather a~natural step: to try exploiting some generalized
Schr\"{o}dinger's basis at analyzing the~problem of $SU(2)$
multiplet  in the~non-Abelian monopole field, if by no reason than
curiosity or  search of some  points of unification.

There exists else one line justified the~interest to just the~aforementioned
approach: the~Shr\"{o}dinger's tetrad basis and Wigner's $D$-functions are
deeply  connected with what is called the~formalism of spin-weight harmonics
[68-70]  developed in the~frame of the~Newman-Penrose method of light
(or isotropic))  tetrad. Some relationships between spin-weight and spinor
monopole harmonics have been examined in the literature [71-73],
 the~present work   follows the~notation used in [67].

There is an~additional reason for  special attention just to
the~Scr\"{o}dinger's  basis on the~background of non-Abelian monopole matter.
As will be seen subsequently,  that basis  can be associated  with the~unitary
isotopic gauge in the~non-Abelian monopole problem\footnote{There is  something
rather enigmatic in such a~relation between those, apparently not-touching
each other,  matters. In the~same time, those always will be attractive points
for theoreticians}; in Sec.2 the latter fact will be discussed in full detail.

Thus, the present work is, in its working mathematical language, somewhat
reminiscent in contrast to prodigious  modern investigations based on
the~abstruse geometrical theory underlying the monopole matter; it exploits rather conventional
(if not ancient)  mathematical and physical methods and  intends to trace some
unifying  points among them (tetrad approach, Wigner's functions, selection of
allowable wave function, non-Abelian theories, and monopole area).
After all these  opening  and  general  statements,
some more concrete  remarks  referring  to  our  further
work and designated to delineate its content are to be given.

  Sec.2 determines explicitly  all the technical facts necessary to follow
the~subsequent content of the  work in full detail; in that sense, it plays
a~subsidiary role.  Here, in the~first place, the~question of intrinsic
structure  of the~t'Hooft-Polyakov potentials
$(\Phi^{a}(x), W^{a}_{\beta})$
is reexamined (to be more exact, the~dyon ansatz of Julia-Zee
[74] is taken initially). The~main  guideline consists in  the~following:
It is well-known that the usual Abelian monopole potential
generates a~certain non-Abelian potential being a~solution of
the~Yang-Mills ($Y-M$) equations. First, such a~specific non-Abelian
solution was found out in [75].  A~procedure itself of that
embedding the~Abelian 4-vector $A_{\mu }(x)$
in the non-Abelian scheme:
$  A_{\mu }(x) \rightarrow  A^{(a)}_{\mu }(x)  \equiv
   ( 0 , 0 , A^{(3)}_{\mu }$ = $ A_{\mu }(x) )$
ensures automatically that   $A^{(a)}_{\mu }(x)$    will satisfy
the free $Y-M$ equations. Thus, it may be readily verified that the
vector
         $A_{\mu }(x) = (0, 0, 0, A_{\phi } = g\cos\theta )$
obeys the Maxwell general covariant equations in every curved
space-time  with  the~spherical symmetry:
$$
dS^{2}=[e^{2\nu }(dt)^{2} - e^{2\mu }(dr)^{2} -
r^{2} ((d\theta )^{2} +\sin^{2}\theta  (d\phi )^{2}) ] \; ;
A_{\phi } =  g\cos\theta  \rightarrow
F_{\theta \phi } = - g \sin\theta;
$$

\noindent here we will face a~single equation that coincides
with the~Abelian one. In turn, the~non-Abelian tensor
$F^{(a)}_{\mu \nu }(x)$ defined by
$
\; F^{(a)}_{\mu \nu }(x) =  \nabla _{\mu } \; A^{(a)}_{\nu } \; - \;
\nabla _{\nu }\; A^{(a)}_{\mu } \; +\; e\;  \epsilon _{abc} \; A^{(b)}_{\mu }
A^{(c)}_{\nu }\;$
and associated with the $A^{(a)}_{\mu }$ above  has a~very  simple  isotopic
structure: $ F^{(3)}_{\theta \phi } = - g \sin\theta $   and  all
other $F^{(a)}_{\nu \mu }$ are equal to zero. So, this substitution
$
F^{(a)}_{\nu \mu } =
( 0 , 0, F^{(3)}_{\theta \phi } = - g \sin \theta)
$
leads the~$Y-M$ equations to the~single  equation of  the~Abelian case.
Therefore, strictly speaking, we cannot state that $A^{(a)}_{\mu }(x)$
obeys a~certain set of really nonlinear equations (it satisfies linear
rather than nonlinear ones). Thus, this potential
may be interpreted as a~trivially non-Abelian solution  of $Y-M$
equations. Supposing that such a~sub-potential is presented in the
well-known monopole solutions:
$$
 \Phi ^{(a)}(x) = x^{a} \Phi (r) , \qquad
 W^{(a)}_{0}(x) = x^{a} F(r)     , \qquad
 W^{(a)}_{i}(x) = \epsilon_ {iab} x^{b} K(r)
\eqno(1.6)
$$

\noindent we can try to establish explicitly that constituent structure.
The use of the~spherical coordinates and special gauge transformation
enables us to separate the~trivial and non-trivial
(in other terms, Abelian and genuinely non-Abelian) parts of the~potentials
(1.6) into  different isotopic components
(see the formula (2.7b) ). In the~process of this  rearrangement of
the monopole's constituents, heuristically useful concepts of three gauges:
 Cartesian (associated with
the~representation (1.6)), Dirac and Schwinger's (both latter are unitary
ones) in isotopic  space are defined.  In  order  to  avoid  possible
confusion with all the~different bases we  will  label  the~functions  and
operators by signs of used gauges. The abbreviations  $S.,  D.,  C. $
will be associated, respectively,  with the~Schwinger, Dirac, and Cartesian
gauges  in the~isotopic space, whereas the~abbreviation  {\it sph.} and
{\it Cart.} are referred to
the~spherical and Cartesian  tetrads, respectively\footnote{Below, for
simplicity, those latter are often omitted.}.

Also, in Sec.2, we briefly review several, the~most important for our work,
facts concerning the~generally relativistic Dirac's equation.
Besides, for
convenience of the~readers, some information about the~aforementioned
Pauli's criterion is given.

Sec.3 begins  analyzing  the~doublet-monopole problem. Starting from
Schwinger unitary gauge
$(\;\Phi_{(a)\beta}^{S.},\; W_{(a)\beta}^{S.}\;)$
and the~spherical
tetrad-based matter equation (1.1), the problem of obtaining and partly
analyzing the relevant radial equations is studied here all over again.
At the~correlated choice of frames in both Lorentzian and isotopic space,
an explicit form of the total momentum operator\footnote{Another essential feature of the~given
frame is the~appearance of  a~very simple expression for
the~term which mixes up together two distinct  components  of  the
isotopic doublet (see (3.1)). Moreover, it is evident at once that
both these  features  will be retained,   with   no    substantial
variations, when generalizing this particular problem  to  more
complex ones with other fixed Lorentzian or isotopic spin.}:
$$
J^{S.}_{1} = l_{1} + {{(i \sigma ^{12} + t^{3}) \cos\phi}
                       \over {\sin \theta }}  ,            \qquad
J^{S.}_{2} = l_{2} + {{(i \sigma ^{12} + t^{3}) \sin\phi}
                       \over {\sin \theta }} ,              \qquad
J^{S.}_{3}= l_{3}
\eqno(1.7)
$$

\noindent  characterizes this particular composite frame
as of Schr\''{o}dinger's type; so, the~above general technique is quite applicable.
The $(\theta, \phi)$-dependence in the~relevant wave function
is described by $D$-functions of three kinds: $D^{j}_{-m,m'}, \; m' =
0, -1, +1$ (also see in (3.3)):
$$
\Psi _{\epsilon jm}(x)\; = \; {{e^{-i\epsilon t} } \over r} \;
 [\;  T_{+1/2} \otimes  F(r,\theta,\phi) \;  +  \;
 T_{-1/2} \otimes  G(r,\theta,\phi)\; ] ,
$$
$$
F  = \left ( \begin{array}{l}
                     f_{1}(r) D_{-1} \\
                     f_{2}(r) D_{\; 0} \\
                     f_{3}(r) D_{-1} \\
                     f_{4}(r) D_{\; 0}
                    \end{array} \right )  , \;\;
G  = \left ( \begin{array}{l}
                     g_{1}(r) D_{\; 0} \\
                     g_{2}(r) D_{+1} \\
                     g_{3}(r) D_{\; 0} \\
                     g_{4}(r) D_{+1}
                    \end{array} \right )  \; ,
T_{+1/2}= \pmatrix{1 \cr 0} \qquad ,\;\;  T_{-1/2} = \pmatrix{0 \cr 1} \; .
\eqno(1.8)
$$

\noindent $D_{\sigma} \equiv D_{-m, \sigma}^{j}(\phi, \theta, 0)$.
The~Pauli criterion (see sec.2) allows here
all positive integer values for $j : j = 0, 1, 2, 3,\ldots$

\noindent The separation of variables in the equation  is  accomplished
by the~conventional $D$-function recursive relation techniques.
Moreover, it turns out that only two  relationships  from  the~enormous
$D$-function apparatus are really needed in doing this separation [66]:
$$
{{\partial} \over {\partial \beta }} \; D^{j}_{mm'}(\alpha ,\beta ,\gamma ) =
+{1\over 2} \; \sqrt {(j+m')(j-m'+1)} \;  e^{-i\gamma } D^{j}_{m,m'-1} \; -
$$
$$
 {1\over 2} \; \sqrt {(j-m')(j+m'+1)} \; e^{+i\gamma } D^{j}_{m,m'+1} \;\; ;
$$
$$
{{m-m' \cos \theta} \over {\sin \theta}} D^{j}_{mm'}(\alpha ,\beta ,\gamma ) =
-{1\over 2} \; \sqrt {(j+m')(j-m'+1)} \; e^{-i\gamma } D^{j}_{m,m'-1} \; -
$$
$$
 {1\over 2}  \sqrt {(j-m')(j+m'+1)}\; e^{+i\gamma } D^{j}_{m,m'+1} \; .
\eqno(1.9)
$$

As known, an important case in theoretical  investigation  is
the electron-monopole system at the minimal value of  the  quantum
number $j$; so, the~case $j = 0$  should  be  considered  especially
carefully, and we  do  this.  In  the~chosen  frame,  it  is
the~independence on $\theta ,\phi $-variables that  sets  the~wave
functions  of minimal $j$  apart  from  all  other  particle
multiplet states
(certainly, functions $f_{1}(r)$ , $ f_{3}(r)$ , $ g_{2}(r) $ ,  $g_{4}(r)$
in the substitution (1.8) must be equated to zero at once).
Correspondingly, the~relevant  angular  term in  the~wave equation
will be  effectively eliminated.

The~system of radial equations found by separation of
variables ($4$ and $8$ equations in the cases of $j = 0$ and $j > 0$,
respectively) are rather complicated.
They are simplified by searching a~suitable
operator that could be diagonalized simultaneously with
$ \vec J ^{2} , J_{3}$.
The~usual  space  reflection ($P$-inversion) operator
for a~bispinor doublet field has to be followed by a~certain discrete
transformation in the~isotopic space, so that  a~required quantity
could be constructed. The~solution of this problem which
has been established to date (see, fore example, in  [1, 11-16, 76-85])
 is not general
as much as possible. For this reason, the~question of reflection symmetry  in
the~doublet-monopole system is reexamined here all over again.
As a~result we find out\footnote{And this is a~crucial moment in subsequent
construction of the~present work.}  that there are two different
possibilities depending on what type of external monopole  potential
is taken. So, in case of the~non-trivial potential, the~composite
reflection operator with required properties is
(apart from an~arbitrary numerical factor)
$$
\hat{N}^{S.} \; = \;  \hat{\pi } \otimes \hat{P}_{bisp.} \otimes  \hat{P} ,
\qquad   \hat{\pi } = + \sigma _{1}
\eqno(1.10)
$$

\noindent here, the quantities  $ \hat{\pi }$ and $\hat{P}_{bisp.}$ represent
fixed matrices acting in the isotopic and bispinor space, respectively,
and changing simultaneously with any variations of relevant bases
(see (3.8a)). A~totally  different situation occurs in case of the~simplest
monopole potential.  Now, a~possible additional operator, suitable
for separating the~variables, depends on an~arbitrary
complex numerical parameter $A$ ($\Delta = e^{iA}\; , \;
 e^{iA} \neq  0 , \infty $ ):
$$
\hat{N}^{S.}_{A} = \hat{\pi }_{A} \otimes \hat{P}_{bisp.} \otimes
\hat{P} , \qquad     \hat{\pi }_{A} = e^{iA \sigma _{3}} \sigma _{1} \; .
\eqno{1.11a}
$$

\noindent The same quantity  $A$ appears  also  in  expressions
for the~corresponding wave functions
$ \Psi ^{A}_{\epsilon jm\delta}(t,r,\theta ,\phi)$
(the eigenvalues $ N_{A} = \delta  (-1)^{j+1} ; \delta  = \pm  1$) :
$$
\Psi ^{A}_{\epsilon jm\delta}(x) = [\; T_{+1/2} \otimes  F(x) \; + \;
                       \delta\; e^{iA}\; T_{-1/2} \otimes  G(x) \; ] \;
\eqno(1.11b)
$$

\noindent the additional limitations (3.10a) are imposed on the~radial
functions in $F$ and $G$. Further, throughout all the sections 4-11, we look into the~fermion doublet
just in this simplest monopole field, and all results  and discussion  concern only
this particular system unless the~inverse is indicated.

Sec.4 , in the~first place,  finished the~work on searching a~remaining
operator from a~supposedly  complete set:
$\{\; \hat{H}, \;\vec{J}^{2},\; J_{3}, \hat{N}_{A}, \; \hat{K} = ?  \}$ .
That $\hat{K} $  is determined as a~natural
extension of the~well-known (Abelian) Dirac operator to the~non-Abelian
case. Correspondingly, the~set of  radial equations is
eventually reduced to a~set of two ones; the~latter is well known  and
coincides with that relating to the~Abelian electron-monopole system; which
has  been studied by many authors.
Then, on simple comparing the~non-Abelian doublet functions with
the~Abelian ones, we arrive at an~explicit factorization of the~doublet
functions by Abelian ones and isotopic basis vectors (see (4.4)).
The~relevant decompositions have been found  for the~composite states
with all values of $J$, including the~minimal
one $J_{min.}=0$ too. As known, the~case of minimal $J$ in the~Abelian theory
supplies some unexpected and rather singular features: in particular,
it gives a~candidate for a~possible bound state in the~electron-monopole
system, it significantly touches the~Hamiltonian self-conjugacy property
and some others. Thus, as evidenced by the~factorization,
all those purely Abelian peculiarities, concerning the~$J_{min.}$-state,
likewise turn out to be represented, in a~practically unchanged form, in
the~non~Abelian theory (it should be
remembered that here the~case of special non-Abelian monopole field is meant).

Else one fact associated with the~above decomposition should be noted.
It is matter that the~$A$-ambiguity in determining~the discrete operator
 $\hat{N}_{A}$ ranges from zero to infinity:  $0 < \mid e^{A} \mid < \infty $.
Therefore, the~two distinct
isotopic doublet components (see (1.11b)), proportional respectively to
$T_{-1/2}$ and $T_{+1/2}$, cannot be eigenfunctions of the
$\hat{N}_{A}$  whatever  the~values of $A$-parameter may be.
In other words,  the~above  $\hat{N}_{A}$ is to be considered as
a~specifically non-Abelian operator, and
the~parameter $e^{iA}$  itself  may  be  regarded  as
a~quantity  measuring  violation  of   Abelicity   in the~composite
non-Abelian  wave  function.
Of course, these two  Abelian-like
doublet states can formally  be obtained  from (1.11b)  too:  it
suffices to put $e^{iA} = 0$ or $ \infty$, but those  singular  cases  are  not
covered by the~above-mentioned complete set of operators. In that sense,
the~two bound values  $A =   0$  and  $\infty$ represent singular transition
points   between th~Abelian and non-Abelian theories.

Sec.~5 concerns distinctions between Abelian and non-Abelian monopoles.
We consider the~question: what in the~stated above (Sections 2-4) is dictated
solely by presence of the~external field and what
is determined, in turn, only  by isotopic multiplet's structure.
To this end, we compare the~fermion doublet multiplet, being subjected to
the~monopole effect,  with a~free one. We draw attention to  the~fact that
these two systems   have their  spherical
symmetry  operators $\vec{J}^{2}, J_{3} , \hat{N}$  identically coincided.
Correspondingly, the relevant wave functions
do  not  vary  at  all  in  their dependence on angular variables
 $\theta , \phi $; instead, a~single difference appears in one  parametric
function entering the~systems of radial equations.
These  non-Abelian  wave
functions' property sharply contrasts with the Abelian one. Indeed, as well
known, particle's  wave functions and all
spherical symmetry operators undergo  substantial changes
(see (1.4) and (1.5) as the~external monopole field is in effect.
To clarify  and spell out all the~significance of such
a~`minor' alteration in (1.5) as the simple displacement in a~single index,
we look at just one mathematical  characteristic of the~$D$-functions involved
in the~particle wave functions: namely, their  boundary
properties  at the~points $\theta = 0$ and $\theta = \pi$
 (see Tables 1-3 in Sec.4). On comparing those characteristics
for  $D^{j}_{-m, \pm1/2}(\phi, \theta,0)$ and
$D^{j}_{-m, eg \pm1/2}(\phi, \theta,0)$
we can conclude
that these sets of $D$-functions provide us with the~bases in
different  functional spaces $\{ F^{eg=0}(\theta,\phi) \}$ and
$\{ F^{eg \neq 0}(\theta,\phi) \}$. Every  of those functional
spaces  is characterized by its own behavior at limiting points, which is
irreconcilable with that of any other space.

So, from the very~beginning, in the~Abelian monopole situation, we face
a~fact being crucial one by its  further implications: the~space of
quantum-mechanical states of a~particle in the~monopole field is quite
a~contrasting one to a~free particle's space.

This circumstance implies
a~lot of hampering implications. In particular, we discuss some relations
 of them to   quantum-mechanical superposition principle. Also, else one
awkward question of that kind is: what is the~meaning
of the~relevant scattering theory, if even at infinity itself, some
manifestation of the~magnetic charge  presence does not vanish [17-19,86-94]
(just because  of the~given $\theta, \phi$-dependence).

By contrast, in the non-Abelian theory, there not exist any problems of
such a~kind. Even more, we may  state   that   one  of
the~substantial features characterizing the~non-Abelian
monopole is that such a~potential, does  not  destroy
the~isotopic angular structure  of  the~particle  multiplet.
From this point of view, this potential  represents a~certain analog
of a~spherically symmetrical  Abelian potential
$A_{\mu }  = ( A_{0}(r), 0,0,0,)$
rather than of the~Abelian {\it monopole} one. In this connection,
 one additional remark might be useful:
one should give attention to the~fact  that
the~designation {\it monopole} in the~non-Abelian  terminology,
anticipates tacitly interpretation of $W^{a}_{\mu }(x)$  as
ones carrying, in a~new      situation,  the~essence  of  the~well-known
Abelian   monopole, although really, as evidenced by the~above arguments
and some other, the~real degree of their similarity may be probably less
than one might expect. Also, the following point should
be stressed. Though as was mentioned above, certain close relationships
between the~non-Abelian doublet wave functions and Abelian fermion-monopole
functions occur (see the formulas (4.4a,b)), the~non-Abelian
situation, in reality, is intrinsically non-monopole-like ($=$ non-singular
one). The following aspect is meant: in the~non-Abelian case, the~totality
of possible transformations (upon the~relevant wave functions) which bear
the~gauge status are materially different from ones that there are in effect
in the~purely Abelian theory. As a~consequence of this, the~non-Abelian fermion
doublet wave functions (1.8) can be readily transformed, by carrying
out together the~gauge transformations in Lorentzian and isotopic spaces
($S. \; \rightarrow \; C.\;$ and $\;sph. \; \rightarrow \; Cart.$),
into the form (see the formulas (A.9) and (A.10)) where they will be
single-valued  functions of spatial points.  In the~Abelian monopole situation,
the~representation for particle-monopole functions can by no means
be translated  to any single-valued one.

So, in a~sense, the~whole multiplicity of Abelian monopole manifestations
seems to  be much more problematical than non-Abelian monopole's.
Furthermore, as it appears to be likely, examination of the non-Abelian
case does not lead up to solving some purely Abelian problems. These two
mathematical and physical theories should be only associated heuristically.

As else one confirmation to this general view, in Sec.5 , we will compare
two different situations  relating to discrete symmetry problems
tied up respectively  with the~Abelian and non-Abelian models (see also in
[1,11-16,76-85]).
We explain carefully how the~Abelian $P$-symmetry problem (to be exact,
its violation  by a~magnetic charge) is embedded in the~non-Abelian
model and the~way Abelian $P$-violation results in the~ discrete composite
symmetry in the~non-Abelian theory. The~main ideas are as follows:
In virtue of the~well-known Abelian monopole
$P$-violation, the usual bispinor particle $P$-inversion operator
$\hat{P}_{bisp.} \otimes \hat{P}$
does not commute with the Hamiltonian $\hat{H}^{eg}$. The way of how to obtain
 a~certain formal covariance of the~monopole-containing system  with
respect to $P$-symmetry there has been a~subject of special interest in
the~literature.
All the suggestions represent, in the essence, a~single one:
the~magnetic charge characteristic is to be considered as a~pseudo scalar
quantity\footnote{One should take into account that this, as it is,
applies only to the~Schwinger basis; the~use of the~Dirac gauge or any other,
except Wu-Yang's, implies quite definite modifications  in representation of
the~$P$-operation on the monopole 4-potential; see, for example, in [78]}.
For the~subject under consideration, this assumption implies that
one ought to accompany the~ordinary $P$-transformation with a~formal
operator  $\hat{\pi}$ changing the~parameter $g$ into $-g$.
Correspondingly, the~composite Abelian discrete operator
$$
\hat{M} = \hat{\pi}_{Abel.} \otimes \hat{P}_{bisp.} \otimes \hat{P}
$$

\noindent will commute with the~relevant Hamiltonian indeed. Besides, this
$\hat{M}$ can be diagonalized\footnote{It  should be emphasized that
some unexpected peculiarities with that procedure, in reality, occur as
we turn to the states of minimal values of $j$; for more detail see in [67]}
 on the~functions
 $\Psi^{eg}: \; \hat{M} \; \Psi^{eg}_{\epsilon jm \delta} =
\delta \; (-1)^{j+1} \; \Psi^{eg}_{\epsilon jm \delta}$.
However, as evidenced in Sec.5, this operator $\hat{M}$ does not result in
a~basic structural condition
$$
\Phi (t,-\vec{x}) \; = \;
( 4\times 4-matrix) \;  \Phi (t,\vec{x})
\eqno(1.12a)
$$

\noindent which would  guarantee indeed the~existence of certain
selection rules with respect to a~discrete operator. In place of (1.12a), there
exists just the~following  one
$$
\Phi^{+eg}_{\epsilon jm \delta } (t,-\vec{x}) \; = \;
\delta (-1)^{j+1} \; \hat{P}_{bisp.}\; \Phi ^{-eg}_{\epsilon jm \delta }(t,\vec{x})
\eqno(1.12b)
$$
\noindent take notice of change in the~sign at  $eg$ parameter; this minor
alteration\footnote{This observation can be conceptualized in more formal
mathematical terms: this inversion-like operator $\hat{M}$ turns out to be
a~non-self-adjoint one; therefore it might not follow all the familiar
patterns of behavior for self-adjoint quantities.} is completely detrimental
to the~possibility of producing any selection rules: those do not exist
whatever. Therefore, no discrete symmetry-based  selection rules
in presence of Abelian  monopole are  possible; those will be really
achieved  only if any relation with
the~general structure ((1.12a), apart from modification due to the~type
of particle, which would influence a~matrix involved) exists.

However, a~relation of required structure
there occurs in the~non-Abelian model:
$$
\Psi_{\epsilon jm \delta } (t,-\vec{x}) \; = \;
\delta (-1)^{j+1} (\sigma^{2} \otimes \hat{P}_{bisp.}) \;
 \Psi_{\epsilon jm \delta }(t,\vec{x})
\eqno(1.12c)
$$
\noindent Correspondingly, the~relevant selection rules with respect to
 that composite $N$-parity can be established, and we did it.

Sec.6 concerns some technical details related to explicit expressions of
the~doublet wave functions and  discrete operator $\hat{N}_{A}$ in the two
other gauges: Dirac's unitary and Cartesian ones\footnote{Among other thing,
this material is designated to ease understanding in full for readers
preferring the~use of these gauges.}. The following expression for
the~above non-Abelian $\hat{\pi}_{A}$-operation, now referring to the~Cartesian
gauge, has been calculated:
$\hat{\pi }^{C.}_{A} = (- i ) \;
\exp ( i A \vec{\sigma} \vec{n}_{\theta ,\phi } )$ , where
$\vec{n}_{\theta ,\phi }$ stands for the ordinary radial unit vector.
If $A=0$ then the~form of the operator $N_{A=0}^{C.}$  (it does not involve
any isotopic transformation) might be a~source of some speculation about
an~extremely significant role of the~Abelian $P$-symmetry  in the~non-Abelian
model. Some subtle considerations  related to this matter are discussed.
In particular, it should  be remembered that a~genuinely Abelian fermion
$P$-symmetry implies both a~definite explicit expression for $P$-operation
and definite properties of  the~corresponding wave functions.
In this connection,  the~relevant decomposition of fermion doublet wave
functions in terms of Abelian fermion-like functions and unit isotopic
vectors is given. From   that it is evident that
the~usual Abelian fermion wave functions
and non-Abelian doublet ones belong to substantially  different types
(see (6.9) at $A=0$), so that the~$P$-inversion operator plays only
 a~subsidiary role in forming the~composite  functions.

Sec.7 turns to the question of how the above complex parameter $A$  can
manifests   itself  physically in matrix elements.  On that line, as a~natural
illustration,  the problem of parity selection rules is looked into again,
but now depending on the $A$-background. The~notions of
a~composite $N_{A}$-scalar and $N_{A}$-pseudoscalar   related  to some
quantities with non-trivial isotopic structure are given; then
the~corresponding selection rules are found.
For every fixed $A$,  these rules imply   their own  special
limitations on composite  scalars  and  pseudoscalars, which  are
individualized by this  $A$;  correspondingly, selection   rules
arising in sequel for matrix elements (if a quantity belongs
to  the~scalars or pseudoscalars)  differ basically from each other.

Sec.8 is interested in the~question: Where does the above $A$-ambiguity
come from?  As shown, the~origin of such a~freedom lies in the~existence of
an~additional (one parametric) operation $U(A)$ that leaves the~doublet-monopole
Hamiltinian invariant. Just this operation $U(A)$ changes $\hat{N}_{A=0}$
into   $\hat{N}_{A}$.
Different values for $A$ lead to the~same whole functional
space; each fixed $A$ governs only the basis states
$\Psi^{A} _{\epsilon jm\delta }(x)$ of it, and the~symmetry operation acts
transitively on those states: $\Psi ^{A'}_{\epsilon jm\delta}(x) =
U(A'-A) \Psi ^{A}_{\epsilon jm\delta }(x)$. An analogy between that isotopic
symmetry and a~more familiar
example of Abelian chiral ($\gamma ^{5}$) symmetry in massless Dirac
field theory [95,96] is drawn\footnote{So, the~author suggests  the~term
`isotopic chiral symmetry'. Besides, to be terminologically exact, one should
split the~notion of $\gamma^{5}$ complex chiral symmetry into two ones; properly
$\gamma^{5}$-symmetry (when $A$ is real number) and conformal symmetry (when
$A$ is purely imaginary number); but for simplicity, we will use the~term
`complex chiral symmetry'.}. The~role of the~Abelian
$\gamma^{5}$-matrix is taken by the~isotopic $\sigma_{3}$-matrix: its form
in the $S.$-gauge is $U^{S.}(A) = exp(A/2) \; exp(i {A \over 2} \sigma_{3})$.
Some additional technical details touching this operation
are given; in particular, we find expressions for $U(A)$
in the~Cartesian gauge.
 In Cartesian  frame, this symmetry transformation takes the~form
$$
U_{C.}(A) = e^{+iA/2}
\exp  [\; -i\; {A\over 2} \;\vec{\sigma }\; \vec{n}_{\theta ,\phi }\; ]
\eqno(1.13)
$$

\noindent where the~second factor represents a~2-spinor
local transformation from  the~ 3-dimensional complex  rotation   group
$SO(3.C)$.  The~explicit coordinate dependence  appearing in Cartesian gauge
results from the~non-commutation $\sigma_{3}$ with a~gauge transformation
involved into transition from Shwinger's to Cartesian isotopic basis.
In the~analogous Abelian situation, the form of the~chiral transformation
remains the~same because $\gamma^{5}$ and the~relevant gauge matrix (that
belongs to the~bispinor local representation of the group $SL(2.C)$) are
commutative with each other.

In Sec.9, we look into some qualitative peculiarities of the~$A$-freedom
placing special notice  to the~division of $A$-s values into the real and
complex ones. All those  values  (complex as well as real) are equally
permissible: they only govern bases ot the~same Hilbert space of quantum
states, which can be related to each other by the use of the ordinary
superposition principle. However, a~material distinction  between real and
complex $A$-s will appear, if one turns to  the~orthogonality properties
of those basis states $\Psi^{A}_{\epsilon jm\delta}(x)$.
As will be seen, at $A^{*} \neq A$, the~states
$\Psi^{A}_{\epsilon jm, -1}(x)$      and
$\Psi^{A}_{\epsilon jm, +1}(x)$
are not orthogonal to each other. Such specific (non-orthogonal) bases,
though not  being of very common use and having
a~number of peculiar features, are allowed to be exploited in conventional
quantum theory.

Else one fact associated with the~above (real-complex) division of $A$-s is that
the~discrete operator $\hat{N}_{A}$ represents a~non-self-adjoint quantity
as  $A^{*} \neq A$. In this point, there are two possibilities to  choose
from: whether we restrict ourselves to the~real $A$-values (correspondingly,
no problems with self-conjugacy there arise) or we exploit the~complex
$A$-values  as well as the~real ones, and thereby, the non-orthogonal bases
and non-self-adjoint character of the~discrete operator,
are allowed in the~theory.
We have chosen to accept and look into the~second possibility.

Further, we  consider narrowly such a~specific nature of the~$\hat{N}_{A}$
since, as the~complex $A$-values are allowed, we will violate the~well-known
quantum-mechanical regulation about the self-adjointness of measurable
physical quantities. The~main guideline ideas in clearing up the~problem faced
us here is as follows.
One should notice the~fact that the~single
relation   $(\hat{N}_{A})^2 = I $ is abundantly sufficient one to produce
real proper values: $+1$ and $-1$. Furthermore, as it was stressed above, just
 {\it real values} are not material here whatever; instead,
the~only required consequence of this symmetry is the~mere distinction
between two different quantum possibilities. In the~light of this,
the~automatical  incorporation of all discrete operators into
class of self-adjoint ones does not seem inevitable. But accepting this, there is
a~problem to solve: what is the meaning of complex expectation values of such
non-self-adjoint operators. We carefully explain how one may
interpret all such complex values as being physically measurable ones.

Finally, we  devote Sec.10 to clearing up else one, and rather important
from the~physical viewpoint, peculiarity of the~doublet-monopole system.
It is matter that if the~parameter $A$ is real one, then the~matrix
translating  $\Psi ^{A=0}_{\epsilon jm\delta }(x)$ into
$\Psi ^{A}_{\epsilon jm\delta }(x)$
coincides (apart from a~phase factor $e^{iA/2}$) with a~matrix lying
in the~group $SU(2)$. However, the~group $SU(2)$ has the~status of gauge one for
the~system under consideration. So, the point of view might be brought to light:
one could claim that the~two functions
$\Psi ^{A}(x)$   and $\Psi ^{A'}(x)$ ,
referring respectively to the~different values $A$ and $A'$,  represent
in reality only the transforms of each other in the sense of $SU(2)$ gauge
theory. And further, as a~direct consequence, one could  insist on
the~impossibility in principle to observe indeed any physical distinctions
between those wave functions. If the~above transformation gets estimated so, then
ultimately one will conclude that the~above $N_{A}$-parity selection rules
(explicitly depended on $A$ which is real in that case) are the~mathematical
fiction only, since the~transformation $U(A)$ is not physically observable.

For answering this question, we carefully follow some interplay  between
the~quantum-mechanical superposition principle and  the~concepts of gauge
and non-gauge symmetries. As will be seen, the general outlook prescribing
to interpret the transformation $U(A)$ as exclusively a~gauge one
contradicts with some basic regulations stemming from
the~superposition principle. Even more, as shown,
starting from  the~exclusively gauge understanding of that transformation,
one can arrive at the~requirement of {\it physical identification} of the~two
states  $\Psi ^{A}_{\epsilon jm\delta =-1}(x)$   and
$\Psi ^{A}_{\epsilon jm\delta = +1}(x)$. However, in a~sense, this is
equivalent to the effective  returning into the~Abelian scheme, that
hardly can be desirable effect.
Nevertheless,  the matrix $U(A)$ belongs to $SU(2)^{gauge}_{loc.}$
(apart from $U(1)$ factor). In order not to reach a~deadlock,
in author's  opinion,
there exists just one and very simple way out of  this  situation, which
consists in the following: The complete symmetry group  of  the~system
under consideration is  of the~form
$
\hat{F}(A) \otimes  SL(2.C)^{loc.}_{gauge}
\otimes  SU(2)^{loc.}_{gauge}
$.
This group, in particular, contains  the~gauge  and  non-gauge
symmetry operations which  both have the~same  mathematical  form
but different physical status.

Finally, in Sec.11, we briefly discuss extension of the~present analysis to
other situations, with different values of isotopic and Lorentzian spin, and
gauge groups. It is argued that some facts discerned
in the~present work for $SU(2)$-model might bear upon similar aspects of
other gauge group-based theories.

In Supplement A, we consider some additional relationships between
explicit forms of the~fermion-monopole functions
in the~bases of spherical and Cartesian tetrads.

\subsection*{2 Dirac and Schwinger gauges in isotopic space}

This  section deals with some representation of the
non-Abelian monopole potential,  which  will  be  the  most
convenient one to formulate and analyze the problem of isotopic
multiplet in this field.
Let us begin describing in detail this matter. The well-known
form  of the monopole solution  introduced  by t'Hooft and Polyakov
([42-44]; see also Julia-Zee [74]) may be taken as a starting point.
The field   $W^{(a)}_{\alpha }$   represents a~covariant vector with
the~usual transformation law
$W^{(a)}_{\beta } = (\partial x^{i} / \partial x^{\beta }) W^{(a)}_{i})$
and our first step  is the~change of variables in 3-space.
Thus, the~given  potentials
$( \Phi ^{(a)}(x), W^{(a)}_{\alpha })$ convert into
$ (\Phi ^{(a)}(x), W^{(a)}_{t}, W^{(a)}_{r}$,
$W^{(a)}_{\theta }, W^{(a)}_{\phi})$.
Our second step is a~special  gauge  transformation  in  the
isotopic space. The~required gauge matrix can be determined  (only
partly) by the~condition
$\;(O_{ab} \Phi ^{b}(x) ) = ( 0 , 0 , r \Phi (r)\; )$.
This equation has a~set of solutions  since the~isotopic
rotation by every angle about the~third axis   $(0, 0, 1)$   will  not
change the~finishing vector $( 0, 0, r \Phi (r) )$. We fix
such  an~ambiguity  by  deciding  in  favor  of   the~simplest
transformation matrix. It will be convenient to utilize  the~known
group $SO(3.R)$ parameterization through the Gibbs  3-vector\footnote{The author highly
recommends the book [97] for many further details
developing  the Gibbs approach to groups
$SO(3.R), SO(3.C), SO_{0}(3.1)$, etc.}
$$
O = O( \vec{c}) =  I + 2 \;{ \vec{c}^{\times } + ( \vec{c}^{\times })^{2} \over
 1 + \vec{c}^{2}} \;\; ,  \qquad
 (\vec{c}^{\times })_{ac} = - \epsilon _{acb} \; c_{b} \; .
\eqno(2.1)
$$

\noindent According to [98], the simplest rotation above is
$\vec{B} = O(\vec{c}) \vec{A} ,\; \vec{c} =
 [\vec{B} \vec{A} ] /  (\vec{A} +\vec{B} ) \vec{A}\;$, therefore,
$$
\hbox{if}\;\;\;\; \vec{A} = r \Phi (r) \;
\vec{n}_{\theta, \phi} \;  , \;
\vec{B} = r \Phi (r) ( 0 , 0 , 1 )\; , \; \;
\hbox{then}  \;\;
\vec{c} = {{\sin \theta }\over {1 + \cos \theta }} \;
( + \sin \phi  , - \cos \phi  , 0 )    \; .
\eqno(2.2)
$$

\noindent Together with  varying the~scalar  field $\Phi ^{a}(x)$, the~vector
triplet $W^{(a)}_{\beta }(x)$  is to be transformed from one isotopic
gauge  to another under the~law  [99]
$$
W'^{(a)}_{\alpha }(x)\; = \; O_{ab}(\vec{c}(x)) \; W^{(b)}_{\alpha }(x) \; + \;
{1\over e} \; f_{ab}( \vec{c}(x))\; {{\partial c_{b} } \over
 { \partial x^{\alpha }}} \;\; ,  \qquad
f(\vec{c}) = - 2\; {{1 + \vec{c}^{\times }} \over
{1 + \vec{c}^2}} \;\; .
\eqno(2.3)
$$

\noindent With  the  use  of (2.3), we obtain  the~new
representation
$$
\Phi^{D.(a)}= r \Phi (r)
\left ( \begin{array}{c}
   0 \\ 0 \\ 1
\end{array} \right )\; , \qquad
              W^{D.(a)}_{\theta } = (r^{2} K + 1/e)
\left ( \begin{array}{c}
              - \sin \phi \\
              + \cos \phi  \\
                0
\end{array} \right ) \; ,  \qquad
W^{D.(a)}_{r} = \pmatrix{0\cr0\cr0}    \; ,
$$
$$
W^{D.(a)}_{t} =
\left ( \begin{array}{c}
              0  \\
              0  \\
              rF(r)
\end{array} \right ) \; , \qquad
                          W^{D.(a)}_{\phi } =
                        \left ( \begin{array}{c}
-(r^{2}K + 1/e) \sin\theta \cos\phi \\
-(r^{2}K + 1/e) \sin\theta \sin\phi \\
        {1 \over e} (\cos\theta - 1)
\end{array} \right )   \; .
\eqno(2.4)
$$

\noindent It should be noticed that the~factor $(r^{2} K(r) + 1/ e)$  will
vanish when $K = -1 / e r^{2}$. Thus, only  the~delicate fitting of
the~single proportional   coefficient  (it  must  be
taken as $-1/ e)$  results in the~actual  formal  simplification  of
the~non-Abelian monopole potential.

There exists close connection between $W^{D.(a)}_{\phi }$  from (2.4) and
the~Dirac's expression for the~Abelian monopole potential
(supposing that $\vec{n} = (0, 0 ,-1)$):
$$
A^{\beta }_{D.} = g \;\left ( \; 0 ,\; {{[\; \vec{n} \; \vec{r}\; ]} \over
{(r + \vec{r} \; \vec{n} )\; r }}\; \right ) \;\; , \qquad or \qquad
A^{D.}_{\phi } = - g \; ( \cos \theta  - 1 ) \; .
\eqno(2.5)
$$

\noindent So, $W^{triv.}_{(a)\alpha }(x)$  from (2.4)
 (produced by setting   $K = - 1 / e r^{2}$)
can be thought of  as  the~result of embedding the~ Abelian
potential (2.5) in the~non-Abelian gauge scheme:
$W^{(a)D.}_{\alpha }(x) \equiv  ( 0 , 0 , A^{D.}_{\alpha }(x))$.
The quantity $W^{(a)D.}_{\alpha }(x)$  labelled with symbol  $D.$
will be  named
after its Abelian counterpart; in other words, this potential will
be treated as relating to the Dirac's non-Abelian   gauge  in
the~isotopic space.

In Abelian  case,  the  Dirac's  potential $A^{D.}_{\alpha }(x)$
 can  be
converted into the~Schwinger form $A^{S.}_{\alpha }(x)$
$$
A^{S.}_{\alpha } = \left ( 0 ,\; g \; {{[\; \vec{r}\; \vec{n}\; ] \;
( \vec{r} \; \vec{n})} \over {(r^{2} \; - \; ( \vec{r}\; \vec{n})^{2})r}} \right ) \; ,
\qquad  or \qquad A^{S.}_{\phi } = g \; \cos \theta
\eqno(2.6)
$$

\noindent by means of the~following transformation
$$
A^{S.}_{\alpha }\; = \;A ^{D.}_{\alpha } \; +  \;
{{\hbar c} \over{ie}} \; S \; {{ \partial} \over { \partial x^{\alpha }}} \; S^{-1},
\qquad
S(x) = \exp  (-i{eg \over \hbar c} \phi  ) \;\; .
$$

\noindent It is  possible  to  draw  an~analogy  between  the~Abelian  and
non-Abelian   models.   That is, we    may introduce the~Schwinger
non-Abelian basis in the~isotopic space:
$$
( \Phi ^{D.(a)}, W^{D.(a)}_{\alpha }) \qquad \stackrel{\vec{c}~'} \rightarrow
 \qquad
 ( \Phi ^{S.(a)}, W^{S.(a)}_{\alpha } ) , \qquad
\vec{c}~' = ( 0 , 0 , - \tan  \phi /2 ) \;  ;
\eqno(2.7a)
$$

\noindent where
$$
O(\vec{c}~') = \pmatrix{ \cos \phi & \sin \phi & 0 \cr
                        -\sin \phi & \cos \phi & 0 \cr
                         0         &   0       & 1 }  \;  .
$$

\noindent Now an~explicit form of the~monopole potential is given by
$$
               W^{S.(a)}_{\theta} =
\left ( \begin{array}{c}
              0  \\( r^{2} K + 1/e ) \\ 0
\end{array} \right ) \; , \qquad
 W^{S.(a)}_{\phi } =
\left ( \begin{array}{c}
             -(r^{2} K + 1/e)  \\ 0  \\ {1\over e} \cos \theta
\end{array} \right )\; ,
$$
$$
W^{S.(a)}_{r} = \pmatrix{0\cr0\cr0}, \qquad
W^{S.(a)}_{t} = \pmatrix{ 0 \cr 0 \cr rF(r)} , \qquad
 \Phi ^{S.(a)} = \pmatrix{0 \cr 0 \cr r\Phi (r)}
\eqno(2.7b)
$$

\noindent where the symbol  $S.$  stands for the~Schwinger gauge.

Both $D.$- and $S.$-gauges (see (2.4) and (2.7b))  are  unitary
ones in the~isotopic space due to the~respective  scalar  fields
$\Phi ^{D.}_{(a)}(x)$   and $\Phi ^{S.}_{(a)}(x)$  are
 $x_{3}$-unidirectional,  but  one  of  them
(Schwinger's) seems simpler than another (Dirac's).

For the following it  will  be  convenient  to  determine
the~matrix $0(\vec{c}~'')$ relating the~Cartesian   gauge  of
isotopic space with Schwinger's:
$$
O(\vec{c}~'') = O(\vec{c}~') O(\vec{c}) =
\pmatrix{ \cos\theta  \cos\phi & \cos\theta \sin\phi & -\sin\theta \cr
         -\sin\phi             & \cos\phi            &   0         \cr
          \sin\theta  \cos\phi & \sin\theta \sin\phi &  \cos\theta } ,
$$
$$
\vec{c}~'' = (+ \tan\theta /2  \tan\phi /2, - \tan\theta /2, -\tan\phi /2).
\eqno(2.8)
$$

This matrix  $O(\vec{c}~'')$  is also well-known in
other  context  as  a~matrix  linking  Cartesian  and  spherical  tetrads
in  the space-time of special relativity  (as well as in a~curved  space-time
of spherical symmetry)
$$
x^{\alpha } = (x^{0}, x^{1}, x^{2}, x^{3}) \; , \;\;
dS^{2}= [(dx_{0})^{2} - (dx_{1})^{2} - (dx_{2})^{2} - (dx_{3})^{2}] \; , \;\;
e^{\alpha }_{(a)}(x) = \delta ^{\alpha }_{a}
\eqno(2.9a)
$$

\noindent and
$$
x'^{\alpha } = ( t , r , \theta  , \phi  ) \; , \qquad
dS^{2} = [dt^{2} - dr^{2} - r^{2}(d\theta ^{2} +
\sin ^{2} \theta  d\phi ^{2})] \; ,
$$
$$
e^{\alpha'}_{(0)} = ( 1 , 0 , 0 , 0 )\; ,  \qquad
e^{\alpha'}_{(1)} = ( 0, 0 , 1/r, 0 )\; ,
$$
$$
e^{\alpha'}_{(2)} = ( 0 ,0 , 0 ,1/ r \sin \theta)\; ,  \qquad
e^{\alpha'}_{(3)} = ( 0 , 1 , 0 , 0 ) \;\;  .
 \eqno(2.9b)
$$

Below we  review briefly   some   relevant  facts  about
the~tetrad formalism. In  the~presence of an~external  gravitational
field, the~starting Dirac equation
 $(i \gamma ^{a} \partial /\partial x^{a} - m ) \Psi (x) = 0$
is generalized into  [55-63]
$$
[\; i \gamma ^{\alpha}(x) \; (\partial_{\alpha} \; + \;
 \Gamma _{\alpha }(x) ) \;  - \;  m \; ] \; \Psi (x)  = 0
\eqno(2.10)
$$

\noindent where $\gamma ^{\alpha }(x) = \gamma ^{a} e^{\alpha }_{(a)}(x)$, and
$e^{\alpha }_{(a)}(x)$,
 $\Gamma _{\alpha }(x) = {1\over 2}
  \sigma ^{ab} e^{\beta }_{(a)} \nabla _{\alpha }(e^{\alpha }_{(b)\beta })$,
$\nabla _{\alpha }$ stand for a~tetrad, the~bispinor connection, and
the~covariant derivative symbol, respectively. In the~spinor basis:
$$
                      \psi (x)=
\left ( \begin{array}{c}
                  \xi (x) \\  \eta (x)
\end{array} \right ) ,   \qquad
\gamma ^{a} =
\left ( \begin{array}{cc}
                   0  & \bar{\sigma}^{a} \\
                   \sigma^{a}  &       0
\end{array} \right ) , \qquad
\sigma ^{a} = (I ,+ \sigma ^{k}), \qquad
\bar{\sigma }^{a}  = (I, -\sigma^{k})
$$

\noindent ($\sigma ^{k}$ are the two-row Pauli spin matrices; $k = 1,2,3$) we
 have  two equations
$$
i \sigma ^{\alpha }(x) \;[\; \partial_{\alpha} \; + \;
  \Sigma _{\alpha }(x)\; ] \; \xi (x) =  m \;\eta (x),  \qquad
i \bar{\sigma}^{\alpha }(x)\; [\; \partial_{\alpha } \; + \;
 \bar{\Sigma}_{\alpha }(x)\; ] \; \eta (x) = m \; \xi (x)
\eqno(2.11)
$$

\noindent where the~symbols
$\sigma ^{\alpha }(x), \bar{\sigma }^{\alpha }(x),
\Sigma _{\alpha }(x), \bar{\Sigma }_{\alpha }(x)$
 denote respectively
$$
\sigma ^{\alpha }(x)= \sigma ^{a} e^{\alpha }_{(a)}(x),\qquad
\bar{\sigma}^{\alpha }(x)= \bar{\sigma}^{a} e^{\alpha }_{(a)}(x),
$$
$$
\Sigma _{\alpha }(x) =
{1\over 2} \Sigma ^{ab} e^{\beta }_{(a)}
\nabla _{\alpha }(e_{(b)\beta }) ,   \qquad
\bar{\Sigma}_{\alpha }(x) = {1\over 2} \bar{\Sigma}^{ab} e^{\beta }_{(x)}
\nabla _{\alpha }(e_{(b)\beta }) ,
$$
$$
\Sigma ^{ab} = {1\over 4}(\bar{\sigma}^{a} \sigma^{b} -
                          \bar{\sigma}^{b} \sigma^{a}),     \qquad
\bar{\Sigma}^{ab} = {1\over 4} (\sigma^{a} \bar{\sigma}^{b} -
                                \sigma^{b} \bar{\sigma}^{a})  \; .
$$

\noindent Setting $m$ equal to zero, we obtain Weyl equation for neutrino
$\eta (x)$  and anti-neutrino $\xi (x)$, or the~Dirac equation for a~massless
particle (the latter will  be  used  further  in  Sec.8).

The form of equations (2.10), (2.11)  implies  quite definite
their symmetry properties. It is common,  considering  the~Dirac
equation in the~same space-time, to use  some   different  tetrads
$e^{\beta }_{(a)}(x)$ and $e'^{\beta }_{(b)}(x)$, so that  we have
the~equation  (2.10)  and
an~analogous one with a~new tetrad mark.  In  other  words,  together
with  (2.10)  there  exists  an~equation  on $\Psi'(x)$, where
quantities $\gamma'^{\alpha }(x)$ and $\Gamma'_{\alpha}(x)$,
in contrast with $\gamma^{\alpha }(x)$ and $\Gamma_{\alpha}(x)$,
are based on a~new tetrad
$e'^{\beta }_{b)}(x)$ related  to
 $e^{\beta }_{(a)}(x)$  through a~certain local Lorentz matrix
$$
e'^{\beta }_{(b)}(x) \; =  \; L^{\;\;a}_{b}(x) \; e^{\beta }_{(a)}(x) \; .
\eqno(2.12a)
$$

\noindent It may be shown that these two Dirac equations on
 functions $\Psi (x)$ and $\Psi'(x)$
 are related to each other  by  a~definite  bispinor transformation:
$$
\xi'(x)  = B(k(x)) \xi (x), \qquad
\eta'(x) = B^{+}(\bar{k}(x)) \eta (x) \; .
\eqno(2.12b)
$$

\noindent Here, $B(k(x)) = \sigma ^{a} k_{a}(x)$ is a~local matrix
 from the $SL(2.C)$ group;
4-vector $k_{a}$ is the well-known parameter on this group [100].
The~matrix $L^{\;\;a}_{b}(x)$ from (2.12a)  may  be  expressed as a~function of
arguments $k_{a}(x)$  and $k^{*}_{a}(x)$ :
$$
L^{a}_{b}(k, k^{*})  \; = \; \bar{\delta}^{c}_{b} \;
[\; - \delta^{a}_{c} \; k^{n} \; k^{*}_{n} \; + \;  k_{c} \; k^{a*}\;  +  \;
  k^{*}_{c} \; k^{a} \; + \; i \epsilon ^{\;\;anm}_{c}\; k_{n}\; k^{*}_{m} \;]
\eqno(2.12c)
$$

\noindent where  $\bar{\delta}^{c}_{b}$  is a~special Cronecker symbol:
$$
\bar{\delta}^{c}_{b} = 0 \;\;\; if \;\;\;c \neq \; ; \;\;= +1 \;\;\; if \;\; \;
 c = b = 0 \; ; \;\; = -1 \;\;\; if \;\; \;  c = b = 1,2,3 \;\; .
$$

By the way, it is normal practice that some different  tetrads  are
used in examining the~Dirac equation on the~same
Rimannian space-time background. If there is a~need to analyze some
correlation between  solutions in those distinct tetrads,  then  it
is important  to know what are the~relevant  gauge
transformations over the~spinor wave functions.
In particular, the~matrix
relating  spinor  wave  functions  in   Cartesian   and  spherical
tetrads (see (2.9)) is as follows
$$
B = \pm \pmatrix{ \cos\theta /2  \; e^{i\phi  /2} &
                   \sin\theta /2 \; e^{-i\phi /2} \cr
                 -\sin\theta /2  \; e^{i\phi  /2} &
                  \cos\theta /2  \; e^{-i\phi /2} }
\equiv  B( \vec{c}~'') =
\pm {{ I  - i \vec{\sigma } \vec{c}~'' } \over
{ \sqrt{1 -  (\vec{c}~'')^{2} }}}   \; .
\eqno(2.12d)
$$

\noindent The vector matrix $L^{\;\;a}_{b}(\theta ,\phi )$ referring to
the spinor's  $B(\theta ,\phi )$
is the same as $O(\vec{c}~'')$  from (2.8). It is significant that the~two
gauge transformations, arising in quite different contexts, correspond so
 closely  with each other.

This basis of spherical tetrad will play a~substantial
role in our subsequent work.
This Schr\"odinger frame of spherical tetrad [64] was  used  with  great  efficiency  by
Pauli [65] when investigating the~problem of  allowed  spherically
symmetrical wave  functions  in  quantum  mechanics. Below, we briefly review
some results of this investigation.
Let the $J^{\lambda }_{i}$   denote
$$
J_{1}= (\; l_{1} + \lambda\; {{\cos  \phi  }\over {\sin \theta}}\; ),\qquad
J_{2}= (\; l_{2} + \lambda\; {{\sin  \phi  }\over {\sin \theta}}\; ),\qquad J_{3} = l_{3} \; .
$$

\noindent At an~arbitrary $\lambda$, as readily  verified, those $J_{i}$
 satisfy the~commutation rules of the~Lie algebra
$SU(2): [ J_{a},\; J_{b} ] = i \; \epsilon _{abc} \; J_{c}$.
As known,  all  irreducible  representations  of  such  an~ abstract
algebra are determined by a~set of weights
   $j = 0, 1/2, 1, 3/2,... \; \; ({\em dim} \;j = 2j + 1)$.
Given the~explicit expressions of  $J_{a}$   above,  we  will
find functions
               $\Phi ^{\lambda }_{jm}(\theta ,\phi )$
on which the~representation of  weight $j$ is realized. In agreement with
the~generally known method, those solutions are to be established by
the~following relations
$$
J_{+} \; \Phi ^{\lambda }_{jj} \; = \; 0 \;\; , \qquad
\Phi ^{\lambda }_{jm} \; = \; \sqrt{{(j+m)! \over (j-m)! \; (2j)! }} \; J^{(j-m)}_{-} \;
\Phi^{\lambda}_{jj} \;\; ,
\eqno(2.13)
$$
$$
J_ {\pm} \; = \; ( J_{1} \pm i J_{2}) \; = \;
e^{\pm i\phi }\; [\; \pm { \partial \over \partial \theta } \; + \;
 i \cot \theta \; { \partial \over  \partial \phi} \; + \;
 { \lambda \over  \sin  \theta }\; ] \;\; .
$$

\noindent From the equations
$J_{+} \; \Phi ^{\lambda }_{jj} \; = \; 0 \;$ and
$\; J_{3} \; \Phi ^{\lambda }_{jj} \;  = \; j \; \Phi ^{\lambda }_{jj}$,
it follows that
$$
\Phi ^{\lambda }_{jj}  =
N^{\lambda }_{jj} \;  e^{ij\phi} \; \sin^{j}\theta \;\;
{( 1 + \cos \theta  )^{+\lambda /2} \over ( 1 - \cos \theta )^{\lambda /2}} ,\;
N^{\lambda }_{jj}  = {1 \over \sqrt{2\pi}} \; { 1 \over 2^{j} } \;
 \sqrt{{(2j+1) \over \Gamma(j+m+1) \; \Gamma(j-m+1)}} \; .
$$

\noindent Further, employing (2.13) we produce the~functions
$\Phi ^{\lambda }_{jm}$
$$
\Phi ^{\lambda }_{jm} \; = \; N^{\lambda }_{jm} \;  e^{im\phi} \;
 {1 \over \sin^{m}\theta } \;\;
{(1 - \cos \theta)^{\lambda/2} \over (1 + \cos \theta)^{+\lambda/2}} \; \times
$$
$$
({ d \over d \cos  \theta})^{j-m} \; [\; (1 + \cos  \theta ) ^{j + \lambda } \;
(1 - \cos  \theta ) ^{j-\lambda } \; ]
\eqno(2.14)
$$

\noindent where
$$
N^{\lambda }_{jm} \;  = \; {1 \over \sqrt{2\pi} 2^{j}} \;
 \sqrt{{(2j+1) \; (j+m)! \over
2(j-m)!  \Gamma(j + \lambda +1) \; \Gamma(j- \lambda +1)}} \;\; .
$$

\noindent The Pauli criterion tells us that the $(2j + 1)$ functions
$ \Phi ^{\lambda }_{jm}(\theta ,\phi ), \; m = - j,... , +j$
so  constructed, are  guaranteed  to be a~basis for a~finite-dimension
representation, providing that the~functions
$\Phi ^{\lambda }_{j,-j}(\theta ,\phi )$,
found by this procedure,  obey the~identity
$$
J_{-} \;\; \Phi ^{\lambda }_{j,-j} \; = \; 0 \; .
\eqno(2.15a)
$$

\noindent After substituting the~function
 $\Phi ^{\lambda }_{j,-j}(\theta ,\phi )$, the~relation (2.15a)  reads
$$
J_{-} \; \Phi ^{\lambda }_{j,-j} \; = \;  N^{\lambda }_{j,-j} \;
e^{-i(j+1)\phi }\; (\sin \theta)^{j+1}  \;
{(1 - \cos  \theta )^{\lambda /2} \over (1 + \cos  \theta )^{\lambda /2}} \; \times
$$
$$
({d \over d \cos \theta})^{2j+1} \; [\; (1 + \cos  \theta )^{j+\lambda } \;
(1 - \cos  \theta )^{j-\lambda } ) \; ]\;  = \;  0
\eqno(2.15b)
$$

\noindent which in turn gives the~following restriction on
 $j$  and $\lambda $
$$
({d \over d \cos  \theta})^{2j+1} \; [\; (1 + \cos  \theta  )^{j+\lambda } \;
(1 - \cos  \theta  )^{j-\lambda } \; ] \;  = \; 0 \; \; .
\eqno(2.15c)
$$

\noindent But the~relation (2.15c) can be satisfied  only  if  the~factor
 $P(\theta )$,
subjected to the~operation of taking derivative
$( d/d \cos \theta ) ^{2j+1}$,
is a~polynomial of degree $2j$  in $ \cos \theta$. So, we have (as a~result
of the~Pauli criterion)

\vspace{5mm}
1.  {\em the} $\lambda$ {\em is allowed to take values}
$, +1/2,\; -1/2,\; +1,\; -1, \ldots$
\vspace{5mm}

\noindent Besides, as the~latter condition is satisfied,
$P(\theta )$  takes different forms depending on
the $(j , \lambda)$-correlation:
$$
P(\theta ) \; = \; (1 + \cos \theta )^{j+\lambda } \;
  (1 - \cos \theta )^{j - \lambda } \; = \;
P^{2j}(\cos \theta ),\qquad if\qquad j = \mid \lambda \mid,
\mid \lambda \mid +1,...
$$
or
$$
P(\theta ) \; = \;
{ P^{2j+1}(\cos \theta ) \over \sin \theta }, \qquad if \qquad
 j = \mid \lambda \mid +1/2, \mid \lambda \mid +3/2,...
$$

\noindent so that the second necessary condition  resulting from
the~Pauli criterion  is

\vspace{5mm}
2.  {\em given } $\lambda$ {\em according to 1.,
    the number j is allowed to take values}
  $j = \mid \lambda \mid, \mid \lambda \mid +1,...$

\vspace{5mm}
\noindent Hereafter, these two conditions: 1   and  2 will  be termed,
respectively, as  the~first and  the~second   Pauli
consequences\footnote{We draw attention to   that
the  Pauli  criterion
$
J_{-} \Phi _{j,-j}(t,r,\theta ,\phi )\; =\; 0
$
affords the~condition that is invariant relative to possible  gauge
transformations. The function $\Phi _{j,m}(t,r,\theta ,\phi )$
may be subjected  to any gauge transformation. But if  all the~components
 $J_{i}$ vary  in a~corresponding way too, then the
Pauli  condition provides the~same result on $(j,\lambda)$-quantization. In contrast
to this, the common  requirement to be a~single-valued function of spatial
points, often applied to produce a~criterion on selection  of
allowable wave functions in quantum mechanics, is  not
invariant under gauge transformations and can easily be destroyed
by a~suitable gauge one.}.
Also, it should be noted  that  the~angular  variable $\phi $  is  not
affected (charged) by the Pauli criterion; instead,
a~variable that  works above is the~$\theta$. Significantly,  in
the~contrast to this, the~well-known procedure [3-10] of deriving
 the~electric charge quantization  condition  from investigating continuity
properties of quantum mechanical wave functions,  such
a~working variable is the $\phi $.

If the~first and second Pauli consequences fail, then we face
rather unpleasant mathema\-ti\-cal and physical problems\footnote{Reader  is
referred  to  the~Pauli article [65] for more detail about
those peculiarities.}. As a~simple illustration,  we  may  indicate
the~familiar case  when $\lambda= 0$; if
 the~second  Pauli  condition is violated, then we will have the~integer and
half-integer  values  of the~orbital angular momentum number
$l = 0, 1/2, 1, 3/2,\ldots\;$

As regards  the~Dirac  electron  with  the  components  of  the  total
angular momentum in the~form (1.2),
we have to employ the~above Pauli  criterion in  the~constituent
form owing to $\lambda $  changed into $\Sigma _{3}$.
Ultimately, we obtain the~allowable set $J = 1/2, 3/2, \ldots$.

A~fact of primary  practical importance to  us  is  that  the~functions
$\Phi ^{\lambda }_{jm}(\theta, \phi )$ constructed  above  relate
directly  to  the~known Wigner $D$-functions [66]:
$
\Phi ^{\lambda }_{jm}(\theta , \phi ) \; =  \;
(-1)^{j-m} \; D^{j}_{-m, \lambda}(\phi, \theta, 0)
$.

\subsection*{3. Separation of variables and
                 a~composite inversion operator}

We will utilize  the general relativity  covariant  formalism
when a~fundamental  {\em Dirac} equation is (1.1).
In the~spherical tetrad basis (2.9b) and  the~Schwinger unitary gauge of
the~monopole  potentials   (2.7b),
the~matter equation (1.1)  takes  the form
$$
\left [\;\gamma ^{0} \; ( i \; \partial _{t} \; +\; e\;  r F(r) \; t^{3})\; + \;
i \gamma ^{3}\; ( \partial _{r} \;+ \; {1\over r}) \; +\;
{1\over r} \; \Sigma ^{S.}_{\theta ,\phi } \; + \;    \right.
\eqno(3.1)
$$
$$
\left. {{e r^{2}K(r) + 1 } \over r} \; (\gamma ^{1} \otimes  t^{2}\;  - \;
                              \gamma ^{2} \otimes  t^{1})\; - \;
 ( \; m  \; + \; \kappa \; r\; \Phi(r)\; t^{3} )\;  \right ] \; \Psi ^{S.} = 0 \;\;  ,
$$
$$
\Sigma ^{S.}_{\theta ,\phi } =
\left [ \; i \; \gamma ^{1} \; \partial _{\theta } \;+  \;
  \gamma ^{2}\; {{ i\partial _{\phi } \;+ \; (i \sigma ^{12} \; + \; t^{3})\cos \theta}
  \over {\sin \theta }}\; \right ]
\eqno(3.2)
$$

\noindent here $t^{j} = (1/2)\; \sigma ^{j}$.  The equation's representation
(3.1) itself is remarkable: the~choice of working basis automatically
produces a~required rearrangement of its terms. It is useful to look at all
the~particular ones in (3.1) and further to trace their respective and rather
distinctive contributions; as will be seen, each of them has its practical side
in the~subsequent formation of the~doublet-monopole system's properties.
In particular, just one term in (3.1), proportional to
$(e r^{2} K(r) + 1)$, mixes up together  the~components of
the~multiplet and this term  vanishes  in case of  the~simplest monopole
potential.   The~peculiarity of both $ e\; r\; F(r) \; t^{3}$ and
$\kappa \; r\; \Phi(r) \;t^{3} $ terms, at least as far as they are really
touched    in the~present work, will be brought to light when we turn to
the~diagonalization of a~composite (isotopic-Lorentzian) discrete operator.
In the~given basis, the~components of total conserved momentum
are determined by (1.7), and correspondingly,
the~starting  doublet wave function $\Psi _{\epsilon jm}(x)$ is
as in (1.8).

An~important  case  in  theoretical  investigation  is
the~electron-monopole system at the~minimal value of quantum number $j$.
The~allowed values for $j$ are $0, 1, 2,\ldots$; the~case of $j = 0$
needs a~careful separate consideratiobn. If $j=0$, then the~used symbols
$D^{0}_{0,\pm 1}$
(in (1.8)) are  meaningless,  and  the~wave  function $\Psi _{\epsilon 0}(x)$
has to be constructed as
$$
\Psi _{\epsilon 0} = {e^{-i\epsilon t} \over r}
\left [\; T_{+1/2} \otimes
             \left ( \begin{array}{l}
  0 \\ f_{2}(r)  \\   0  \\  f_{4}(r)
             \end{array} \right ) \; + \;
T_{-1/2} \otimes
             \left ( \begin{array}{l}
  g_{1}(r)      \\   0   \\  g_{3}(r)  \\   0
             \end{array} \right ) \; \right ]  \; .
\eqno(3.3)
$$

Using the~required recursive relations for Wigner functions (see (1.9))
( $\nu  = {\sqrt{j(j + 1)}}, \omega={\sqrt{(j - 1)(j + 2)}}, j \neq 0$)
$$
\partial _{\theta } D_{-1} = {1\over 2} (\omega  D_{-2} - \nu D_{0}),\qquad
{{m - \cos \theta  } \over{ \sin \theta }} D_{-1} =
{1\over 2} (\omega  D_{-2} + \nu  D_{0}),
$$
$$
\partial _{\theta } D_{0} = {1\over 2} (\nu  D_{-1} - \nu D_{+1}),    \qquad
{m \over {\sin  \theta }} D_{0} =
{1\over 2} (\nu D_{-1} + \nu  D_{0}),
$$
$$
\partial _{\theta } D_{+1} = {1\over 2} (\nu D_{0} - \omega D_{+2}),  \qquad
{{m + \cos \theta } \over{ \sin \theta }} D_{+1} =
{1\over 2} (\nu  D_{0} + \omega  D_{+2})
\eqno(3.4a)
$$

\noindent we find
$$
\Sigma ^{S.}_{\theta ,\phi } \;\Psi ^{S.}_{jm}  = \; \nu \;
\left [\; T_{+1/2} \otimes
             \left ( \begin{array}{l}
-i f_{4} \; D_{-1} \\ +i f_{3} \; D_{0} \\ +i f_{2} \; D_{-1} \\ -i f_{1}\; D_{0}
             \end{array} \right ) \;  + \;
T_{-1/2} \otimes
             \left ( \begin{array}{l}
-i g_{4} \; D_{0} \\ +i g_{3} \; D_{+1} \\ +i g_{2} \; D_{0} \\ -i g_{1} \; D_{+1}
             \end{array} \right ) \;\right ]  \; .
\eqno(3.4b)
$$

Further, let us write down the~expression for the term that mixes up
the~isotopic components
$$
{{e r^{2}K(r)  + 1} \over r} \;
(\gamma ^{1}\otimes  t^{2} \; - \; \gamma ^{2} \otimes  t^{1}) \; \Psi _{jm} =
\;{{e r^{2}K(r) + 1 } \over 2 r} \; \times
$$
$$
\left [ \; T_{+1/2} \otimes
             \left ( \begin{array}{l}
   0 \\  +i g_{3} D_{0} \\   0   \\ -i g_{1} D_{0}
             \end{array} \right ) \; +  \;
T_{-1/2} \otimes
             \left ( \begin{array}{l}
-i f_{4} D_{0} \\   0   \\ +i f_{2} D_{0} \\   0
             \end{array} \right ) \;\right ] \; .
\eqno(3.5)
$$

After a  simple  calculation  one   finds  the~system  of  radial
equations (for shortness we  set
$W\equiv (e\; r^{2}\; K(r)\;  +\; 1)/2 \; , \;
 \tilde{F} \equiv  e\; r \; F(r)/2 \; , \;
 \tilde{\Phi } \equiv \kappa\; r \; \Phi (r)/2\;$)
$$
(- i {d\over dr} + \epsilon + \tilde{F} ) f_{3} -
i{\nu \over r} f_{4}  - ( m + \tilde{\Phi } ) f_{1} = 0
$$
$$
(+ i {d\over dr} + \epsilon  + \tilde{F}) f_{4} +
i{\nu \over r} f_{3} + i{W \over r} g_{3} - ( m + \tilde{\Phi }) f_{2} = 0
$$
$$
(+ i {d\over dr} + \epsilon  + \tilde{F} ) f_{1} +
i{\nu \over r} f_{2} - ( m + \tilde{\Phi } ) f_{3} = 0
$$
$$
(- i {d\over dr} + \epsilon  + \tilde{F} ) f_{2} -
i{\nu \over r} f_{1} - i{W\over r} g_{1} - ( m + \tilde{\Phi } ) f_{4} = 0
$$
$$
(- i{d\over dr} + \epsilon  - \tilde{F} ) g_{3} - i{\nu \over r} g_{4} -
i{W\over r} f_{4} - ( m - \tilde{\Phi } ) g_{1} = 0
$$
$$
(+ i {d\over dr} + \epsilon  - \tilde{F} ) g_{4} + i{\nu \over r} g_{3}
    - ( m - \tilde{\Phi } ) g_{2} = 0
$$
$$
(+ i{d \over dr} + \epsilon  - \tilde{F} ) g_{1} +
i{\nu \over r} g_{2} + i{W\over r} f_{2} - ( m - \tilde{\Phi } ) g_{3} = 0
$$
$$
(-i{d\over dr} + \epsilon  - \tilde{F} ) g_{2} -
i{\nu \over r} g_{1} - ( m - \tilde{\Phi } ) g_{4} = 0  \; .
\eqno(3.6)
$$

\noindent When $j$ takes on  value $0$ (then
$\Sigma _{\theta ,\phi } \Psi _{\epsilon 0} \equiv  0$), the~radial system
is
$$
( + i {d\over dr} + \epsilon  + \tilde{F} ) f_{4} +
i{W\over r} g_{3} - ( m + \tilde{\Phi } ) f_{2} = 0
$$
$$
( - i {d\over dr} + \epsilon  + \tilde{F} ) f_{2} -
i{W\over r} g_{1} - ( m + \tilde{\Phi } ) f_{4} = 0
$$
$$
( - i{d\over dr} + \epsilon  - \tilde{F} ) g_{3} -
i{W\over r} f_{4} - ( m - \tilde{\Phi } ) g_{1} = 0
$$
$$
( + i {d\over dr} + \epsilon  - \tilde{F} ) g_{1} +
i{W\over r} f_{2} - ( m - \tilde{\Phi } ) g_{3} = 0  \; .
\eqno(3.7)
$$

\noindent Both  these  systems  (3.6)   and   (3.7)   are   sufficiently
complicated. To proceed further in a~situation like  that,  it  is
normal practice to search  a~suitable  operator  which  could  be
diagonalized additionally. It is known that the~usual $P$-inversion operator
for a~bispinor field cannot be  completely  appropriate  for  this
purpose  and  a~required quantity  has  to  be  constructed  as
a~combination of  bispinor $P$-inversion operator and a~certain discrete
transformation  in  the~isotopic space.
Indeed, considering that the~usual $P$-inversion operator
for a~bispinor field (in the~basis of Cartesian  tetrad, it is
$\hat{P}_{bisp.}^{Cart.} \otimes \hat{P} =
i \gamma^{0} \otimes \hat{P}$, where $\hat{P}$  causes the~usual
$P$-reflection
of  space coordinates)   is determined in the~given (spherical) basis as
$$
\hat{P}_{bisp.}^{sph.} \otimes \hat{P} =
\left ( \begin{array}{rrrr}
             0  &  0  &  0  &  -1 \\
             0  &  0  & -1  &   0 \\
             0  & -1  &  0  &   0 \\
            -1  &  0  &  0  &   0
\end{array} \right ) \otimes \hat{P} =
  - ( \gamma ^{5} \gamma ^{1} ) \otimes  \hat{P}
\eqno(3.8a)
$$

\noindent and  it acts upon the wave function $\Psi _{jm}(x)$ as follows
(the factor $e^{i\epsilon t} /r$ is omitted)
$$
(\hat{P}_{bisp.}^{sph.} \otimes \hat{P}) \; \Psi _{\epsilon jm}(x) =
(-1)^{j+1}
\left [\; T_{+1/2} \otimes
             \left ( \begin{array}{l}
f_{4} \; D_{0} \\ f_{3} \; D_{+1} \\ f_{2} \; D_{0} \\ f_{1} \; D_{+1}
             \end{array} \right ) \; + \;
T_{-1/2} \otimes
             \left ( \begin{array}{l}
g_{4} \; D_{-1} \\ g_{3} \; D_{0} \\ g_{2} \; D_{-1} \\ g_{1} \; D_{0}
             \end{array} \right )\;\right ] \; .
\eqno(3.8b)
$$

\noindent This relationship points the~way  towards  the~search  for
a~required discrete operator: it would  have the~structure
$$
\hat{N}^{S.}_{sph.} \equiv \hat{\pi}^{S.} \otimes
\hat{P}_{bisp.}^{sph.} \otimes \hat{P} \; , \qquad
\hat{\pi}^{S.} = ( a \; \sigma^{1} \; + \; b \; \sigma^{2} ) , \qquad
\hat{\pi}^{S.} \; T_{\pm 1/2} = ( a \; \pm \; i b ) \; T_{\mp 1/2}\; .
\eqno(3.9)
$$

\noindent The total multiplier at the~quantity
 $\hat{\pi }^{S.}$  is  not  material one for
separating the~variables, below one sets
$(\hat{\pi }^{S.})^{2} = ( a^{2} \; +\;  b^{2} ) = + 1$.

From  the~equation
$\hat{N}^{S.}_{sph.} \Psi _{jm}  = N \Psi _{jm}$  one   finds   two  proper
values $N$  and corresponding limitation on  the~functions $f_{i}(r)$
and $g_{i}(r)$:
$$
N = \delta \; (-1)^{j+1} \; , \;\;  \delta = \pm \; 1 : \qquad
g_{1} = \delta \; (a + i b) \; f_{4} \; , \qquad
g_{2} = \delta \; (a  + i b) \;  f_{3} \; ,
$$
$$
g_{3} = \delta \; (a + i b) \; f_{2} \; ,\qquad
g_{4} = \delta \; (a  + i b)\; f_{1}    \; .
\eqno(3.10a)
$$

\noindent Taking into account the relations (3.10a),  one  produces
the~equations ( $\Delta  \equiv (a + i b )$ )
$$
(-i{d\over dr} + \epsilon  + \tilde{F} ) f_{3} -
{\nu \over r} f_{4}  - ( m + \tilde{\Phi}) f_{1} = 0
$$
$$
(+i{d\over dr} + \epsilon  + \tilde{F} ) f_{4} +
{\nu \over r} f_{3} + i{W \over r} - \delta  \Delta  f_{2} -
( m  +  \tilde{\Phi} ) f_{2} = 0
$$
$$
(+i{d\over dr} + \epsilon  + \tilde{F} ) f_{1} +
{\nu \over r} f_{2}   - ( m + \tilde {\Phi } ) f_{3} = 0
$$
$$
(-i{d\over dr} + \epsilon  + \tilde{F} ) f_{2} -
{\nu \over r} f_{1} - i{W\over r} \delta  \Delta  f_{4} -
( m + \tilde{\Phi } ) f_{4} = 0
$$
$$
(-i{d\over dr} + \epsilon  - \tilde{F} ) f_{2} -
{\nu \over r} f_{1} - i {W\over r} \Delta ^{-1} \delta  f_{4}-
( m - \tilde{\Phi } ) f_{4} = 0
$$
$$
(+i{d\over dr} + \epsilon  - \tilde{F} ) f_{1} +
{\nu \over r} f_{2}   - ( m - \tilde{\Phi } ) f_{3} = 0
$$
$$
(+i{d\over dr} + \epsilon  - \tilde{F} ) f_{4} +
{\nu \over r} f_{3} + i {W\over r} \Delta ^{-1} \delta  f_{2}-
( m - \tilde{\Phi } ) f_{2} = 0
$$
$$
(-i{d\over dr} + \epsilon  - \tilde{F} ) f_{3} -
{\nu \over r} f_{4}  - ( m - \tilde{\Phi } ) f_{1} = 0  \;\;  .
\eqno(3.10b)
$$

\noindent It is evident at once that the~system (3.10b)  would be compatible
with itself  provided that  $\tilde{F}(r) = 0$  and $\tilde{\Phi }(r) = 0$.
In other words,
the above-mentioned operator $\hat{N}^{S.}$ can be diagonalized on
the~functions $\Psi _{\epsilon jm}(x)$ if and only if $W^{(a)}_{t}= 0$
and $\kappa = 0$; below  we
suppose that these requirements  will be  satisfied.
Moreover, given this limitation satisfied, it is necessary to draw
distinction  between  two cases depending on expression for $W(r)$.
If $W(r) = 0$,  the~difference between $\Delta $ and $\Delta ^{-1}$
in the~equations (3.10b)  is  not
essential in simplifying these equations (because
the~relevant terms just vanish).

Thus, for the first case, the~system
(3.10b) converts into (the symbol $\Delta$ at $\hat{N}$  stands for
 the~$a$- and $b$-dependence):
$$
 W(r) = 0 , \qquad \hat{N}^{S.}_{\Delta } =
(a \;\sigma ^{1} + b \; \sigma ^{2}) \otimes  \hat{P}_{bisp.} \otimes \hat{P}\;\;:
$$
$$
( - i {d\over dr} + \epsilon  ) f_{3} - {\nu \over r} f_{4} - m f_{1} = 0 \; ,
\qquad
( + i {d\over dr} + \epsilon  ) f_{4} + {\nu \over r} f_{3} - m f_{2} = 0 \; ,
$$
$$
( + i {d\over dr} + \epsilon  ) f_{1} + {\nu \over r} f_{2} - m f_{3} = 0 \; ,
\qquad
( - i {d\over dr} + \epsilon  ) f_{2} - {\nu \over r} f_{1} - m f_{4} = 0 \; .
\eqno(3.11)
$$

There exists sharply distinct situation at $W \neq  0$. Here,
the~equations are  consistent  with  each  other  only  if
$\Delta  = \Delta ^{-1}$;
therefore $\Delta  = (a + i b) =  \pm 1$. Combining   this  relation  with
the~normalizing  condition  $(a + i b) (a - i b) = 1$, one  gets
 $a = \pm 1$  and $b = 0$ (for definiteness, let this parameter $a$
 be equal $+1$). The~corresponding set of radial equations, obtained from
(3.10b), is
$$
\hat{N}^{S.}= ( \sigma ^{1} \otimes  \hat{P}_{bisp} \otimes \hat{P}), \;\;
N = \delta  (-1)^{j +1} \;\; :
$$
$$
( -i{d\over dr} + \epsilon ) f_{3} - {\nu \over r} f_{4} - m f_{1} = 0 \; ,
\qquad
( +i{d\over dr} + \epsilon ) f_{4} + {\nu \over r} f_{3} +
i{W\over r} \delta  f_{2} - m f_{2} = 0   \; ,
$$
$$
( +i {d\over dr} + \epsilon ) f_{1} + {\nu \over r} f_{2}  - m f_{3} = 0  \; ,
\qquad
(- i {d\over dr} + \epsilon ) f_{2} - {\nu \over r} f_{1}  -
 i{W\over r} \delta  f_{4} - m f_{4} = 0        \; .
\eqno(3.12)
 $$

In the~same  way, the~case  $j = 0$ can be considered. Here,
the~proper values and limitation  are
$$
N = - \; \delta  ,\;\; \delta  = \pm 1 \; : \qquad
g_{1}(r) \; = \delta \; \Delta \; f_{4}(r) \;,\qquad
g_{3}(r) \; = \delta \; \Delta \; f_{2}(r) \; .
\eqno(3.13a)
$$

\noindent Further, the~quantities  $\tilde{F}$  and $\tilde{\Phi }$
are to be equated  to zero; again there are two  possibilities depending
on $W$:
$$
W(r) = 0 \; : \qquad
( i {d\over dr}  + \epsilon ) f_{4} - m f_{2}  = 0 \;  ,
\qquad
(-i {d\over dr} + \epsilon ) f_{2}  - m f_{4}  = 0 \; ;
\eqno(3.13b)
$$
$$
W(r) \neq  0 \; : \qquad
( i {d\over dr} + \epsilon ) f_{4} - ( m - i {\delta \over r} W ) f_{2} = 0 \; ,
\;\;
(- i {d\over dr} + \epsilon ) f_{2} - ( m + i {\delta\over r} W ) f_{4} = 0 \; .
\eqno(3.13c)
$$

The explicit forms  of  the  wave  functions
 $\Psi _{\epsilon jm\delta }(x)$ and $\Psi _{\epsilon 0\delta }(x)$ are
as follows:

The case $\; W(r) \neq  0 , \; j > 0 \; $,
$$
\Psi _{\epsilon jm}(x) = {{e^{-i\epsilon}t} \over r }
\left [\; T_{+1/2} \otimes
             \left ( \begin{array}{l}
f_{1} \; D_{-1} \\ f_{2} \; D_{0} \\ f_{3} \; D_{-1} \\ f_{4} \; D_{0}
             \end{array} \right )  \; + \;
\delta \; T_{-1/2} \otimes
             \left ( \begin{array}{l}
f_{4} \; D_{0} \\ f_{3} \; D_{+1} \\ f_{2} \; D_{0} \\ f_{1} \; D_{+1}
             \end{array} \right )\; \right ] \; ;
\eqno(3.14a)
$$

The case $W(r) \neq  0 , \;\; j = 0 \;$,
$$
\Psi _{\epsilon 0} = {e^{-i\epsilon t} \over r}
\left [\; T_{+1/2} \otimes
             \left ( \begin{array}{l}
  0 \\ f_{2}(r)  \\   0  \\  f_{4}(r)
             \end{array} \right )  \; + \;
\delta \; T_{-1/2} \otimes
             \left ( \begin{array}{l}
  f_{4}(r)      \\   0   \\  f_{2}(r)  \\   0
             \end{array} \right ) \; \right ]
\eqno(3.14b)
$$

\noindent where $\delta \; T_{-1/2}$  is to be changed for
 $\delta \;  \Delta \; T_{-1/2}$  when $W = 0$.

These  formulas  point to the~non-featureless and  non-formal
union of the~one particle pattern with another (two distinct isotopic
components), but a~structural  and
specific one; at that, the~second term in the~composite doublet wave
function is strictly determined  up to the~whole  phase factor,
by the~first term,  so that  this system is
not a~plain sum of two components without any intrinsic structure.

\subsection*{4 Analyzis of the particular case of simplest
 monopole field}

Now, some added aspects of the~simplest monopole are examined more closely.
The~system of radial equations, specified for this potential, is basically
simpler than in general case, so that the~whole problem including
the~radial functions can be carried out to its complete conclusion.

Actually,  the equation (3.11) admits of some
further simplifications owing to diagonalyzing the  operator
$\hat{K}_{\theta ,\phi }=
- i \gamma ^{0} \gamma ^{5} \Sigma _{\theta ,\phi }$.
From the~equation
$
\hat{K}_{\theta ,\phi } \Psi _{jm}  =
\lambda  \Psi _{jm}$ , it
follows that $\lambda  = - \mu \;  {\sqrt{j(j + 1}}), \; \mu  = \pm 1$  and
$$
f_{4} = \mu \;  f_{1} , \qquad f_{3} = \mu \; f_{2} ,\qquad
g_{4} = \mu \;  g_{1} , \qquad g_{3} = \mu \;  g_{2} \; .
\eqno(4.1)
$$

\noindent Correspondingly, the system (3.11) yields
$$
(+ i {d\over dr}  + \epsilon ) f_{1} + i {\nu \over r} f _{2} -
\mu \; m \; f_{2}  = 0 \;  ,
\qquad
(- i {d\over dr} +  \epsilon ) f_{2}  - i {\nu \over r} f _{1} -
\mu \;  m \; f_{1}  = 0  \; .
\eqno(4.2a)
$$

\noindent The wave function with quantum  numbers
$(\epsilon , j, m, \delta , \mu )$  has  the~form
$$
\Psi _{\epsilon jm\delta\mu}^{\Delta}(x) = {{e^{-i\epsilon}t} \over r }\;
\left [\; T_{+1/2} \otimes
             \left ( \begin{array}{l}
f_{1} \; D_{-1} \\ f_{2} \; D_{0} \\ \mu f_{3} \; D_{-1} \\ \mu f_{4} \; D_{0}
             \end{array} \right ) \;  + \;
\Delta \; \mu\;  \delta\;  T_{-1/2} \otimes
             \left ( \begin{array}{l}
f_{4} \; D_{0} \\ f_{3} \; D_{+1} \\ \mu f_{2} \; D_{0} \\ \mu f_{1} \; D_{+1}
             \end{array} \right ) \; \right ] \;  .
\eqno(4.2b)
$$

We will not consider these systems of two radial equations;
this  would  represent  an~easy problem  concerning  the~well-known  spherical
Bessel functions. Instead, we  relate these functions (4.2b) (also
$\Psi _{\epsilon 0\delta }(x)$) with the~wave functions satisfying  the~Dirac
equation  in  the~Abelian monopole potential. Those latter were investigated
by many authors; below we  will use the~notation
according to [67]).
At $ j > j_{\min }$ these Abelian functions are described as in (1.5) with
taking into account the~additional relation
$$
f_{4} = \mu \;  f_{1} , \qquad f_{3} = \mu \; f_{2} , \qquad \mu = \pm 1 \; .
$$

\noindent For the minimal values $\;j = j_{min.} = \mid eg\mid -1/2 $,
they are\footnote{Just these functions can be referred to the solutions of
third type in terminology used by Kazama, Yang, and Goldhaber; see in [25].}:
$$
eg = + 1/2, + 1, + 3/2, ... \qquad
\Phi ^{(eg)}_{\epsilon 0} (t,r,\theta ,\phi )=
                              \left ( \begin{array}{l}
          f_{1}(t,r) \; D^{j}_{-m,-1/2}(\phi ,\theta ,0) \\
                            0                         \\
          f_{3}(t,r) \; D^{j}_{-m,-1/2}(\phi ,\theta ,0) \\
                            0
                               \end{array} \right )  \;  ;
\eqno(4.3a)
$$
$$
eg = - 1/2, - 1, - 3/2,...\qquad
\Phi ^{(eg)}_{\epsilon 0}(t,r,\theta ,\phi ) =
                              \left ( \begin{array}{l}
                             0                    \\
          f_{2}(t,r) \; D^{j}_{-m,+1/2}(\phi ,\theta ,0) \\
                             0                     \\
          f_{4}(t,r) \; D^{j}_{-m,+1/2}(\phi ,\theta ,0)
                               \end{array} \right )        .
\eqno(4.3b)
$$

\noindent On comparing the formulas (3.14a,b) with (1.5) and ((4.3a,b),
the~following expansions  can be easily  found (respectively, for
$j > 0$ and $j=0$ cases):
$$
\Psi ^{\Delta \delta \mu }_{\epsilon jm}(x) \; = \;
\left [ \; T_{+1/2} \otimes  \Phi ^{eg=-1/2}_{\epsilon jm\mu }(x)
  \;\; + \;\; \mu \; \delta \;  \Delta\;   T_{-1/2} \otimes
 \Phi ^{eg =+1/2}_{\epsilon jm\mu }(x)\;\right ] \; ,
\eqno(4.4a)
$$
$$
\Psi ^{\Delta }_{\epsilon 0\delta } (x) \;= \;
\left [ \; T_{+1/2} \otimes  \Phi ^{eg =-1/2}_{\epsilon 0}(x) \; \; + \;\;
\delta \; \Delta \;\;  T_{-1/2} \otimes  \Phi ^{eg =+1/2}_{\epsilon 0}(x)
\;\right ]\; .
\eqno(4.4b)
$$

\noindent In reference with the formulas (4.4a,b), one additional remark
should be given.  Though,  as evidenced by (4.4a,b),  definite  close
relationships between the~non-Abelian doublet wave functions and
Abelian fermion-monopole
functions can be explicitly discerned, in reality,  the~non-Abelian
situation is intrinsically non-monopole-like (non-singular one).
Indeed,  in the~non-Abelian case, the~totality
of possible transformations (upon the~relevant wave functions) which bear
the~gauge status are very different from ones that there are
in the~purely Abelian theory. In  a~consequence of this, the~non-Abelian
fermion doublet wave functions (1.8) can be readily transformed, by carrying
out the~gauge transformations in Lorentzian and isotopic spaces together
($S. \; \rightarrow \; C.\;$ and $\;sph. \; \rightarrow \; Cart.$),
into the~form (see the formulas (A.9) and (A.10)) where they are
single-valued  functions of spatial points.  In the~Abelian monopole situation,
the~analogous particle-monopole functions can by no means
be translated  to any single-valued ones (see also in Supplement A).

\subsection*{5  On distinction between manifestation of the Abelian
           and  non-Abelian monopoles. Some comments
           about parity selection rules}

The problem that we have discussed so  far  concerned  solely
an~isotopic doublet affected by the~external monopole field. Let us now
pass  on  to ~somewhat another subject, namely, we consider  what  in
the~everything
having stated above was dictated by the~presence of the~non-Abelian external  field
and what was fixed  only  by  the~isotopic  multiplet structure.
To  this  end, it suffices  to  compare  the~doublet-monopole
system with a~free doublet.

A free wave equation is   as follows
$$
\left [\;(\;i \gamma ^{0} \partial _{t} \; + \;  i \;\gamma ^{3} ( \partial _{r} \; +  \;
{1\over r} ) + {{\gamma ^{1}\otimes t^{2} \;  - \; \gamma ^{2}\otimes t^{1}} \over r} \;\; +\right.
$$
$$
\left. {1\over r} \; (\; i \gamma ^{1} \; + \;  \gamma ^{2} \; {{\partial _{\phi } \; +\;
(i\sigma ^{12} \;+\; t^{3}) \cos \theta} \over \sin\theta }\; ) \; - \;
 m\; \right  ]\; \Psi ^{S.}_{sph.} = 0  \; .
\eqno(5.1)
$$

\noindent We draw attention to the~term
$(\gamma ^{1} \otimes  t^{2} - \gamma ^{2} \otimes t^{1})/r$  mixing
both isotopic  components,  which  somewhat  events  out  any  contrast
between these two physical systems, so that everything said  above
and concerned the~case $W \neq  0$   is  valid  for  this  particular
situation too\footnote{Correspondingly, no $A$-freedom in choosing
the~composite inversion-like operator occurs in case of free fermion
 doublet.}. A~single distinction is  the~explicit form of the~factor at
$( \gamma ^{1} \otimes  t^{2} -  \gamma ^{2} \times  t^{1}) /r $-term.

The presence  of such a~mixing term in the~equation referring
to the~free doublet might seem rather surprising fact. Nevertheless, as
can be easily shown, its  origin is due to a~gauge transformation.
Actually, the~corresponding free equation  in the~Cartesian
isotopic  gauge (compare it with (1.1))
$$
[\; i\; \gamma ^{\alpha }(x) (\partial _{\alpha }  \; +  \;
 \Gamma _{\alpha }(x)) \otimes  I \; - \; m \;] \Psi ^{0}_{C.} = 0
\eqno(5.2a)
$$

\noindent takes on the following explicit form in  the  basis  of  spherical
tetrad
$$
\left [ \;i \gamma ^{0} \partial _{t} \; + \; i \gamma ^{3} \; (\partial _{r} \; + \;
{1 \over r} ) \;  +  \;
{1\over r} \; (\; i \gamma ^{1}\; +\; \gamma ^{2}\; {{ i\partial _{\phi }\; + \;
i\sigma ^{12}\cos \theta } \over \sin \theta }  -\; m \;\right ]\; \Psi ^{0}_{C.} = 0 \; .
\eqno(5.2b)
$$

\noindent Applying the~isotopic gauge transformation  to
$\Psi ^{0}_{C.}$ (see (2.12d)):
$\Psi ^{0}_{S.}(x) \; = \; B(\theta, \phi)\; \Psi ^{0}_{C.}(x)$,
one  can bring the equation (5.2b) to the~form
$$
\left [ \;i \gamma ^{\alpha }(x) \; ( \partial _{\alpha } \; +  \;
\Gamma _{\alpha }(x) ) \otimes  I\;  +  \;
i\gamma ^{\alpha }(x) \otimes  ( B \; {{\partial  B^{-1} } \over
{\partial x^{\alpha }}} ) \; - \;  m \;\right ] \;  \Psi ^{0}_{S.}   =  0
\eqno(5.3a)
$$

\noindent where
$$
i\; \gamma ^{\alpha }(x) \otimes  ( B \; {{ \partial B^{-1} } \over
 { \partial x^{\alpha }}} )\; = \;
{1\over r} (\gamma  ^{1}\otimes  t ^{2} - \gamma  ^{2}\otimes  t ^{1}) +
\gamma  ^{2} \otimes  {{t_{3} \cos \theta} \over {r \sin \theta }} \;  .
\eqno(5.3b)
$$

\noindent The first term in (5.3b) will mix  the  isotopic  components,
the second represents a~term being an~essential addition to  angular  operator
$\Sigma _{\theta ,\phi }$;  and  both  of them have arisen out of the~above
gauge  transformation. Their correlated appearing may be regarded as a~formal
mathematical description of efficient linking the radial  functions
through   the~kinematical  coupling  of two  isotopic  components  by
diagonalization of the~total angular momentum operators.
In this context, the~above-mentioned simplification of radial
equations at $W = 0$  can be interpreted as follows:  an  efficient
cinematical mixing (owing  to  the  ordinary  scheme  of angular  momentum
addition) the different isotopic components  is destroyed  through
simple placing  that system into the external  trivial  monopole
field, so that the~angular coupling is conserved but the~efficient
linking through  radial functions no  longer  obtains.  In  other
words, these two factors cancel out each other.

It is significant that in both cases, the~wave  functions
obeying the~free equation and the~equation with external  monopole
potentials, respectively, do not vary at all in their $\theta,\phi$-dependence.
A~single manifestation of the~external monopole field is the~change
in the~single parametric function $W(r)$: the quantity $W^{0}(r) = 1$
is to be replaced with  another $W(r) = ( 1 + e r^{2} K(r) )$.
The~above correlates with the~fact that  the~operators  of  spherical
symmetry $\vec{J}~^{2}, J_{3}, \hat{N}$  of  these  two  different
physical  systems exactly coincide.

Totally  different  from  this   is   the~situation   in   the~Abelian
problem when the~spherical  symmetry  operators
and wave functions are both basically transformed (see (1.3) and (1.4))
in   presence  of  the~Abelian monopole.
The~free basic wave  functions  (setting $eg = 0$  in (1.3))
$\Phi ^{0}_{\epsilon JM\delta }(t,r,\theta ,\phi )$   and the~monopole ones
 $\Phi ^{eg}_{\epsilon jm\mu }(t,r,\theta ,\phi )$    vary
noticeably in their boundary properties at $\theta  = 0, \pi$.
Let us consider this question in some more detail.

To clarify  all the~significance of
the~mere displacement in a~single index at the wave functions in (1.5),
we are going to look at just one mathematical  characteristic of those $D$-functions
 involved in the~particle wave functions: namely, their  boundary
properties  at the~points $\theta = 0$ and $\theta = \pi$. So, the following
Tables can be produced (only some of them are written out):

Table $1a \qquad    D^{j}_{m,+1/2}\;$:
$$
\left. \begin{array}{llcc}
                     &                & \qquad \theta = 0    &  \theta  = \pi    \\
              j=1/2  &                &                      &                   \\
                     &  m=-1/2        & \qquad  0            &  e^{+i\phi/2}     \\
                     &  m=+1/2        & \qquad e^{-i\phi/2}  &   0               \\
              j=3/2  &                &                      &                   \\
                     &  m=-1/2        & \qquad 0             & e^{+i\phi/2}      \\
                     &  m=+1/2        & \qquad e^{-i\phi/2}  & 0                 \\
                     &  m=-3/2        & \qquad 0             & 0                 \\
                     &  m=+3/2        & \qquad 0             & 0                 \\
           j=5/2,... &                &                      &
\end{array} \right.
$$

Table $1b \qquad  D^{j}_{m,-1/2} \;$:
$$
\left. \begin{array}{llcc}
                     &                & \qquad \theta = 0    &  \theta  = \pi     \\                                                         \\
              j=1/2  &                &                      &                    \\
                     &  m=-1/2        & \qquad e^{+i\phi/2}  &   0                \\
                     &  m=+1/2        & \qquad  0            & e^{-i\phi/2}       \\
              j=3/2  &                &                      &                    \\
                     &  m=-1/2        & \qquad e^{+i\phi/2}  &  0                 \\
                     &  m=+1/2        & \qquad 0             & e^{-i\phi/2}       \\
                     &  m=-3/2        & \qquad 0             & 0                  \\
                     &  m=+3/2        & \qquad 0             & 0                  \\
           j=5/2,... &                &                      &
\end{array} \right.
$$

Table $2a \qquad    D^{j}_{m,+1}\;$:
$$
\left. \begin{array}{llcc}
                     &                & \qquad \theta = 0    &  \theta  = \pi    \\
              j=1    &                &                      &                   \\
                     &  m=0           & \qquad  0            &  0                \\
                     &  m=-1          & \qquad  0            &  e^{+i\phi}       \\
                     &  m=+1          & \qquad e^{-i\phi}    &   0               \\
              j=2    &                &                      &                   \\
                     &  m=0           & \qquad  0            &  0                \\
                     &  m=-1          & \qquad 0             & e^{+i\phi}        \\
                     &  m=+1          & \qquad e^{-i\phi}    & 0                 \\
                     &  m=-2          & \qquad 0             & 0                 \\
                     &  m=+2          & \qquad 0             & 0                 \\
            j=3,...  &                &                      &
\end{array} \right.
$$

Table $2b \qquad  D^{j}_{m,-1} \;$:
$$
\left. \begin{array}{llcc}
                     &                & \qquad \theta = 0    &  \theta  = \pi     \\                                                         \\
              j=1    &                &                      &                    \\
                     &  m=0           & \qquad  0            &  0                 \\
                     &  m=-1          & \qquad e^{+i\phi}    &   0                \\
                     &  m=+1          & \qquad  0            & e^{-i\phi}         \\
              j=2    &                &                      &                    \\
                     &  m=0           & \qquad  0            &  0                 \\
                     &  m=-1          & \qquad e^{+i\phi}    &  0                 \\
                     &  m=+1          & \qquad 0             & e^{-i\phi  }       \\
                     &  m=-2          & \qquad 0             & 0                  \\
                     &  m=+2          & \qquad 0             & 0                  \\
            j=3,...  &                &                      &
\end{array} \right.
$$

Table $3a \qquad    D^{j}_{m,+3/2}\;$:
$$
\left. \begin{array}{llcc}
                     &                & \qquad \theta = 0    &  \theta  = \pi    \\
              j=3/2  &                &                      &                   \\
                     &  m=-1/2        & \qquad  0            &  0                \\
                     &  m=+1/2        & \qquad  0            &  0                \\
                     &  m=-3/2        & \qquad  0            &  e^{+i3\phi/2}    \\
                     &  m=+3/2        & \qquad e^{-i3\phi/2} &   0               \\
              j=5/2  &                &                      &                   \\
                     &  m=-1/2        & \qquad  0            &  0                \\
                     &  m=+1/2        & \qquad  0            &  0                \\
                     &  m=-3/2        & \qquad  0            &  e^{+i3\phi/2}    \\
                     &  m=+3/2        & \qquad e^{-i3\phi/2} &   0               \\
                     &  m=-5/2        & \qquad  0            &  0                \\
                     &  m=+5/2        & \qquad  0            &  0                 \\
           j=7/2,... &                &                      &
\end{array} \right.
$$

Table $3b \qquad  D^{j}_{m,-3/2} \;$:
$$
\left. \begin{array}{llcc}
                     &                & \qquad \theta = 0    &  \theta  = \pi     \\                                                         \\
              j=3/2  &                &                      &                    \\
                     &  m=-1/2        & \qquad  0            &  0                \\
                     &  m=+1/2        & \qquad  0            &  0                \\
                     &  m=-3/2        & \qquad e^{+i3\phi/2}  &   0                \\
                     &  m=+3/2        & \qquad  0            & e^{-i3\phi/2}       \\
              j=5/2  &                &                      &                    \\
                     &  m=-1/2        & \qquad  0            &  0                \\
                     &  m=+1/2        & \qquad  0            &  0                \\
                     &  m=-3/2        & \qquad e^{+i3\phi/2}  &  0                 \\
                     &  m=+3/2        & \qquad 0             & e^{-i3\phi/2}       \\
                     &  m=-5/2        & \qquad 0             & 0                  \\
                     &  m=+5/2        & \qquad 0             & 0                  \\
           j=7/2,... &                &                      &
\end{array} \right.
$$

\noindent On comparing such characteristics
for  $D^{j}_{-m, \pm1/2}(\phi, \theta,0)$ and
$D^{j}_{-m, eg \pm1/2}(\phi, \theta,0)$ ,
we can immediately conclude
that these sets of $D$-functions provide us with the~bases in
different  functional spaces $\{ F^{eg=0}(\theta,\phi) \}$ and
$\{ F^{eg \neq 0}(\theta,\phi) \}$  (all various values of the~parameter
$eg$ lead to different functional spaces as well). Each of those
spaces  is characterized by its own behavior at limiting points, which is
irreconcilable with that of any other space.

These peculiarities are rather crucial on their implications.
For example,  that difference leads to  some obvious problems referring  to
the~basic superposition principle of quantum mechanics. Indeed, it is
understandable that any possible series decompositions of particle wave
functions  $\Phi^{eg\neq0}$ by  $\Phi^{eg =0}$ and  inversely:
$$
\Phi ^{eg}_{\epsilon jm\mu }(t,r,\theta ,\phi ) \; = \;
\Sigma  \;C^{\epsilon JM\delta }_{\epsilon jm\mu }\;
\Phi ^{0}_{\epsilon JM\phi }(t,r,\theta ,\phi ) \; ,
$$
$$
\Phi ^{0}_{\epsilon JM\delta }(t,r,\theta ,\phi ) \; = \;
\Sigma \;  C^{\epsilon jn\mu }_{\epsilon JM\delta } \;
\Phi ^{eg}_{\epsilon jm\mu }(t,r,\theta ,\phi )
$$

\noindent cannot be correct at the~whole $x_{3}$-axis.
The~latter leads to a~very interesting, if not serious, question
as to the~physical status of the~monopole potential.
It is matter that conventional (one particle-based) quantum mechanics
presupposes tacitly that any quantum object remains intrinsically
the~same as a~certain identified
entity when this object is placed into an~arbitrary external field.
For example, an~isolated free electron and an~electron in the~Coulomb field,
in both cases, are represented by the~same single entity, {\it{electron}}, just
situated in the~two different conditions. Mathematically, this tacit assumption
is expressed as the~possibility to exploit together the~fundamental
superposition principle and presupposedly identity of those Hilbert spaces:
$\{\Phi^{free}\} \equiv \{\Phi^{ext. field}\}$. So,
we always can obtain extensions of any wave functions of
the~type $\Phi^{ext. field}$ in terms of functions  $\Phi^{free}$,
or inversely\footnote{For instance, there exists
the~momentum representation $\Phi(p)$ for the~Coulomb wave functions,
which may be considered just as an~illustration to the~above.}.

It is easily understandable that limitations  imposed by this condition
on external fields are quite restrictive, and  we  could show
that many commonly used potentials, referring   to  real sources,
satisfy them. Whereas, evidently, the~presupposed existence of a~magnetic
charge is inconsistent with this basic proposition of the~theory.
Indeed,  at~the whole $x_{3}$ axis, some of particle-monopole functions cannot,
in principle, be represented as any linear combinations of free particle
functions: those later $\{ \Phi^{eg=0}(x_{3}.0,0) \}$  do not contain
at all any functions required to describe some representatives from
$\{ \Phi^{eg \neq 0}(x_{3}.0,0) \}$ (see Tables 1-3 and Supplement A).

Furthermore, one might regard the~above criterion as, in a~sense,
a~superselection   principle yielding the~definite separation of any
mathematically  possible potentials  into the~two classes: real and not
real ones (respectively, not changing and changing the relevant Hilbert
spaces).
In such terms, the~Abelian monopole potential should be thought of as
unphysical and forbidden;
whereas the~non-Abelian potential may be regarded as quite allowable.

In this connection, some more remarks might be added.
As a~matter of fact, the~status of monopoles in
physics is, in general, rather peculiar. Indeed, a~modern age in understanding
electricity and magnetism (EM) was ushered in by Dirac [3] who had argued on
the~line of quantum mechanical arguments that electric charge $e$ must be
quantized as an~isolated magnetic charge  $g$ exists, thereby he had gave
an~appreciably  significant piece of EM essentials, now known as the~Dirac's
(electric charge) quantization condition. Certainly, the~electric charge had
been  quantized. However --- to say exactly, from the~very beginning,
this Dirac's condition might be interpreted a~bit differently ---
to say the~least: one could
say that a~new introduced quantity (magnetic charge) must be  quantized
as the~electric charge exists.

In any case, the~problem is just one:
any experimental observation  of a~magnetic charge, unfortunately or may be
luckily, has not been registered to date. Therefore, at the~present time
we have to count  solely on the~theoretical investigation
of those constructs. Furthermore, for the~reason alone of such a~consistently
 invisible character of a~magnetic charge,
at least as it concerns experiments,  there appear  grounds for
special seeking some rationalization of such
a~strange, if not enigmatic, and persistent disinclination of the~monopoles
to be seen experimentally.  Even more, it is the time to have searched
some possible formally conceptualized  grounds for  forbidding, in principle,
this charge from existence in    nature. Therefore, some provisional steps
in this direction might  be made yet  today.

In reference to this, else one peculiarity of the~above property
of the~particle-monopole functions deserves to be  especially noticed:
the~irreconcilable character of monopole-affected and free functions,
respectively, might be interpreted in physical
terms  as the~fundamental impossibility to eliminate the~monopole influence
on the~particle wave functions over all  space  up  to  infinitely
distance points. This property implies
a~lot of hampering implications, in turn giving rise to some awkward
questions. For instance, the~question of that kind is: what is the~meaning
of the~relevant scattering theory, if even at infinity itself, some
manifestation of the~magnetic charge  presence does not vanish (just because
of the~given $\theta,\phi$-dependence). That point finds its natural
corollary in giving rise the~well-known difficulties in the~relevant
scattering theory [17-19,86-94]: the particle-monopole functions, being regarded
at the~asymptotical infinity (far away from the~region of $r=0$),
exhibit such kind of behavior that does not fit it to be appropriate for
a~free quantum-mechanical particle. So, we cannot get rid of the~Abelian
monopole effects up to infinitely distant points, and such a~property
is removed far from what is familiar when  a~situation is less singular
(for instance,  of external electric charge presence).

Now, as a~way of further contrasting the~Abelian and non-Abelian
models, we are going to pass on another subject and consider the~question  of
discrete  symmetry (restricting ourselves to $P$-transformation-involving
operations)  in both these theories (see also in [1,11-16,76-85]).

In the Abelian case  (when $eg \neq  0$),  the~monopole  wave
functions cannot be proper functions of the~usual space reflection
operator for a~particle (for definiteness, the~bispinor field case is meant).
 There exists only  the~following
relationship $( j >j_{min.}$)
$$
(\hat{P}_{bisp.} \otimes \hat{P} ) \; \Phi ^{eg}_{\epsilon jm\mu }(x) =
{{e^{-i\epsilon t} } \over r } \; \mu \; (-1)^{j+1} \;
\left ( \begin{array}{r}
   f_{1}(r) \; D^{j}_{-m,-eg -1/2} \\
   f_{2}(r) \; D^{j}_{-m,-eg+1/2} \\
   \mu   f_{2}(r) \; D^{j}_{-m,-eg -1/2} \\
   \mu   f_{1}(r) \; D^{j}_{-m,-eg+1/2}
\end{array} \right )
\eqno(5.4a)
$$

\noindent take notice of the sign `minus' at the~parameter $eg$.
By contrast, the~required relation  for free wave functions  occurs
$$
(\hat{P}_{bisp.} \otimes \hat{P} ) \; \Phi ^{0}_{\epsilon JM\mu }(x) =  \;
\delta \; (-1)^{J+1} \; \Phi ^{0}_{\epsilon JM\delta }(x)  \; .
\eqno(5.4b)
$$

\noindent It should be emphasized that a~certain,  diagonalized on
the~functions
 $\Phi ^{eg}_{\epsilon jm\mu }$,  discrete operator  can be
obtained through multiplying the~usual $P$-inversion bispinor operator  by
the~formal one $\hat{\pi }_{Abel.}$ that affects the~$eg$-parameter in  the~wave
functions as follows:
$\hat{\pi}_{Abel.} \; \Phi ^{+eg}_{\epsilon jm\mu }(x) =
 \Phi ^{-eg}_{\epsilon jm\mu }(x)\;$. Thus, we have
$$
\hat{M} = \hat{\pi}_{Abel.} \otimes \hat{P}_{bisp.} \otimes \hat{P} \; , \qquad
\hat{M} \; \Phi ^{eg}_{\epsilon jm\mu }(x) =
 \mu \;  (-1)^{j+1} \; \Phi ^{eg}_{\epsilon jm\mu }(x)
\eqno(5.4c)
$$

\noindent but, as  may be seen, the~latter fact does not allow us to obtain
any  $M$-parity selection rules. Actually, a~matrix  element  for  some
physical  observable  $\hat{G}^{0}(x)$  is to be
$$
\int \bar{\Phi}^{eg}_{\epsilon jm\mu }(x) \; \hat{G}^{0}(x) \;
\Phi ^{eg}_{\epsilon j'm'\mu'}(x) dV \equiv
\int r^{2} dr \int f(\vec{x}) \; d \Omega   \;\; .
$$

\noindent First we  examine the case $e g = 0$, in  order  to  compare  it
with the~situation at $eg \neq 0$. Let us relate
 $f(-\vec{x})$ with $f(\vec{x})$.
Considering the~relation (and the~same  with $J'M'\delta '$)
$$
\Phi ^{0}_{\epsilon JM\delta }(-\vec{x} ) =
\delta \; (-1)^{J+1} \;  \hat{P}_{bisp.} \;
\Phi ^{0}_{\epsilon JM\delta }(\vec{x} )
\eqno(5.5a)
$$

\noindent we  get
$$
f(-\vec{x}) = \; \delta \; \delta'\;  (-1)^{J+J'+1} \;
\bar{\Phi}_{\epsilon JM\delta }(\vec{x}) \;
\left [\; \hat{P}^{+}_{bisp.} \; \hat{G}^{0}(-\vec{x}) \;
\hat{P}_{bisp.}\; \right ] \; \Phi _{\epsilon J'M' \delta'}(\vec{x}) \; .
\eqno(5.5b)
$$

\noindent Thus, if the quantity  $\hat{G}^{0}(\vec{x})$  obeys  the~equation
$$
 \hat{P}^{+}_{bisp.} \; \hat{G}^{0}(-\vec{x}) \; \hat{P}_{bisp.} = \;
\omega ^{0} \; \hat{G}^{0}(\vec{x})
\eqno(5.5c)
$$

\noindent here $\omega ^{0}$ defined to be $+1$   or  $ -1$
  relates  to  the  scalar  and
pseudoscalar, respectively, then the~relationship (5.5b)  comes
to
$$
f(-\vec{x}) = \; \omega \; \delta\;  \delta' \; (-1)^{J+J'+1} \; f(\vec{x})
\eqno(5.5d)
$$

\noindent that generates the well-known $P$-parity selection rules.

In contrast to everything just said, the~situation at $eg \neq  0$
is completely different because any  equality in the  form (5.5a)
or (5.5b) does not appear there; instead, the~relations (5.4a) and
 (1.12b) only  occur. So, there not exist any $M$-parity selection
rules in the~presence of the~Abelian monopole. In accordance  with
this,  for instance, the~expectation value for the~usual  operator
of space coordinates $\vec{x}$ need not equal zero and  it follows this
(see,for example, in [79-85]).

Now, let us return to the~non-Abelian problem when there exists
the~relationship  of required form:
$$
\Psi_{\epsilon jm\delta }(-\vec{x} ) =
( \sigma ^{2} \otimes \hat{P}_{bisp.}) \;
\delta \; (-1)^{j+1} \; \Psi _{\epsilon jm\delta }(\vec{x} )
\eqno(5.6a)
$$

\noindent owing to the $N$-reflection symmetry; so that
$$
f(-\vec{x} ) = \; \delta \; \delta' \; (-1)^{j+j'} \;
\bar{\Psi}_{\epsilon jm\delta }(x) \;
\left [\; (\sigma ^{2} \otimes \hat{P}_{bisp.}^{+} ) \; \hat{G}(-\vec{x}) \;
(\sigma ^{2} \otimes \hat{P}_{bisp.} )\; \right ] \;
\Psi _{\epsilon j'm'\delta'}(\vec{x})  \; .
\eqno(5.6b)
$$

\noindent If a~certain quantity
$\hat{G}(\vec{x})$ depending on isotopic coordinates
$$
\hat{G}(\vec{x}) =
\pmatrix{\hat{g}_{11}(\vec{x}) & \hat{g}_{12}(\vec{x})  \cr
         \hat{g}_{21}(\vec{x}) & \hat{g}_{22}(\vec{x})}
\otimes  \hat{G}^{0}(\vec{x})
$$

\noindent obeys  the~following structural condition (what is the~definition
of the~composite scalar and pseudoscalar)
$$
(\sigma ^{2} \otimes \hat{P}_{bisp.}^{+})  \;
\hat{G}(-\vec{x})\;
(\sigma ^{2} \otimes \hat{P}_{bisp.}) = \; \Omega \; \hat{G}(\vec{x})
\eqno(5.6c)
$$

\noindent where $\Omega $ defined to be $+1$   or   $-1$, the~relationship
(5.6b) converts  into
$$
f(-\vec{x}) = \; \Omega \; \delta \;  \delta' \; (-1)^{j+j'} \; f(\vec{x})
\eqno(5.6d)
$$

\noindent that  results  in  the  evident $N$-parity  selection  rules.   For
instance, applying these rules to $\hat{G}(\vec{x}) \equiv  \vec{x}$,
we found out
$
<\Psi _{\epsilon jm \delta }(x) \mid \vec{x} \mid
 \Psi _{\epsilon jm \delta }(x) > \;\;  \sim  \;
[\; 1 \; - \; \delta ^{2} \; (-1)^{2j}\;  ] \; \equiv \;  0 \;  .
$

\subsection*{6. Some additional remarks on $\hat{N}_{A}$ operator
                in Cartesian gauge}

Now we proceed further with  studying  the~reflection  symmetry  for
the~fermion-monopole system and  consider the~question  of
explicit form of the~discrete  operator    $\hat{N}_{\Delta }$
in  several  other  gauges.
All our calculations so far (in sections 3-5)  have been  tied with
the~Schwinger isotopic frame, now let us turn  to  the~unitary Dirac
and the~Cartesian (both isotopic)  gauges.  Simple calculations result in
$$
\hat{\pi }^{D.}_{\Delta } \; = \;
        \pmatrix{    0   & -i\; ( a - i \; b)\;  e^{-i \phi }  \cr
               + i \;( a + i\; b) \; e^{+i\phi }  &  0   }    \;  ,
\eqno(6.1a)
$$
$$
\hat{\pi }^{C.}_{\Delta } =
                 \left ( \begin{array}{cc}
(-i \;\; {\Delta + \Delta^{-1} \over 2} \; + \; i \; {\Delta - \Delta^{-1} \over 2}\cos\theta)&
\; + \; i \sin\theta\; e^{-i\phi} \; {\Delta - \Delta^{-1} \over 2}     \\
+ \; i\; \sin\theta\; e^{+i\phi} \; {\Delta - \Delta^{-1} \over 2}      &
(- i \;\; {\Delta + \Delta^{-1} \over 2} \; - \; i{\Delta - \Delta^{-1} \over 2} \; \cos\theta)
                   \end{array} \right )
\eqno(6.1b)
$$

\noindent and the~corresponding wave functions
$$
\Phi ^{D.\Delta}_{\epsilon jm\delta } =  \;
{e^{- i \epsilon t} \over r} \;
\left [\; e^{+i\phi /2} \; T_{+1/2} \otimes F \; + \; \Delta \; \delta \;
  e^{-i\phi /2} \;  T_{-1/2} \otimes G \;\right ] \; ,
\eqno(6.2a)
$$
$$
\Psi ^{C.\Delta}_{\epsilon jm\delta } (x) =    \;
{e^{- i \epsilon t} \over r} \;
\left [ \;  \pmatrix{ \cos\theta /2 & e^{-i\phi /2} \cr
                           \sin\theta /2 & e^{+i\phi /2}}  \otimes  F\; + \;
 \; \Delta \; \delta \;
                 \pmatrix{ \- \sin\theta /2 & e^{-i\phi /2} \cr
               \cos\theta /2 & e^{+i\phi /2}} \otimes  G \;\right ] \; .
\eqno(6.2b)
$$

\noindent It is convenient  for  our  further  work  to  rewrite  the
expression (6.1b) for the~matrix $\hat{\pi }^{C.}_{\Delta }$  in the~form
$$
   \hat{\pi }^{C.}_{\Delta } =
\left [\; -i \; {\Delta + \Delta ^{-1} \over 2} \; +
 \; i  \; {\Delta - \Delta ^{-1} \over 2} \;
( \vec{\sigma } \; \vec{n}_{\theta ,\phi } )\;\right ] \; .
\eqno(6.3)
$$

\noindent Setting  $\Delta = 1$,  it follows from  (6.3) that
$\hat{\pi }^{C.}_{\Delta } = - i\; I$.
Therefore, the~above $N$-reflection  operator  in  the~Cartesian
gauge takes on the form  (at $\Delta = 1$)
$$
\hat{N}^{C.} = \; ( - i \; I) \otimes \hat{P}_{bisp.} \otimes \hat{P}
\eqno(6.4)
$$

\noindent thus, the $\hat{N}^{C.}$ does not involve any transformation
on the~isotopic coordinates.  In other words,  the~ordinary $P$-reflection
operator for a~bispinor field can be diagonalized upon the~composite
(doublet) wave  functions.  But  this  fact   is  not  of  primary
or conceptualizable importance; mainly because it is not gauge invariant.
Therefore,  relying on this relationship,  we  cannot
come to  the~conclusion  that  non-Abelian  problem  of  monopole
discrete symmetry amounts to the~Abelian (monopole free) problem of
discrete symmetry.

Furthermore, that non-Abelian theory's discrete symmetry
features  have no relationship to the~Abelian monopole case.
The clue to understanding  this  is  that
the~Abelian fermion-monopole wave functions $F(x)$  and $G(x)$ (see  in (6.2b))
are represented in the~non-Abelian functions
 $\Psi ^{C.\Delta }_{\epsilon jm\delta }(x)$
  only  as constructing elements
$$
\Psi ^{C. \Delta }_{\epsilon jm\delta } (x) = \;
{e^{-i\epsilon t}\over r}\;
\left [\; T_{+1/2} \otimes \left (\; \cos \theta /2\; e^{-i\phi /2} \;F(x) \; + \;
\; \Delta \;  \delta \;\sin \theta /2 \;e^{-i\phi /2} G(x)\; \right )\;\; + \right.
$$
$$
\left. T_{-1/2} \otimes \left ( \;\sin \theta /2 \;e^{+i\phi /2}\; F(x)\; + \;
  \Delta \; \delta \; \cos \theta /2 \; e^{+i\phi /2} \;G(x)\; \right ) \;\right ]
\eqno(6.5)
$$

\noindent but the~multiplying functions  at
$( T_{+1/2} \otimes)$   and $(T_{-1/2} \otimes )$ , in themselves,
cannot be obtained by any $U(1)$-gauge transformation from the~real
Abelian particle-monopole functions $F(x)$  and $G(x)$,
and  we  should  set  a~higher
value on this than on the~form of $\hat{N}^{C.}$ in (6.4).
In other words, at the~price of  the~gauge transformation used above
 ( $S. \rightarrow  C.$),   we  only  have
carried the non-null action upon isotopic coordinates  (generated  by
$N^{S.}$-inversion) into a~null action  upon  these coordinates  and
a~concomitant  vanishing  of  all  individual  Abelian-like qualities
(belonging solely to the $F(x)$ and $G(x)$).

The relation (6.5) is interesting from another standpoint: it is convenient
to produce some factorizations of the~doublet-fermion functions by
the~Abelian fermion functions and the~isotopic vectors $T_{\pm 1/2}$.
Indeed, taking into account the~known recursive relations  [66]
$$
\cos{\beta \over 2} \; e^{i(\alpha + \gamma)/2} \;
D ^{j}_{m+1/2,m'+1/2}(\alpha,\beta, \gamma) \; =
{ \sqrt{(j+m+1/2)(j+m'+1/2)} \over 2j+1} \;
D^{j-1/2}_{m,m'}(\alpha,\beta, \gamma)\; + \;
$$
$$
{ \sqrt{(j-m+1/2)(j-m'+1/2)} \over 2j+1} \;
D^{j+1/2}_{m,m'}(\alpha,\beta, \gamma) \; ;
$$
$$
\sin{\beta \over 2} \; e^{i(\alpha - \gamma)/2}
D ^{j}_{m+1/2,m'-1/2}(\alpha,\beta, \gamma)  =
-\; { \sqrt{(j+m+1/2)(j-m'+1/2)} \over 2j+1}
D^{j-1/2}_{m,m'}(\alpha,\beta, \gamma)\; + \;
$$
$$
{ \sqrt{(j-m+1/2)(j+m'+1/2)} \over 2j+1} \;
D^{j+1/2}_{m,m'}(\alpha,\beta, \gamma) \;
$$

\noindent the~representation (6.5) can be transformed into the~form:
$$
\Psi ^{C. \Delta }_{\epsilon jm\delta } (x) =
\; {e^{-i\epsilon t}\over r}\; \times
$$
$$
\left [\; \; T_{+1/2} \otimes \;  \; {\sqrt{j+m} \over 2j+1} \;
\left ( \begin{array}{r}
(\; \sqrt{j+1} \; f_{1} \; + \; \delta \; \Delta \; \sqrt{j}\; f_{4} \; ) \\                         \\
(\;\sqrt{j}\; f_{2}  \;    + \; \delta \; \Delta \; \sqrt{j+1}\; f_{3}\;) \\
+ \; \delta \; \Delta \;
(\; \sqrt{j} \; f_{2}\; +\; \delta \; \Delta^{-1}\; \sqrt{j+1}\; f_{3}\;)  \\
+\; \delta \; \Delta \;
(\; \sqrt{j+1} \; f_{1}\; +\; \delta \; \Delta^{-1}\; \sqrt{j}\; f_{4}\;)
\end{array} \right ) \; \left ( D^{j-1/2}_{-m+1/2}\right ) ;\; +   \right.
$$
$$
T_{+1/2} \otimes \; {\sqrt{j-m+1} \over 2j+1} \;
\left ( \begin{array}{r}
(\;\sqrt{j}\; f_{1} \; - \; \delta \; \Delta \; \sqrt{j+1} \; f_{4}\;  )  \\
(\; \sqrt{j+1} \; f_{2} \; -\; \delta \; \Delta \; \sqrt{j} \; f_{3}\; ) \\
-\; \delta \; \Delta \;
 (\; \sqrt{j+1} \; f_{2} \; - \; \delta \; \Delta^{-1} \; \sqrt{j} \; f_{3}\; )  \\
-\; \delta \; \Delta \;
(\; \sqrt{j}\; f_{1}\; -\; \delta \; \Delta^{-1}\; \sqrt{j+1} \; f_{4}\; )
\end{array} \right ) \; \left ( D^{j+1/2}_{-m+1/2}\right ) ;\;  +
$$
$$
T_{-1/2} \otimes \; {\sqrt{j-m} \over 2j+1} \;
\left ( \begin{array}{r}
(\;- \sqrt{j+1} \; f_{1} \; + \; \delta \; \Delta \;\sqrt{j}\; f_{4}\; )  \\
(\;-\sqrt{j}\; f_{2} \; + \; \delta \; \Delta \; \sqrt{j+1}\;  f_{3})     \\
-\; \delta \; \Delta \;
(\;- \sqrt{j} \; f_{2} \; + \; \delta \; \Delta^{-1} \; \sqrt{j+1}\; f_{3}\;) \\
-\;\delta \; \Delta \;
 (\; -\sqrt{j+1} \; f_{1} \; + \; \delta \; \Delta^{-1}\;  \sqrt{j}\; f_{4})
\end{array} \right ) \;\left ( D^{j-1/2}_{-m-1/2}\right ) \;\;  +
$$
$$
\left. T_{-1/2} \otimes \; {\sqrt{j+m+1} \over 2j+1} \;
\left ( \begin{array}{r}
(\;\sqrt{j}\; f_{1} \; + \; \delta \; \Delta \; \sqrt{j+1} \; f_{4}\; )  \\
(\; \sqrt{j+1} \; f_{2} \; +\; \delta \; \Delta \; \sqrt{j} \; f_{3}\; ) \\
+\; \delta \; \Delta \;
(\; \sqrt{j+1} \; f_{2} \; + \; \delta \; \Delta^{-1} \; \sqrt{j} \; f_{3}\; ) \\
+\; \delta \; \Delta \;
(\; \sqrt{j}\; f_{1}\; +\; \delta \; \Delta^{-1}\; \sqrt{j+1} \; f_{4}\; )
\end{array} \right )  \;\left ( D^{j+1/2}_{-m-1/2}\right ) \;  \right ]
\eqno(6.6)
$$

\noindent where, for the sake of brevity,  the~angular dependence in the~wave functions is described
with the~use of the~following symbolic designation (here there are four
different possibilities)
$$
D^{j \pm 1/2}_{-m \pm 1/2} \; = \;
\left ( \begin{array}{c}
         D^{j \pm 1/2}_{-m \pm 1/2,-1/2} \\
         D^{j \pm 1/2}_{-m \pm 1/2,+1/2} \\
         D^{j \pm 1/2}_{-m \pm 1/2,-1/2} \\
         D^{j \pm 1/2}_{-m \pm 1/2,+1/2}
\end{array} \right )        \; .
$$

Now, remembering the~equality $(a^{2} + b^{2}) = 1$, let us   introduce
a~new variable  $A$  defined by  $\cos A = a$  and  $\sin A = b$,  so
the~above operator $\hat{\pi }^{C.}_{\Delta }$ is expressed as
$$
\hat{\pi }^{C.}_{A} =
 (- i ) \; \exp [ \;i A \; \vec{\sigma } \; \vec{n}_{\theta ,\phi }\; ]
\eqno(6.7)
$$

\noindent here the  $A$ is a~complex parameter  due to  the~quantities  $a$
and $b$ are complex ones.  Correspondingly, using the~notation according to
$$
\sqrt{j+1}\; f_{1} \; + \; \delta \; e^{iA} \; \sqrt{j}\; f_{4}\; = \;
K^{A}_{\delta}  \; , \qquad
\sqrt{j} \; f_{2} \; + \; \delta \; e^{iA} \; \sqrt{j+1}\;  f_{3}\; = \;
L^{A}_{\delta}  \; ,
$$
$$
\sqrt{j}\; f_{1} \; - \; \delta \; e^{iA} \; \sqrt{j+1} \; f_{4}\; =  \;
M^{A}_{-\delta} \; , \qquad
\sqrt{j+1} \; f_{2} \; - \; \delta \; e^{iA} \; \sqrt{j} \; f_{3}\;  =\;
N^{A}_{-\delta}
$$

\noindent the~representation (6.6) can be rewritten as follows
$$
\Psi ^{C. A}_{\epsilon jm\delta }(x)\; =
\; {e^{-i\epsilon t}\over r}\; \times
$$
$$
\left [\;  T_{+1/2} \otimes \; {\sqrt{j+m} \over 2j+1} \;
\left ( \begin{array}{r}
K^{+A}_{\delta}  \\
L^{+A}_{\delta}  \\
\delta \; e^{iA} \; L^{-A}_{\delta}   \\
\delta \; e^{iA} \; K^{-A}_{\delta}
\end{array} \right )\; \left ( D^{j-1/2}_{-m+1/2}\right )  \;\; +    \right.
$$
$$
T_{+1/2} \otimes \; {\sqrt{j-m+1} \over 2j+1} \;
\left ( \begin{array}{r}
M^{+A}_{-\delta}   \\
N^{+A}_{-\delta}   \\
-\; \delta \; e^{iA}\; N^{-A}_{-\delta}    \\
-\; \delta \; e^{iA} \; M^{-A}_{-\delta}
\end{array} \right )\; \left ( D^{j+1/2}_{-m+1/2}\right ) \; \; +
$$
$$
T_{-1/2} \otimes \; {\sqrt{j-m} \over 2j+1} \;
\left ( \begin{array}{r}
-K^{+A}_{-\delta}   \\
-L^{+A}_{-\delta}   \\
\delta \; e^{iA} \; L^{-A}_{-\delta}  \\
\delta \; e^{iA} \; K^{-A}_{-\delta}
\end{array} \right ) \;\left ( D^{j-1/2}_{-m-1/2}\right ) \;\; +
$$
$$
\left. T_{-1/2} \otimes \; {\sqrt{j+m+1} \over 2j+1} \;
\left ( \begin{array}{r}
M^{+A}_{\delta}  \\
N^{+A}_{\delta}  \\
\delta \; e^{iA} \;  N^{-A}_{\delta}    \\
\delta \; e^{iA} \;  M^{-A}_{\delta}
\end{array} \right ) \;\left ( D^{j+1/2}_{-m-1/2}\right ) \; \;\right ] \; .
\eqno(6.8)
$$

\noindent When $A = 0$, then all the~formulas in (6.8) will be significantly
simplified,
so that the~familiar sub-structure  of electronic wave functions with fixed
P-parity can easily be seen (see (8.4)). This is what one might expect because,
if $A=0$,  then the~operator $N_{A=0}^{C.}$   does not involve
any isotopic transformation.   The~latter might be a~source of some speculation
about an~extremely significant role of the~Abelian $P$-symmetry in
the~non-Abelian model. However, one should remember that a~genuinely
Abelian fermion   $P$-symmetry implies
both a~definite explicit expression for $P$-operation and definite properties
of  the~corresponding wave functions. The~above decomposition (6.8)
of fermion doublet wave functions in terms
of Abelian fermion functions and unit isotopic vectors shows that
the~usual Abelian fermion particle wave functions
and non-Abelian doublet ones belong to substantially  different classes.
(see (6.9) at $A=0$); so that the~Abelian like $P$-operation plays   only
a~subsidiary role in forming the~whole composite wave functions; besides,
as was mentioned above, this role will be completely negated in other
isotopic gauges.

By the way, an~analogous principle of checking
the~property of functional space (in particular, as to whether or not
the~relevant functions  are single-valued ones,
apart from any possible gauge transformations) might serve as a~guideline
argument to prevent some serious discussion (if not speculation)
on fermion interpretation for bosons as well as boson-like interpretation
for fermions [20-23] when, in the~Abelian model,  the~monopole is in effect
 and the~case of half-integer $eg$-values is realized.
Of course, such possibilities seem striking
and attractive for every physicist, however they are correct and quite
satisfactory  ones only at first
glance. Evidently, such monopole-based fermions or bosons being produced
from the usual bosons and fermions respectively, turn out to be tied with
functional spaces which are absolutely different from those used in reference
with the~usual (fermion or boson) particles. In addition, the~Lorentz
group-based transformation characteristics of such new `fermions and boson',
in reality, will completely negate their new nature and will be dictated
by their old classification  assignment.

Also (in the~author's opinion), the vastly discussed
producing `spin from isospin' [20-23] belongs to the same class of striking
but hardly realized possibilities.  Many arguments against might be formulated;
the~simplest one is as follows: the~correct understanding of the~meaning of
Lorentzian spin presupposes quite definite properties of this characteristic
under Lorentz group transformations; however, such a~`spin', being produced
from isospin, does not obey these regulations. Else one theoretical criterion
might  be given: placing such an~object into the background of any Rimannian
(curved) space-time. Evidently, the gravitational field will ignore any
boson-fermion inverting based on the~above mechanism.

\subsection*{7.  Free parameter at discrete symmetry \\ and
                 $N_{A}$-parity selection rules}

Now we proceed with analyzing the~totality of the~discrete
operators $\hat{N}_{A}$, which all are suitable  for  separation  of  variables.
What is the~meaning of the~parameter $A$? In other words,  how
can this $A$ manifest itself and why does such an~unexpected  ambiguity
exist?

The $A$   fixes  up  one  of  the~complete
set of operators  $\{\; i\; \partial _{t} ,\; \vec{J}^{2} ,
\; J_{3} ,\;  \hat{N}_{A} ,\; \hat{K} \;\}$,
and correspondingly this~$A$ also labels all  basic
wave functions.
It is obvious, that this parameter  $A$  can manifest itself in
matrix elements of physical  quantities. To see this, it
suffices  to  look  at the general structure
of the~relevant  expectation value  of   those
observables\footnote{To be exact, any variations
in this $A$ will lead to alteration in normalization conditions for
the relevant wave function $\Psi ^{A}_{\epsilon jm \delta \mu }$
(see in Sec.9), so that this circumstance should
be taken into account; but for simplicity, we pass over those alterations.}
(the $\epsilon , J, M$  are omitted):
$$
\bar{G} \; = \; < \Psi ^{A}_{\delta \mu }
\mid \; \hat{G} \; \mid  \Psi ^{A}_{\delta \mu } > \; = \;
< T_{+1/2} \otimes  \Phi^{(+)} _{\mu }(x) \; \mid \hat{G}
\mid \;  T_{+1/2} \otimes \Phi^{(+)} _{\mu }(x) > \; +
$$
$$
\mid e^{iA}\; \mid \;\;< T_{-1/2} \otimes  \Phi^{(-)} _{\mu }(x) \mid \;
\hat{G} \;
\mid  T_{-1/2} \otimes  \Phi^{(-)} _{\mu }(x) > \;\;  +
$$
$$
2 \; \delta \;  \mu \;\; Re  \;
\left [\; e^{iA}\; < T_{+1/2} \otimes  \Phi^{(+)} _{\mu}(x) \mid \;
 \hat{G} \; \mid  T_{-1/2} \otimes  \Phi^{(-)} _{\mu }(x) > \;\right ] \;  .
\eqno(7.1)
$$

\noindent  If such a~$\hat{G}$  has
the~ diagonal isotopic structure
$$
\hat{G}(x)=
\left ( \begin{array}{cc}
               \hat{g}_{11}(x)  &  0  \\
                  0   &       \hat{g}_{22}(x)
\end{array} \right )
\otimes  \hat{G}^{0}(x)
$$

\noindent then the third term in (7.1) vanishes and the~matrix element  only
depends on $\mid e^{iA}\mid$.

As  a~simple  example  let  us  consider  a~new   form of
the~above-mentioned selection rules depending on the~$A$-parameter. Now,
the~matrix element examined is
$$
\int \bar{\Psi} ^{A}_{\epsilon JM\delta \mu }(x) \; \hat{G}(x)\;
     \Psi ^{A}_{\epsilon J'M'\delta'\mu'}(x) \;  dV \; \equiv   \;
\int r^{2} dr \int f^{A}(\vec{x}) \; d \Omega
$$

\noindent then
$$
f^{A}(-\vec{x} ) = \; \delta \;  \delta'\; (-1)^{J+J'} \;
                \bar {\Psi}^{A}_{\epsilon JM\delta \mu }(x) \; \times
$$
$$
\left [\;(a^{*} \sigma ^{1} + b^{*} \sigma ^{2}) \otimes \hat{P}_{bisp.}
         \hat{G}(-\vec{x}) \;
 (a \sigma ^{1} + b \sigma ^{2}) \otimes \hat{P}_{bisp.} \; \right ]  \;
\Psi ^{A}_{\epsilon J'M'\delta'\mu'} (\vec{x}) \; .
\eqno(7.3a)
$$

\noindent If this $\hat{G}$  obeys the condition
$$
\left [\;(a^{*} \sigma ^{1} + b^{*} \sigma ^{2}) \otimes
\hat{P}_{bisp.}\;] \; \hat{G}(-\vec{x}) \; [\; (a \sigma ^{2} + b \sigma ^{1})
 \otimes \hat{P}_{bisp.}\; \right ]\; = \; \Omega ^{A} \; \hat{G}(\vec{x})
\eqno(7.3b)
$$

\noindent which is equivalent to
$$
\left ( \begin{array}{cc}
 e^{i(A-A^{*})}  \; \hat{g}_{22} (-\vec{x})     &   \;
 e^{-i(A+A^{*})} \; \hat{g}_{21}(-\vec{x})  \\
 e^{i(A+A^{*})}  \; \hat{g}_{12} (-\vec{x})  &      \;
 e^{-i(A-A^{*})} \; \hat{g}_{11}(-\vec{x})
\end{array}  \right ) \otimes
$$
$$
\left [ \; \hat{P}_{bisp.} \; \hat{G}^{0}(-\vec{x})\; \hat{P}_{bisp.} \;
 \right ] \;
 = \;
\Omega^{A}  \left ( \begin{array}{cc}
 \hat{g}_{11}(\vec{x})     &
 \hat{g}_{12}(\vec{x})     \\
 \hat{g}_{21} (\vec{x}) &
 \hat{g}_{22}(\vec{x})
\end{array}  \right ) \otimes  \hat{G}(\vec{x})
\eqno(7.3c)
$$

\noindent where $\Omega ^{A} = + 1$ or  $-1$,
 then the relationship (7.3a) comes to
$$
f^{A}(-\vec{x})\;  = \;  \Omega ^{A}\; \delta \;  \delta'\; (-1)^{J+J'} \;
 f^{A}(\vec{x}) \; .
\eqno(7.3d)
$$

\noindent Taking into account (7.3d), we bring the matrix element's
 integral above to the~form
$$
\int \bar{\Psi}^{A}_{\epsilon JM\delta \mu }(x) \; \hat{G}(x) \;
\Phi ^{A}_{\epsilon J'M'\delta'\mu'}(x) \;dV \; = \;
\left [\; 1\; + \; \Omega ^{A} \; \delta \; \delta'\; (-1)^{J+J'} \; \right ] \;
\int_{V_{1/2}} f^{A}( \vec{x} ) \;dV
\eqno(7.4a)
$$

\noindent where the~integration in the~right-hand side is done on the~half-space.  This
expansion  provides the following selection rules:
$$
M E \equiv  0  \qquad \longleftrightarrow  \qquad
\left [\; 1\; +\; \Omega^{A} \; \delta \; \delta' \; (-1)^{J+J'}\; \right ]
 \; =\; 0 \;\; .
\eqno(7.4b)
$$

\noindent It is to be especially  emphasized  that  the~quantity
 $\Omega ^{A}$,
defined to be $+ 1$ or  $- 1$, is not the~same as the~analogous  that
$\omega $ in (5.6d).  These $\omega$  and $\Omega^{A}$  involve  their own
particular limitations on composite scalar or pseudoscalar because they imply
respective (and rather specific) configurations of their isotopic parts,
obtained by delicate  fitting all the~quantities $\hat{g}_{ij}$.
 Therefore, each of those  $A$  will  generate  its own distinctive
selection rules.

\subsection*{8. Existence of the parameter $A$
                  and isotopic chiral symmetry}

Where does this $A$-ambiguity come from and what is the meaning
of this parameter $A$? To proceed further with this problem, one is
to realize that the~all different values for $A$  lead to  the  same
whole functional space; each fixed value for $A$  governs  only
the~basis states $\Psi ^{A}_{\epsilon JM\delta \mu }(x)$ associated with $A$,
but  with  no   change in~the whole space. Connection  between any two sets
of functions $\{ \Psi(x)  \}^{A}$ and $\{ \Psi(x) \}^{A'}$
is characterized by
$$
\Psi ^{A'\; S.}_{\epsilon JM\delta \mu } =
U_{S.}(A'-A) \;\Psi ^{A \;S.}_{\epsilon JM\delta \mu }(x) \; , \qquad
U _{S.}(A'-A) =  \; e^{-iA} \;
\left ( \begin{array}{cc}
 e^{iA}  & 0 \\
  0  &  e ^{iA'}
\end{array}   \right ) \otimes  I    \; .
\eqno(8.1a)
$$

\noindent Besides, it is readily verified that the operator $\hat{N}^{S.}_{A}$
(depending on $A$) can be obtained  from the~operator $\hat{N}^{S.}$  as follows
$$
\hat{N}^{S.}_{A} \;  = \; U_{S.}(A) \;\; \hat{N}^{S.} \; U^{-1}_{S.}(A) \; .
\eqno(8.1b)
$$

\noindent The matrix $U_{S.}(A'-A)$  is so simple only  in  the~Schwinger
basis; after translating that into Cartesian one we will have
$$
\Psi ^{A'\;C.}_{\epsilon JM\delta \mu } (x) \; = \;
U_{C.} (A'-A) \; \Psi ^{A\; C.}_{\epsilon JM\delta \mu }(x)  \; ,
\eqno(8.1c)
$$
$$
S_{C.}=  {1 \over \Delta}
                       \left ( \begin{array}{cc}
(\Delta  \cos ^{2} \theta /2 \; + \; \Delta' \sin ^{2}\theta /2 ) &
{1\over 2} \; (\Delta \; - \; \Delta')\; \sin \theta \; e^{-i\phi } \\
{1\over 2} \; (\Delta \; - \; \Delta' ) \; \sin \theta \; e^{+i\phi }  &
(\Delta' \cos ^{2}\theta /2 \; +\; \Delta  \sin ^{2} \theta /2 )
\end{array} \right ) \otimes I
$$

\noindent and $S_{C.}(A'-A)$  satisfies the equation
$$
\hat{N}^{C.}_{A} \; = \; U_{C.} (A) \;\; \hat{N}^{C.}  \;
\hat{U}^{-1}_{Cart.} (A) \; .
\eqno(8.1d)
$$

In connection with everything said above on parity  selection
rules  and  just  `unexpectedly'  established
relationship (8.1b) (or (8.1d)), we  need  to  think  this  over
again before finding a~conclusive answer.
Let us begin from some  generalities.  As  well  known,  when
analyzing any Lie group problems (or  their  algebra's)  there
indeed exists a~concept of equivalent representations:
 $U \; M_{k} \; U^{-1} \; = \; M'_{k} \; \rightarrow \;  M_{k} \; \sim \; M'_{k}$.
In this context,  the two  sets  of
operators $\{ J^{S.}_{i}, \; \hat{N}^{S.} \}$  and
$\{ J^{S.}_{i}, \; \hat{N}^{S.}_{A} \}$  provide   basically  just the same
representation of the $O(3.R)$-algebra
$$
\{ J^{S.}_{i}, \; \hat{N}^{S.}_{A} \} \; = \; U_{S.}(A) \;
\{ J^{S.}_{i}, \; \hat{N}^{S.} \} \; U^{-1}_{S.}(A) \; .
\eqno(8.2a)
$$

\noindent The totally different situation  occurs  in the context of the use
of those two operator sets as physical observables concerning the  system
with the~fixed Hamiltonian
$$
\{ \vec{J}^{2}_{S.}, \; J^{S.}_{3}, \; \hat{N}^{S.} \}^{\hat{H}}
\qquad and \qquad
\{ \vec{J}^{2}_{S.}, \; J^{S.}_{3}, \; \hat{N}^{S.}_{A} \}^{\hat{H}} \; .
\eqno(8.2b)
$$

\noindent Actually, in this case the two operator sets  represent different
observables at the  same  physical  system:  both  of  them  are
followed by the~same Hamiltonian $\hat{H}$  and also  lead  to  the~same
functional space, changing  only its basis vectors
 $\{ \Psi _{\epsilon JM\delta \mu }(x) \}^{A}$.
Moreover,
in the~quantum mechanics it  seems always possible to  relate
two  arbitrary  complete  sets  of  operators  by   some   unitary
transformation:
$$
\{ \hat{X}_{\mu } , \; \mu  = 1, \ldots \}^{\hat{H}}
 \;\;  \rightarrow  \;\;
\{ \hat{Y}_{\mu } , \; \mu  = 1, \ldots \}^{\hat{H}}    \; ,
\{ \Phi _{x_{1} \ldots x_{s}} \}  \;\;  \rightarrow  \;\;
\{ \Phi _{y_{1} \ldots y_{s}} \} \; .
$$

\noindent But arbitrary transformations $U$ cannot generate, through
converting
$U \; \{ \hat{X}_{\mu } \} \; U^{-1} \; = \; \hat{Y}_{\mu }$, a~new
complete  set  of  variables;
instead, only some Hamiltonian symmetry's operations are suitable for
this: $U \; \hat{H} \; U^{-1} = H$.

In this connection, we may recall  a more  familiar  situation
for Dirac massless field [95,96]. The wave equation for  this  system
was earlier mentioned   (see  Sec.~2) and that  has the~form
$$
i \bar{\sigma}^{\alpha}(x)\; (\partial _{\alpha } +
\bar{\Sigma}_{\alpha })\; \xi (x) = 0 \;\; , \qquad
i \sigma ^{\alpha }(x) \; (\partial _{\alpha }  + \Sigma _{\alpha })\; \eta (x) = 0 \; .
\eqno(8.3a)
$$

\noindent If the function $\Phi (x) = ( \xi (x) , \; \eta (x) )$ is subjected
to the~transformation
$$
\pmatrix{\xi'(x) \cr \eta'(x)} =
\pmatrix{I  & 0 \cr 0 & z\;I}  \pmatrix{\xi (x) \cr \eta (x)}
\eqno(8.3b)
$$

\noindent where $z$ is an arbitrary complex number,  then  the  new  function
$\Phi'(x) = ( \xi'(x), \eta'(x) )$ satisfies again the equation in the form
(8.3a). This manifests the Dirac massless  field's  symmetry  with
respect to  the~transformation
$$
\hat{H}' = U \; \hat{H} \;  U^{-1}  = \hat{H} , \qquad
\Phi'(x) = U \; \Phi (x)   \; .
\eqno(8.3c)
$$

\noindent The existence of the~symmetry raises the~question  as  to
whether this symmetry affects  determination of  complete set  of
diagonalized  operators   and  constructing spherical wave solutions.
These  solutions,  conformed  to  diagonalizing   the~usual
bispinor $P$-inversion operator, in addition to
$\vec{j}^{2}$ and $j_{3}$, are as in (5.4a) at $eg =0$.
In the same time, other spherical solutions,  together
with corresponding diagonalized discrete operator, can be produced:
$$
\Phi^{z}_{\epsilon jm \delta } = {e^{-i\epsilon t}\over r}
\left ( \begin{array}{r}
                  f_{1} \; D^{j}_{-m, -1/2} \\
                  f_{2} \; D^{j}_{-m, +1/2} \\
    z \; \delta \; f_{2} \; D^{j}_{-m, -1/2} \\
    z \; \delta \; f_{1} \; D^{j}_{-m, +1/2}
\end{array} \right ) \;  ,
\eqno(8.4a)
$$
$$
U \;\left (\; \hat{P}^{sph.}_{bisp.} \otimes \hat{P} \; \right ) \; U^{-1}\;  = \;
\left [\; {1\over 2} \; ( z + {1\over z} )\; (- \gamma^{5} \gamma^{1}) \; + \;
  {1\over 2} \; ( z - {1\over z} ) \; (- \gamma^{1}) \;\right ] \otimes  \hat{P} \; .
\eqno(8.4b)
$$

\noindent Introducing  another  complex  variable $A$   instead  of
 the~parameter $z :\;  z = (\cos A + i \sin A) = e^{iA}$; so that
 the~operator from (8.4b) is  rewritten in the~form
$$
( \cos  A \; + \; i \sin  A \; \gamma ^{5} )  \;
(- \gamma ^{5} \; \gamma ^{1}) \otimes \hat{P} \;
\equiv e^{+iA \gamma ^{5}} \; \hat{P}^{sph.}_{bisp.} \otimes \hat{P}
\eqno(8.4c)
$$

\noindent (8.3b) may be expressed  as follows
$$
\Phi'(x) \; = \; e^{+iA/2} \; exp(+i \gamma ^{5} {A\over 2}) \; \Phi(x)
\eqno(8.4d)
$$

\noindent Evidently, that  translation of the basis of spherical tetrad into
Cartesian tetrad's will preserve the~general structure of (8.4c):
$
e^{+iA \gamma ^{5}} \;\hat{P}^{Cart.}_{bisp.} \otimes \hat{P}\;
$,
since the~gauge  matrix $S(k(x),\bar{k}^{*}(x))$
 and matrix $\gamma ^{5}$ are commutative with  each  other.
In contrast to this, translation of the~isotopic Schwinger frame into
the~Cartesian  that does change the~form  $\hat{N}_{A}$: the~initial one is
$$
\hat{N}^{S.}_{A} \; = \; (e^{-i A \; \sigma^{3}} \hat{\pi }^{S.})
\otimes \hat{P}_{bisp.} \otimes \hat{P}
\eqno(8.5a)
$$

\noindent and the finishing form is
$$
\hat{N}^{C.}_{A} \; =\; (-i) \; \exp \left [\; - i \; A \; \vec{\sigma }\;
 \vec{n}_{\theta ,\phi } \right ]
\otimes \hat{P}_{bisp.} \otimes \hat{P}  \; .
\eqno(8.5b)
$$

\noindent The appearance of this dependence on variables $\theta,\phi$ comes
from noncommutation of the~gauge transformation $B(\theta ,\phi )$  and matrix
$\sigma^{3}$ (the~latter plays the~role  of $\gamma ^{5}$ in case of Abelian
chiral symmetry (8.4d)).

The transformation $U(A)$, after translating it to the~Cartesian basis
(see (8.1c)), can be brought to the~form
$$
 U ^{C.}(A)\; = \;\left [\; {1 + e^{iA} \over 2} \; + \; {1 - e^{iA} \over 2} \;\vec{\sigma} \;
        \vec{n}_{\theta ,\phi } \;\right ]   \; .
$$

\noindent Separating out the factor $e^{iA/2}$  in the right-hand  side  of
this formula, we can rewrite the $U^{C.}$  in the~form
$$
U^{C.} \; = \; e^{iA/2} \; \exp \left [ -\; i \; {A\over 2}\; \vec{\sigma }\;
 \vec{n}_{\theta ,\phi }\right ]
\eqno(8.6)
$$

\noindent where the second factor lies in the~(local) spinor representation
of  the~3-dimensional complex rotational group $SO(3.C)$.
This matrix  provides a~very  special
transformation upon the~isotopic fermion doublet and can be thought
of  as  an~analogue of the~Abelian chiral symmetry transformation; it may be
also termed  as  the~transformation  of {\em isotopic (complex)  chiral} symmetry.
This symmetry leads  to the $A$-ambiguity (8.5) and
permits to choose an~arbitrary reflection operator from the~totality
$\{ \hat{N}_{A} \}$.

\subsection*{9. Complex values of the $A$ and interplay between
               the~quantum  mechanical superposition principle and
               self-conjugacy requirement}

In this section, let us look closely at  some  qualitative peculiarities
of the~above considered  $A$-freedom  placing  special notice to the~division
of $A$-s into the~real and complex values. It is convenient to work at
this matter in the Schwinger unitary basis. Recall that the $A$-freedom
tell us  that  simultaneously  with
$\hat{H}, \vec{j}^{2} , \hat{j}_{3}$,
else one discrete operator $ \hat{N}_{A} $, that  depends  generally
on a~complex number $A$, can be diagonalized on the wave  functions.
Correspondingly, the basis functions associated with the complete set
 ( $ \hat{H} ,\vec{j}^{2}, \hat{j}_{3},\hat{N}_{A}$ )
besides being certain determined functions of the~relevant quantum
numbers $(\epsilon , j, m, \delta  )$, are subject to the~$A$-dependence.

In other words, all different values of this  $A$ lead to different
quantum-mechanical bases of the~system. There exists a~set of
possibilities, but one can relate every two of them  by means of
a~respective linear transformation. For example, the~states
$\Psi ^{A}_{\epsilon jm\delta }(x)$
decompose into the following linear combinations of the~initial  states
$\Psi ^{A=0}_{\epsilon jm\delta }(x)$
(further,  this  $A=0$ index  will  be omitted):
$$
\Psi ^{A}_{\epsilon jm\delta }(x) \; =  \;
  \left [ \; {{1 + \delta  e^{iA} } \over 2} \; \Psi _{\epsilon jm,+1} \; +  \;
  {{1 - \delta  e^{iA} } \over 2} \; \Psi _{\epsilon jm,-1}\; \right ]   \; .
\eqno(9.1)
$$

\noindent One should give heed to that, no matter  what an $A$   is
(either real or complex one), the~new states (9.1), being linear
combinations of the initial states, are permissible as well as
old ones. This added aspect of the~allowance of the~complex values
for $A$ conforms to the quantum-mechanical superposition principle:
the~latter presupposes that arbitrary complex coefficients $c_{i}$
in a~linear combination of some basis states
$\; \Sigma  c_{i} \Psi _{i}\;$   are acceptable.

However, an~essential and subtle distinction between real and
complex $A$-s comes straightforward to light  as we turn to
the~matter of normalization and orthogonality for
$\Psi ^{A}_{\epsilon jm\delta }(x)$. An~elementary calculation gives
$$
< \Psi^{A}_{\epsilon jm,\delta} \mid \Psi^{A}_{\epsilon jm,\delta}  >
= { {1 + e^{i(A-A^{*}) }} \over 2}  \Psi_ {\epsilon j m,+1 } \; ; \;\;
< \Psi^{A}_{\epsilon jm,\delta} \mid \Psi^{A}_{\epsilon jm,-\delta} >
= { {1 - e^{i(A-A^{*}) }} \over 2}
\eqno(9.2)
$$

\noindent i.e. if $A \neq  A^{*}$  then the~normalizing condition
for   $\Psi ^{A}_{\epsilon jm\delta }(x)$  does
not coincide with that for   $\Psi _{\epsilon jm\delta }(x)$,
and what is more, the~states
$\Psi ^{A}_{\epsilon jm,-1}(x)$  and  $\Psi ^{A}_{\epsilon jm,+1}(x)$
are not mutually orthogonal. The~latter means that we face here
the~non-orthogonal basis in Hilbert space and the pure imaginary part of
the~$A$ plays a~crucial role in the~description of its non-orthogonality
property.

The {\em oblique} character of the  basis $\Psi ^{A}_{\epsilon jm\delta }(x) $
 (if $A \neq A^{*}$ )
exhibits its very essential qualitative distinction from
{\em perpendicular} one for $\Psi _{\epsilon jm\delta }(x)$.
 However, those  specific bases
in quantum mechanics, though not  being of very common use and having
a~number of peculiar features, are allowed to be exploited in conventional
quantum theory. Even more, in a~sense,
the~existence itself of the~non-orthogonal bases in the~Hilbert space
represents a~direct consequence of the~quantum-mechanical superposition
principle\footnote{For this reason, a~prohibition against complex $A$-s  could
be  partly a~prohibition against the conventional superposition
principle too (narrowing it); since all complex values for $A$,
having forbidden,
imply specific limitations on two coefficients in (9.1); but those are not
presupposed by the~superposition principle itself.}.

Up to this point, the~complex $A$-s seem to be good as well as
the~real ones. Now, it is the~moment  to  point  to some clouds
handing over this part of the~subject. Indeed, as readily
verified, the operator $\hat{N}_{A}$ does not represent a~self-conjugated
(self-adjoint) one\footnote{The author is grateful to Dr. E.A.Tolkachev
for pointing out that it is so.}
$$
< \hat{N}_{A} \Phi (x) \mid  \Psi (x) > \; = \;
< \Phi (x) \mid \;  e^{i(A-A^{*})\sigma_{3}} \; \hat{N}_{A} \; \Psi (x) >.
$$

\noindent It is understandable that this (non-self-conjugacy) property
correlates with the~above-mentioned nonorthogonality
conditions: as well known, a~self-conjugated operator entails
both real its eigenvalues and the~orthogonality of its
eigenfunctions. As already noted, the~eigenvalues of $\hat{N}_{A}$  are
real ones and this conforms to the~general statement that
all inversion-like operators possess the~property of the~kind: if
$\hat{G}^{2} = I$  then $\lambda $  is a real number, as
 $\hat{G} \; \Phi _{\lambda } = \lambda  \; \Phi _{\lambda } )$.

So, we have got into a~point to choose: whether one has  to  reject  all
complex values  for   $A$   and  thereby  narrow (if not violate)
the~one  quantum
mechanical principle of major generality (of superposition) or whether it
is remain to accept all complex $A$-s as well as real ones and
thereby, in turn, stretch another  quantum-mechanical  regulation
about  the~self-adjoint character of {\em physical} quantities.

We have chosen to accept and look into the~second possibility.
In the author's opinion,
one should accord the~primacy of the~general superposition
principle over the~self-adjointness requirement. In  support of this
point of view, there exist clear-cut physical grounds.

Indeed, recall the~quantum-mechanical status of all inversion-like quantities:
they serve always to distinguish two quantum-mechanical states. Moreover,
to those  quantum  variables there not correspond any classical variables;
the~latter correlates with that any classical apparatus  measuring those
discrete variables does not exist whatsoever. In contrast to this, one
should recollect  why  the~self-adjointness  requirement  had been
imposed on physical quantum operators. The reason is that such operators
imply all their eigenvalues to be real. Besides, that limitation
on physical quantum  variables had been put, in the first place,
for quantum variables having their classical counterparts (with
the continuum of classical  values measured). And after this,  in
the~second place, the~discrete quantities  such  as $P$-inversion
and like it were tacitly incorporated into a~set of self-adjoint
mathematical operations, as a~{\em natural} extrapolation.
But one should notice (and the~author inclines to place a~special emphasis
on this) the~fact that the~single relation  $\hat{N}^{2}_{A} = I$
is completely sufficient that the~eigenvalues of $\hat{N}_{A}$ to be real. In
the~light of this, the~above-mentioned automatic incorporation  of
those discrete operators into a~set of self-adjoint ones does not
seem inevitable.
But admitting this, there is  a~problem to
solve: what is the~meaning of complex expectation  values  of  such
non self-adjoint discrete operators; since,  evidently,
the~conventional formula
 $<\Psi  \mid  \hat{N}_{A} \mid  \Psi  >$
provides  us  with  complex values. Indeed, let
$\Psi (x)$  be  $\Psi (x) = [ m \Psi _{+1}(x) + n \Psi _{-1}(x)]$, then
$$
< \Psi \mid \hat{N}_{A} \mid \Psi > \; = \;
< m \; \Psi_{+1}(x) + n \; \Psi_{-1}(x) \mid m \; \Psi_{+1}(x) - n \; \Psi _{-1}(x) > \; = \;
$$
$$
\left [\; ( m^{*} m - n^{*} n )\; {{1+e^{i(A-A^{*})}} \over 2} \; + \;
  (   n^{*} m - n m^{*} )\; {{1-e^{i(A-A^{*})}} \over 2}\;\right  ] \; .
\eqno(9.3)
$$

\noindent Must one be skeptical about those complex   $ \bar{N}_{A}$ ,
or treat them as physically acceptable quantities? Let us examine this
problem in more detail. It is reasonable to begin with an~elementary
consideration of the~measuring procedure of
the~$\hat{N} = \hat{N}_{A=0}$. Let  a~wave function $\Psi (x)$ decompose
into the~combination
$$
\Psi (x) \; = \;\left [ \; e^{i\alpha } \cos^{2} \Gamma \;  \Psi _{+1}(x) \; + \;
             e^{i\beta  }  \sin^{2} \Gamma \; \Psi _{-1}(x) \; \right ]
\eqno(9.4a)
$$

\noindent where $\alpha $  and $\beta \in  [ 0 , 2 \pi ]$,
and $\Gamma  \in  [ 0 , \pi /2 ]$. For  the $\hat{N}$
expectation value, one gets
$$
  \bar{N} = \; < \Psi  \mid  \hat{N} \mid  \Psi  > \; =
(-1)^{j+1}\; (\cos^{2} \Gamma  - \sin ^{2} \Gamma ) =
(-1)^{j+1} \; \cos 2\Gamma     \; .
\eqno(9.4b)
$$

\noindent From (9.4b), one can  conclude  that  $\bar{N}$, after having  measured,
provides us only with the~information about the~parameter $\Gamma $  at (9.4a),
but does not furnish any information on the~phase  factors $e^{i\alpha }$
and $e^{i\beta }$ (or  their  relative  factor $e^{i(\alpha -\beta )}$).
  This interpretation of measured  $\bar{N}$  as receptacle of
the~quite definite information about superposition coefficients in
the~decomposition
(9.4a), represents one and only physical meaning of  the~$\bar{N}$.

Now, returning to the case of $\hat{N}_{A}$ operation, one  should  put
an~analogous question  concerning the~$\bar{N}_{A}$.
The~material question is: what kind of information about $\Psi (x)$  can be
extracted from the~measured   $\bar{N}_{A}$. It is convenient to rewrite
the~above function  $\Psi (x)$ as a~linear
combination of functions $\Psi ^{A}_{\epsilon jm,+1}$ and
$\Psi ^{A}_{\epsilon jm,-1}$.  Thus inverting the~relations (9.1), we get
$$
\Psi _{\epsilon jm,+1} \; = \;
\left [ \;{{1 + e^{-iA}} \over 2} \; \Psi ^{A}_{\epsilon jm,+1} \; + \;
 {{1 - e^{-iA}} \over 2} \; \Psi ^{A}_{\epsilon jm,-1} \; \right ] \;    ;
$$
$$
\Psi _{\epsilon jm,-1} \; = \;
\left [ \; {{1 - e^{-iA}} \over 2} \; \Psi ^{A}_{\epsilon jm,+1} \; + \;
 {{1 + e^{-iA}} \over 2} \; \Psi ^{A}_{\epsilon jm,-1} \; \right ]
$$

\noindent and then $\Psi (x)$ takes the~form  (the~fixed quantum  numbers
$\epsilon ,j,m$  are omitted)
$$
\Psi (x) = \left  [\; \left (\; e^{i\alpha} \cos \Gamma  \; {{1 + e^{-iA}} \over 2} \;+  \;
             e^{i\beta } \sin \Gamma \; {{1 - e^{-iA}} \over 2} \; \right )
\; \Psi ^{A}_{+1}(x) \; +   \right.
\eqno(9.5a)
$$
$$
\left.  \left (\; e^{i\alpha} \cos \Gamma \; {{1 - e^{-iA}} \over 2} \; + \;
             e^{i\beta } \sin \Gamma \; {{1 + e^{-iA}} \over 2} \;\right ) \;
\Psi ^{A}_{+1}(x) \;\right ] \; =  \;
\left [\; m \; \Psi ^{A}_{+1}(x) \; + \; n \; \Psi ^{A}_{-1}(x) \;\right ] \; .
$$

\noindent Although  the~quantity  $A$   enters the~expansion (9.5a),
but really
$\Psi (x)$ only contains three arbitrary parameters: those are
$\Gamma  , e^{i\alpha }$,
and $e^{i\beta }$. After simple calculation one gets
$$
\bar{N}_{A} = \; < \Psi  \mid  \hat{N}_{A} \mid  \Psi  >\; = (-1)^{j+1} \;
( \rho \cosh g + i \sigma \sinh g  )  \; ,
\eqno(9.5b)
$$
$$
\rho = \cos 2 \Gamma \cos f + \sin 2\Gamma \sin f \sin (\alpha -\beta ) \; , \;\;
\sigma = - \cos 2 \Gamma \sin f + \sin 2\Gamma \cos f \sin (\alpha -\beta )
$$

\noindent where  $f$  and $g$  are real  parameters defined by $A = f + i \; g$ .
Examining this expression, one may single out four particular
cases for separate consideration. Those are:
$$
1. \qquad g = 0 \; , f = 0 \; ,\qquad
\bar{N}_{A} = (-1)^{j+1} \cos 2\Gamma
\eqno(9.6a)
$$

\noindent here, the  $\bar{N}$ only fixes $\Gamma $, but $e^{i(\alpha -\beta )}$
remains indefinite.
$$
2. \qquad  g = 0  , f \neq  0 ,\qquad  \bar{N}_{A} =
(-1)^{j+1} \left (\cos 2\Gamma \cos f +
 \sin 2\Gamma \sin f \sin (\alpha -\beta ) \right )
\eqno(9.6b)
$$

\noindent here, the  measured  $\bar{N}_{A}$  does not fix
$\Gamma $ and $(\alpha - \beta )$, but only imposes a~certain limitation  on
both these parameters.
$$
3. \qquad  g \neq  0 , f = 0  ,\qquad  \bar{N}_{A} =
(-1)^{j+1} \left ( \cos 2 \Gamma \cosh g +
 i \sin 2 \Gamma \sin (\alpha -\beta) \sinh g \right )
\eqno(9.6c)
$$

\noindent here, the  $\bar{N}_{A}$  determines both $\Gamma $  and
$(\alpha - \beta )$;  and  thereby
this complex  $\bar{N}_{A}$  is the~physical quantity being quite
interpreted one. Finally, for the~fourth case
 $ (g \neq  0 , f \neq  0 ) $,   it follows
$$
4. \qquad \cos 2\Gamma = ( \rho \cos f - \sigma \sin f ) , \qquad
\sin 2\Gamma \sin (\alpha -\beta ) = ( \rho \cos f + \sigma \sin f )
\eqno(9.6d)
$$

\noindent i.e. the complex $\bar{N}_{A}$  also gives some information
about $\Gamma $   and $(\alpha - \beta )$ and therefore has character of
a  physically  interpreted  quantity.

\subsection*{10. Why $A$-freedom is not a gauge one?
On logical collision between concepts of gauge and non-gauge symmetries}

There  exists  else  one   cloud  over  the~subject   under
consideration\footnote{The author is grateful to E.~A.~Tolkachev,
L.~M.~Tomil'chik, and Ya.~M.~Shnir for the~fruitful discussion on this matter}.
Indeed, if the~parameter $A$ is a~real number, then the~matrix $S(A)$
translating $\Psi _{\epsilon jm\delta }(x)$ into
$\Psi ^{A}_{\epsilon jm\delta }(x)$
coincides (apart from a~phase factor $e^{iA/2}$) with a~matrix lying
in the group $SU(2)$:
$$
\hat{F}(A) \equiv  e^{-iA/2} S(A)  \in  SU(2)_{loc.}\; , \;\;
\Psi'^{A}_{\epsilon jm\delta }(x) \equiv
\hat{F}(A) \Psi _{\epsilon jm\delta }(x) =
e^{-iA/2} \Psi ^{A}_{\epsilon jm\delta }(x) \; .
\eqno(10.1)
$$

\noindent However,  the group $SU(2)_{loc.}$  has the~status of gauge one for
this system. So, else one point of  view  could
be brought to light: one could claim that two functions
$\Psi _{\epsilon jm\delta }(x)$
and $\Psi'^{A}_{\epsilon jm\delta }(x)$  (at $A^{*} = A$)  are  related
  by  means  of  a~gauge transformation:
and therefore the $\Psi'^{A}_{\epsilon jm\delta }(x)$
 exhibits  in  other  ways  the~same
physical state $\Psi _{\epsilon jm\delta }(x)$.  And further, as
 a~direct  consequence,
one could insist on the~impossibility  in  principle  to  observe
indeed  any  physical  distinctions  between  the~wave  functions
$\Psi _{\epsilon jm\delta }(x)$  and $\Psi'^{A}_{\epsilon jm\delta }(x)$.
 If the~$\hat{F}(A)$ transformation gets estimated
so, then ultimately one concludes  that the~above  $N_{A}$-parity  selection
rules (explicitly depended on  $A$  which is the real  for  this
case) are only a~mathematical  fiction  since  the~transformation
$\hat{F}(A)$  is not physically observable.

In this point we run  across a~problem of material physical
significance, in which one could perceive the~tense interplay  of
the~quantum-mechanical superposition principle and subtle
distinction between the~concepts of gauge and non-gauge symmetries.
In examining of this phenomenon, one should accord the~primacy of
careful coordination  of foregoing quantum-mechanical
generalities over all other considerations.

So,  a~question   of   principle   is   either   the $\hat{F}(A)$
transformation provides us with a~gauge  one  or  not?  The~same
question can be reformulated as follows:  is  the~fact $\hat{F}(A) \in
SU(2)_{loc.}$   sufficient  to  interpret $\hat{F}(A)$  exclusively as
the~transformation with gauge status?

For the moment let us suppose that the $\hat{F}(A)$  is  exclusively
a~gauge transformation and no other else. Then all functions
$\Psi'^{ A}_{\epsilon jm\delta }(x) \equiv
\hat{F}(A) \Psi _{\epsilon jm\delta }(x)$   represent the~same physically
identified state which had been described already by the~initial
function $\Psi _{\epsilon jm\delta }(x)$. In other words, the~function
$\Psi _{\epsilon jm\delta }(x)$   and the~following
$$
\Psi'^{A}_{\epsilon jm\delta } =
\left [\; {{e^{-iA/2} + \delta e^{iA/2} } \over 2} \; \Psi _{\epsilon jm,+1} \;+ \;
  {{e^{-iA/2} - \delta e^{iA/2} } \over 2}\; \Psi _{\epsilon jm,-1}\; \right ]
\eqno(10.2)
$$

\noindent are both only different representatives of the~same physical
state. However, such an~outlook is not acceptable on several physical
grounds. For clearing up this matter it is sufficient to have recourse
again to the~quantum-mechanical superposition principle and its
concomitant requirements. Indeed, the~possibility not to~accompany
the~transition of form
$\Psi _{\epsilon jm\delta }(x) \rightarrow \Psi'^{A}_{\epsilon jm\delta}(x)$
 by
the~similarity  transformation  on  all   physical
operators  ( it is meant $\hat{G} \rightarrow  \hat{G}' =
\hat{F}(A) \hat{G} \hat{F}^{-1}(A) )$  is
generally supposed to be an~essential constituent part in
understanding the~conventional superposition principle. Evidently,
the~essence of the~superposition principle in quantum mechanic
consists in just this assertion but not in a~simple fixation and
reminding  of  the~linearity property of the~matter equation.
In contrast  to  this, the~gauge-like interpretation of the~transformation $\hat{F}(A)$
makes us accompany the~change
$\Psi _{\epsilon jm\delta }(x) \rightarrow
\Psi '^{A}_{\epsilon jm\delta }(x)$   by
a~similarity transformation on $\hat{G} $.
Thus,  a~general  outlook  prescribing  to   interpret   the
transformation $\hat{F}(A)$ as exclusively a gauge one, contradicts  with
regulations stemming from the~superposition principle.

One could suggest that any contradiction does not arise
 here if  all  genuine  observable operators are invariant
under the similarity transformation above and all other
(not obeyed it) operators are unphysical fictions.
In this connection, let us look more closely  at
the~character of limitations imposed on $\hat{G}$  by  this  condition
$\hat{G} = \hat{F}(A) \hat{G} \hat{F}^{-1}(A)$;
an~elementary  analysis  shows  that
the~$\hat{G}$  is to be of diagonal isotopic structure,
 i.e. $\hat{g}_{12}(x)$
and $\hat{g}_{21}(x)$ must be equated to zero.
All  other possibilities  for $\hat{G}$  are associated with
the~assertion that a~certain  physical  distinction  between
$\Psi _{\epsilon jm\delta }(x)$ and $\Psi'^{A}_{\epsilon jm\delta }(x)$
is observable. But there are no grounds for the~use of physical
operators with this diagonal structure only, and  all  the  more,
for imposing the~limitation of the~form $A^{*} = A$.

Besides, the~simultaneous acceptance of operators (with a~status of
physical ones) of diagonal isotopic structure only and  the~added
limitation in the~form $A^{*} = A$, are  indissolubly tied up with
a~very definite conceiving of the~particle doublet  itself.  Indeed,
this can be physically  interpreted  as  an  exclusively  additive
character  of the~particle doublet. In other words, it can be
considered as follows: one must, in the~first place, measure
separately the quantities $\hat{g}_{11}(x)$  and $\hat{g}_{22}(x)$
and after, in the~second place, one can sum up both results. In author's
opinion, having supposed such an attitude for the particle doublet essence,
which in turn presupposes a~quite definite measurement procedures as
allowed, it is a~mystic and fruitless outlook that further
we can hope to have found any real correlations of quantum-mechanical
nature between  components
$T_{+1/2}\otimes  \Phi ^{+}(x)$ and $T_{-1/2}\otimes  \Phi ^{-}(x)$.
If such a~point of view  is  recognized  as
a~truly physical one, then they automatically give the understanding  of
particle doublet as an~entity to some mystic powers which are
not controlled by the~quantum-mechanical mathematical  formalism.
Instead, in  author's opinion, a~truly quantum-mechanical nature  of
particle  doublet  conception  envisages   that    some   physical
operators of non-diagonal form must exist really.

By the~way, if one insists on a~diagonal form only as
possible form of physical operators, one should consider a~further
dimension to the~problem under consideration: what is the~meaning
of the $A$-freedom.  Indeed, let us turn back to (10.1) and (10.2) again  and
set $e^{iA} = 1$ ($A = \pi $), then these relations, in particular,  give
$$
\hat{F} (A =   \pi ) \; \Psi _{\epsilon jm,-1}(x) \; \equiv  \;
\Psi _{\epsilon jm,+1}(x).
$$

\noindent The latter shows that if one decides in favor of the~gauge
character only of the~$A$-freedom, then one faces a~very strange
case.  Since two consistently distinguishable thus for and
linearly independent of  each  other  solutions
$\Psi _{\epsilon jm,-1}(x)$   and $\Psi _{\epsilon jm,+1}(x)$
turn out to be only different representatives of a~single invariant
state. But then the~natural and legitimate
question arises:  what  is  the~meaning of such a~physical
situation. Recalling that, generally speaking, the quantum doublet
states $\Psi ^{A}_{\epsilon jm\delta \mu }(x)$ (above the number
 $\mu $  was  often  omitted)  bear
five quantum numbers in place of four ones ($\epsilon ,j,m,\mu$ )
in  the Abelian case, and  that distinction (4 from 5)
between Abelian and  non-Abelian  situations  seems  to  be  quite
understandable and natural, as a result of addition by hands a~new
degree of freedom at going over to the~non-Abelian case. In the light
of this, it is easy to realize  that the~{\em physical identification}
of the functions  $\Psi _{\epsilon jm,-1,\mu }(x)$  and
$\Psi _{\epsilon jm,+1,\mu }(x)$ ,
being  effectively generated  (through  the~transformation $\hat{F}(A)$)
just  from the~gauge  understanding  of  the~$A$-freedom, represents
a~return to the~Abelian  scheme  again.  But what is the meaning of such
a~strange reversion? Thus,  seemingly, the~interpretation of
the~$A$-freedom as exclusively a~gauge one is  not justified since this
leads to a~logical collision with the~quantum superposition principle
and also entails the~return to the~Abelian scheme.

However, the matrix $F(A) \in  SU(2)^{gauge}_{loc.}$. In author's  opinion,
  there exists just one and very simple way out of  this  situation  which
consists in the following: The complete symmetry group  of  system
under consideration is  (apart from a rotational  symmetry  related
with $\vec{j}^{2}, j_{3}$  that has non-gauge character)  of the form
        $\hat{F}(A) \otimes  SL(2.C)^{loc.}_{gauge}
        \otimes  SU(2)^{loc.}_{gauge}$.
This group, in particular, contains  the  gauge  and  non-gauge
symmetry operations which  both have the~same  mathematical  form
but different physical status. Only such a~way of understanding
allows us not to reach a~deadlock.

\subsection*{11. Discussion and some generalities}

In conclusion, some additional  general  notices are to be given.
The specific analysis implemented in the~above study may play a
~part in considering  analogous  situations  for  more  complicated
gauge groups [1], serving as some guidelines. Since the case of freedom
in choosing an~explicit form of certain discrete operators,
which has its roots in the fact that a certain subgroup $G'$ of
a~complete gauge group $G$ commutes with a~Hamiltonian  can  appear.
Then those  symmetry  operations $G'$  will  generate  some  linear
transformations in a  set  of  basis  functions,  which  in  their
mathematical form will coincide  with  a~matrix  (independent  of
space coordinates) lying formally in the gauge group G.
It appears that analogous studying such Abelian monopole manifestations
on the~background of other (big) gauge groups $G$ is feasible and would
require   no large departures from the~present scheme. Those latter,
seemingly would be completely determined by the~inner structure of the~relevant Lie
algebras, in particular, their respective Cartan's sub-algebras. The~number
of elements in those sub-algebras would coincide with the~number  of
(one-parametric) generalized chiral symmetry transformations.

Else one remark may be given. Existence of the~above isotopic chiral symmetry
is not relevant to whether a~particle multiplet carries the~isotopic spin
$T=1/2$  and the~Lorentzian spin $S=1/2$. The general structure of the matter
equation  (see (3.1)) will remain the~same if one extends the~problem to
 any other values of $T$ and $S$ (to retain the formal similarity, one ought
to exploit the~first order wave equation formalism)\footnote{For instance,
the author has carried out all required calculations for $T=1$
case.  Technically, this is some more laborious task but considered
conclusions are in part similar. The~main difference arisen is that now
there exist two independent one-parametric symmetries of the~triplet-monopole
Hamiltonian (instead of the~one discerned by the~present work for $T=1/2$
case), these symmetry operations  vary substantially in their
mathematical forms and physical manifestations.  An account of this has
appeared to be rather unwieldy, so that it has not been included in
the~present paper.}. Moreover, all the~problem can be easily extended
to an~arbitrary curved space-time of spherical symmetry.

\section*{Acknowledgments}

I would like to express my gratitude to Prof. A.~A.~Bogush, who has
looked through the manuscript and contributed many suggestions for
corrections and additions. I am also grateful to Dr E.~A.~Tolkachev,
whose strong criticism of an initial variant of the article  makes me
revise  some ideas before  sending the conclusive variant to the
journal, and also to Prof. L.~M.~Tomil'chik for contributing comments
and friendly advice. The author is especially indebted to.  Dr V.~V.~Gilewsky
for his wholehearted support and agreeing to help in preparing
the \LaTeX  file on the article.

\newpage
\subsection*{Supplement A. Connection between electron-monopole functions
             in spherical and Cartesian bases}

Let us consider
relationships between fermion-monopole functions in spherical
and Cartesian bases. First, we look at connection between
$D$-functions used above  and the~so-called spinor monopole harmonics.
To this ent, one  ought to  perform subsequently two translations:
from the~spherical  tetrad  and
2-spinor (by Weyl) frame in bispinor space  into,  respectively,
the~Cartesian  tetrad and the~so-called Pauli's (bispinor) frame.
In the first place, it is convenient to accomplish those translations
for a~free electronic function; so as, in the second place, to follow this
pattern further in~the monopole case.

So, subjecting that free electronic  function
to  the~local  bispinor  gauge   transformation
(associated with  the~change $sph. \;\;  \rightarrow  \;\; Cart.$)
$$
\Phi _{Cart.} =
\left ( \begin{array}{cc}
     U^{-1}  & 0 \\ 0  & U^{-1}
\end{array} \right )    \; \Psi_{sph.} , \qquad U^{-1} =
\left ( \begin{array}{lr}
  \cos \theta /2 \; e^{-i\phi /2}  &  - \sin \theta /2 \; e^{-i\phi /2}  \\
  \sin \theta /2 \; e^{+i\phi /2}  &    \cos \theta /2 \; e^{+i\phi /2}
\end{array} \right )
$$

\noindent and further, taking the~bispinor  frame  from  the~Weyl  2-spinor
form into the~Pauli's
$$
\Phi ^{Pauli.}_{Cart.} = \left ( \begin{array}{c}
                 \varphi   \\ \xi
\end{array} \right  ) , \qquad
\Phi^{Weyl}_{Cart.} = \left ( \begin{array}{c}
             \xi   \\ \eta
\end{array} \right ) , \qquad
\varphi  = { \xi  + \eta  \over \sqrt{2}} ,\;\;
\chi     = { \xi -  \eta \over \sqrt{2}}
$$

\noindent we get
$$
\varphi  = \left [\; {f_{1} + f_{3} \over \sqrt{2} } \;
\left ( \begin{array}{c}
    \cos \theta /2 \; e^{-i\phi /2}  \\
    \sin \theta /2 \; e^{+i\phi /2}
\end{array} \right ) \; D_{-1/2} \; +  \;
{ f_{2} + f_{4} \over \sqrt{2}}
\left ( \begin{array}{c}
   -\sin \theta /2 \; e^{-i\phi /2} \\
    \cos \theta /2 \; e^{+i\phi /2}
\end{array} \right )  \; D_{+1/2} \; \right ]  ;
\eqno(A.1a)
$$
$$
\chi  =\left [\; {f_{1} - f_{3} \over \sqrt{2} } \;
\left ( \begin{array}{c}
    \cos \theta /2 \; e^{-i\phi /2}  \\
    \sin \theta /2 \; e^{+i\phi /2}
\end{array} \right ) \; D_{-1/2} \; +  \;
{ f_{2} - f_{4} \over \sqrt{2}}
\left ( \begin{array}{c}
   -\sin \theta /2 \; e^{-i\phi /2} \\
    \cos \theta /2 \; e^{+i\phi /2}
\end{array} \right )  \; D_{+1/2}\; \right ]  ;
\eqno(A.1b)
$$

\noindent Further, for the~above solutions with fixed proper  values  of
$P$-operator, we produce
$$
P=(-1)^{j+1} :
\Phi ^{Pauli}_{Cart.} =
{e^{-i\epsilon t} \over r \sqrt{2}} \;
\left ( \begin{array}{c}
(f_{1} + f_{2}) ( \chi _{+1/2} \; D_{-1/2} + \chi _{-1/2} \;D_{+1/2} ) \\
(f_{1} - f_{2}) ( \chi _{+1/2} \; D_{-1/2} - \chi _{-1/2} \; D_{+1/2})
\end{array} \right )
\eqno(A.2a)
$$
$$
P = (-1)^{j} : \;\;\; \Phi ^{Pauli}_{Cart.} = { e^{-i\epsilon t} \over  r \sqrt{2}} \;
\left ( \begin{array}{c}
(f_{1} - f_{2}) ( \chi _{+1/2} \; D_{-1/2} - \chi _{-1/2} \; D_{+1/2}) \\
(f_{1} + f_{2}) ( \chi _{+1/2} \; D_{-1/2} + \chi _{-1/2} \; D_{+1/2} )
\end{array} \right )
\eqno(A.2b)
$$

\noindent where $\chi _{+1/2}$   and $\chi _{-1/2}$   designate   the~columns
of   matrix $U^{-1}(\theta ,\phi )$ (in the literature they are termed as
helicity spinors)
$$
\chi _{+1/2} =
\left ( \begin{array}{c}
    \cos \theta /2 \;  e^{-i\phi /2} \\
    \sin \theta /2 \;  e^{+i\phi /2}
\end{array} \right ) , \qquad
\chi _{-1/2} =
\left ( \begin{array}{c}
          -\sin \theta /2  \; e^{-i\phi /2}  \\
           \cos \theta /2  \; e^{+i\phi /2}
\end{array} \right )  .
\eqno(A.2c)
$$

\noindent Now, using the  known extensions  for  spherical  spinors
$\Omega ^{j\pm 1/2}_{jm}(\theta ,\phi )$  in terms of
$\chi _{\pm 1/2}$  and $D$-functions  [66]:
$$
\Omega ^{(+)}_{jm} = (-1)^{m+1/2} \sqrt{(2j+1)/8\pi}\;
( + \chi _{+1/2} \; D_{-1/2} \; + \; \chi _{-1/2} \; D_{+1/2}) ,
$$
$$
\Omega ^{(-)}_{jm} = (-1)^{m+1/2} \sqrt{(2j+1)/8\pi} \;
( - \chi _{+1/2} \; D_{-1/2} \; + \; \chi _{-1/2} \; D_{+1/2})
$$

\noindent we   eventually   arrive   at   the~common   representation   of
the~spinor spherical solutions:
$$
P = (-1)^{j+1} : \qquad \Phi ^{Pauli}_{Cart.}= {e^{-i\epsilon t} \over  r} \;
\left ( \begin{array}{r}
   + f(r) \; \Omega ^{(+)}_{jm} (\theta ,\phi ) \\
   - i\;g(r)\; \Omega ^{(-)}_{jm}(\theta ,\phi )
\end{array} \right )      ;
\eqno(A.3a)
$$
$$
P = (-1)^{j} : \qquad \Phi ^{Pauli.}_{Cart.} = {e^{-i\epsilon t} \over r} \;
\left ( \begin{array}{r}
-i\; g(r) \; \Omega ^{(-)}_{jm} (\theta ,\phi ) \\
     f(r) \; \Omega ^{(+)}_{jm}(\theta ,\phi )
\end{array} \right )       .
\eqno(A.3b)
$$

The Abelian monopole situation  can  be  considered in the same  way.
As a~result,  we  produce  the~following  representation  of
the~monopole-electron  functions  in  terms  of   `new'   angular
harmonics ($k \equiv e g$)
$$
M = (-1)^{j+1} : \qquad \Phi ^{Pauli.}_{Cart.} = {e^{-i\epsilon t} \over r} \;
\left ( \begin{array}{r}
 + f(r) \; \xi ^{(1)}_{jmk} (\theta ,\phi ) \\
-i \; g(r) \; \xi ^{(2)}_{jmk}(\theta ,\phi )
\end{array} \right )           ;
\eqno(A.4a)
$$
$$
M = (-1)^{j} : \qquad \Phi^{Pauli}_{(eg)Cart.} ={ e^{-i\epsilon t} \over  r} \;
\left ( \begin{array}{r}
-i \; g(r); \xi ^{(1)}_{jmk}(\theta ,\phi ) \\
+f(r) \; \xi ^{(2)}_{jmk}(\theta ,\phi )
\end{array} \right )   .
\eqno(A.4b)
$$

\noindent Here, the two column functions $\xi ^{(1)}_{jmk} (\theta ,\phi )$
and   $\xi  ^{(2)}_{jmk} (\theta ,\phi )$ denote  the~special  combinations
of $\chi _{\pm 1/2}(\theta ,\phi )$ and
$D_{-m,eg/hc\pm 1/2}(\phi ,\theta ,0)$:
$$
\xi ^{(1)}_{jmk} = ( + \chi _{-1/2}\; D_{k+1/2} + \chi _{+1/2} \;D_{k-1/2} ) \; ,
\qquad
\xi ^{(2)}_{jmk} = ( + \chi _{-1/2} \; D_{k+1/2}- \chi _{+1/2} \; D_{k-1/2} )
\eqno(A.5)
$$

\noindent compare them with  analogous  extensions  for
$\Omega ^{j\pm 1/2}_{jm}(\theta ,\phi )$.  These
2-component and $(\theta ,\phi )$-dependent functions
 $\xi ^{(1)}_{jmk}(\theta ,\phi)$  and $\xi ^{(2)}_{jmk}(\theta ,\phi)$
just  provide  what is called spinor  monopole harmonics.
It should be useful to write down the~detailed explicit form of these
generalized harmonics.  Given  the  known  expressions  for $\chi $-  and
$D$-functions, the formulae  yield  the following
$$
\xi ^{(1,2)}_{jmk}(\theta ,\phi ) =  \left [ \;
e^{{\rm im}\phi } \;
\left ( \begin{array}{r}
-\sin \theta /2  \; e^{-i\phi /2}  \\
 \cos \theta /2  \; e^{+i\phi /2}
\end{array} \right ) \;
 d^{j}_{-m,k+1/2} (\cos \theta ) \;   \pm  \right.
$$
$$
\left. e^{im\phi } \;
\left ( \begin{array}{c}
\cos \theta /2 \; e^{-i\phi /2} \\
\sin \theta /2 \; e^{+i\phi /2}
\end{array} \right ) \;
d^{j}_{-m,k-1/2} (\cos \theta ) \; \right ]
\eqno(A.6)
$$

\noindent here, the signs  $+$ (plus)  and  $-$ (minus) refer  to
$\xi ^{(1)}$ and $\xi ^{(2)}$, respectively.
One can equally work  whether in  terms  of  monopole  harmonics
$\xi ^{(1,2)}(\theta ,\phi )$ or directly in terms of $D$-functions,
but  the  latter alternative has an~advantage over the former because of
the~straightforward  access  to the `unlimited' $D$-function
apparatus, instead  of proving and  producing just disguised old  results.

Above, at translating the electron-monopole functions into the
Cartesian tetrad and Pauli's spin frame, we had overlooked the case
of minimal $j$. Turning to it, on  straightforward  calculation
we find (for $k\; <\; 0$ and $k\; > \; 0$ , respectively)
$$
k > 0 \; : \qquad
\Phi ^{(eg)Cart.}_{j_{min.}} \; = \; {e^{-i\epsilon t} \over \sqrt{2} r} \;
\left ( \begin{array}{c}
   ( f_{1} +  f_{3})  \\    ( f_{1} -  f_{3})
\end{array} \right ) \;
\chi _{+1/2}(\theta ,\phi)\;
D^{\mid k \mid -1/2}_{-m,k-1/2} (\theta ,\phi ,0)  ;
\eqno(A.7a)
$$
$$
k < 0 \; : \qquad
\Phi ^{(eg)Cart.}_{j_{min.}} \; = \; {e^{-i\epsilon t} \over \sqrt{2} r} \;
\left ( \begin{array}{c}
   ( f_{2} +  f_{4})    \\
   ( f_{2} -  f_{4})
\end{array} \right ) \;
\chi _{-1/2}(\theta ,\phi)\;
D^{\mid k \mid -1/2} _{-m,k+1/2} (\theta ,\phi ,0) \; .
\eqno(A.7b)
$$

In addition, now it is a~convenient point to clarify the~way how the~used
gauge transformation $U^{-1}(\theta, \phi)$, not being a~single-valued
matrix-function of spatial points, affects the~continuity (or discontinuity)
properties of the~wave functions under consideration.

Returning to the~$\varphi(x)$  from (A.1a) at the~points $\theta = 0, \pi$
(the functions $\chi(x)$ from (A.1b) are completely analogous ones),
one gets
$$
\varphi (\theta = 0) \; \sim \;  \left [\; {f_{1} + f_{3} \over \sqrt{2} } \;
\left ( \begin{array}{c}
     e^{-i\phi /2}  \\   0
\end{array} \right ) \; D^{j}_{-m,-1/2}(\phi, \theta = 0,0) \; + \right.
$$
$$
\left. { f_{2} + f_{4} \over \sqrt{2}}
\left ( \begin{array}{c}
      0  \\       e^{+i\phi /2}
\end{array} \right )  \; D^{j}_{-m, +1/2}(\phi, \theta = 0,0) \; \right ] \; ;
\eqno(A.8a)
$$
$$
\varphi(\theta = \pi ) \; \sim \;
\left [ \; {f_{1} + f_{3} \over \sqrt{2} } \;
\left ( \begin{array}{c}
         0   \\  e^{+i\phi /2}
 \end{array} \right ) \; D^{j}_{-m,-1/2}(\phi, \theta = \pi, 0)  \; +  \right.
$$
$$
\left. { f_{2} + f_{4} \over \sqrt{2}}
\left ( \begin{array}{c}
    e^{-i\phi /2} \\  0
\end{array} \right ) \; D^{j}_{-m, +1/2}(\phi, \theta = \pi, 0) \; \right ] \; .
\eqno(A.8b)
$$

\noindent Further, allowing for the~relevant relations from
Tables $1a,b$, one produces: $\varphi(\theta = 0)\;$ and
$\varphi(\theta = \pi)\;$ are single-valued functions.

In turn, for the~monopole case, in place of (A.7a,b) one gets (for
definiteness, let $eg = +1/2$)
$$
\varphi^{(eg=+1/2)} (\theta = 0) \; \sim \; = \;
\left [\; {f_{1} + f_{3} \over \sqrt{2} } \;
\left ( \begin{array}{c}
     e^{-i\phi /2}  \\   0
\end{array} \right ) \; D^{j}_{-m,0}(\phi, \theta = 0,0) \; +  \right.
$$
$$
\left. { f_{2} + f_{4} \over \sqrt{2}}
\left ( \begin{array}{c}
      0  \\       e^{+i\phi /2}
\end{array} \right )  \; D^{j}_{-m, +1}(\phi, \theta = 0,0) \; \right ]  \; ;
\eqno(A.9a)
$$
$$
\varphi^{(eg=+1/2)}(\theta = \pi ) \; \sim \;
\left [ \; {f_{1} + f_{3} \over \sqrt{2} } \;
\left ( \begin{array}{c}
         0   \\  e^{+i\phi /2}
 \end{array} \right ) \; D^{j}_{-m,0}(\phi, \theta = \pi, 0) \; +  \right.
$$
$$
\left. { f_{2} + f_{4} \over \sqrt{2}}
\left ( \begin{array}{c}
    e^{-i\phi /2} \\  0
\end{array} \right ) \; D^{j}_{-m, +1}(\phi, \theta = \pi, 0) \; \right ] \; .
\eqno(A.9b)
$$

\noindent From that, allowing for the relations from Tables $2a,b$, one finds that
the~totality of all $\varphi^{(eg=+1/2)}(x)$ consists of
both regular and non-regular (non-single-valued)  functions at the~$x_{3}$
axis; these latter behave like $e^{-i\phi /2}$ and $e^{+i\phi /2}$
at the~half-axes $\theta = 0$ and $\theta = \pi$, respectively.

The case of minimal $j$ follows the~same behavior: for example, if
$eg = \pm1/2$ then one gets
$$
eg = +1/2 \;\; : \qquad
\Phi ^{(eg=+1/2)Cart.}_{j_{min.}} \; = \; {e^{-i\epsilon t} \over \sqrt{2} r} \;
\left ( \begin{array}{c}
   ( f_{1} +  f_{3})  \\    ( f_{1} -  f_{3})
\end{array} \right ) \;
\chi _{+1/2}(\theta ,\phi)\;      \; ;
\eqno(A.10a)
$$
$$
eg = -1/2 \;\; :\qquad
\Phi ^{(eg=-1/2)Cart.}_{j_{min.}} =  {e^{-i\epsilon t} \over \sqrt{2} r} \;
\left ( \begin{array}{c}
   ( f_{2} +  f_{4}) \;   \\
   ( f_{2} -  f_{4})
\end{array} \right ) \;
\chi _{-1/2}(\theta ,\phi) \; .
\eqno(A.10b)
$$

Now, let us turn to the non-Abelian doublet-fermion case. The task is to
translate the~composite functions  $\Psi ^{C. sph.(A)}_{\epsilon jm\delta}$
(see  (6.9)) into the Cartesian tetrad basis:
$$
\Psi^{C. sph.}_{\epsilon jm\delta}  \qquad \rightarrow \qquad
\Psi ^{C. Cart.}_{\epsilon jm\delta} =
\left ( \begin{array}{c}
\Sigma^{(+)}_{\epsilon jm\delta}(x)  \\
\Sigma^{(-)}_{\epsilon jm\delta}(x)
\end{array} \right )
$$

\noindent where the~2-component structure in Lorentzian space is explicitly
detailed. For those composite two-column functions $\Sigma^{(\pm)}_{\epsilon jm\delta}(x)$
one gets
$$
\Sigma^{(\pm)}_{\epsilon jm\delta}(x)  \; = \;
{e^{-i\epsilon t}\over r}\; \times
\eqno(A.11)
$$
$$
\left \{\;  T_{+1/2} \otimes \; {\sqrt{j+m} \over 2j+1} {1 \over \sqrt{2}} \;
\left [\;
(\;K^{A}_{\delta} \; \pm \; \delta\; e^{iA}\; L^{-A}_{\delta}\;)  \;
\chi_{+1/2} \; D^{j-1/2}_{-m+1/2,-1/2} \;\; + \right. \right.
$$
$$
\left. (\; L^{A}_{\delta} \; \pm \; \delta\; e^{iA}\; K^{-A}_{\delta}\;)  \;
\chi_{-1/2} \; D^{j-1/2}_{-m+1/2,+1/2} \;  \right ] \;\; +
$$
$$
T_{+1/2} \otimes \; {\sqrt{j-m+1} \over 2j+1} \;{1 \over \sqrt{2}} \;
\left [ \;
(\;M^{A}_{-\delta} \; \mp \; \delta\; e^{iA}\; N^{-A}_{-\delta}\;)  \;
\chi_{+1/2} \; D^{j+1/2}_{-m+1/2,-1/2} \;\; + \right.
$$
$$
\left. (\; N^{A}_{-\delta} \; \mp \; \delta\; e^{iA}\; M^{-A}_{-\delta}\;)  \;
\chi_{-1/2} \; D^{j+1/2}_{-m+1/2,+1/2} \; \right ] \;\; +
$$
$$
T_{-1/2} \otimes \; {\sqrt{j-m} \over 2j+1} {1 \over \sqrt{2}} \;
\left[ \;
(-\;K^{A}_{-\delta} \; \pm \; \delta\; e^{iA}\; L^{-A}_{-\delta}\;)  \;
\chi_{+1/2} \; D^{j-1/2}_{-m-1/2,-1/2} \;\;  +    \right.
$$
$$
\left. (-\; L^{A}_{-\delta} \; \pm \; \delta\; e^{iA}\; K^{-A}_{-\delta}\;)  \;
\chi_{-1/2} \; D^{j-1/2}_{-m-1/2,+1/2} \;  \right ] \;\; +
$$
$$
T_{-1/2} \otimes \; {\sqrt{j+m+1} \over 2j+1} \;{1 \over \sqrt{2}} \;
\left [\;
(\;M^{A}_{\delta} \; \pm \; \delta\; e^{iA}\; N^{-A}_{\delta}\;)  \;
\chi_{+1/2} \; D^{j+1/2}_{-m-1/2,-1/2} \;\; +     \right.
$$
$$
\left. \left. (\; N^{A}_{\delta} \; \pm \; \delta\; e^{iA}\; M^{-A}_{\delta}\;)  \;
\chi_{-1/2} \; D^{j+1/2}_{-m-1/2,+1/2} \;  \right ] \; \right \} \; .
$$

\noindent With the~use of four formulas [16]
$$
\chi_{\pm1/2} \; D^{j-1/2}_{-m+1/2,\mp1/2} \; = \;
(\; \Omega ^{(+)}_{j-1/2,m-1/2} \; \mp \; \Omega^{(-)}_{j-1/2,m-1/2} \;) \;
{\sqrt{4\pi} \over 2 \sqrt{j}(-1)^{m} } \; ,
$$
$$
\chi_{\pm1/2} \; D^{j+1/2}_{-m+1/2,\mp1/2} \; = \;
(\; \Omega ^{(+)}_{j+1/2,m-1/2} \; \mp \;\Omega^{(-)}_{j+1/2,m-1/2} \;) \;
{\sqrt{4\pi} \over 2 \sqrt{j+1}(-1)^{m} } \; ,
$$
$$
\chi_{\pm1/2} \; D^{j-1/2}_{-m-1/2,\mp1/2} \; = \;
( \; \Omega ^{(+)}_{j-1/2,m+1/2} \; \mp \;\Omega^{(-)}_{j-1/2,m+1/2} \;) \;
{\sqrt{4\pi} \over 2 \sqrt{j}(-1)^{m+1} } \; ,
$$
$$
\chi_{\pm1/2} \; D^{j+1/2}_{-m-1/2,\mp1/2} \; = \;
( \; \Omega ^{(+)}_{j+1/2,m+1/2} \; \mp \;\Omega^{(-)}_{j+1/2,m+1/2} \;) \;
{\sqrt{4\pi} \over  2 \sqrt{j+1} (-1)^{m+1} } \;
$$

\noindent the~above expression (A.11) for  $\Sigma^{(\pm)}_{\epsilon jm\delta}(x)$
can be rewritten in the~form
$$
\Sigma^{(\pm)}_{\epsilon jm\delta}(x)  \; = \;
{e^{-i\epsilon t}\over r}\; \times
\eqno(A.12)
$$
$$
T_{+1/2} \otimes \; B \;\;  \left \{ \;\;  \left  [ \;
(\;K^{A}_{\delta} \; \pm \; \delta\; e^{iA}\; K^{-A}_{\delta}\;) \; +  \;
(\;L^{A}_{\delta} \; \pm \; \delta\; e^{iA}\; L^{-A}_{\delta}\;)\; \right ] \;
\Omega^{(+)}_{j-1/2,m-1/2} \; +  \right.
$$
$$
\left. \left [ \; (\;-K^{A}_{\delta} \; \pm \; \delta\; e^{iA}\; K^{-A}_{\delta}\;) \; +  \;
(\;L^{A}_{\delta} \; \mp \; \delta\; e^{iA}\; L^{-A}_{\delta}\;) \;\right ]\;
\Omega^{(-)}_{j-1/2,m-1/2} \;\;  \right \} \;\; +
$$
$$
T_{+1/2} \otimes \; C \;\;  \left \{ \;\;
\left [ \;(\;M^{A}_{-\delta} \; \mp \; \delta\; e^{iA}\; M^{-A}_{-\delta}\;) \; +  \;
(\;N^{A}_{-\delta} \; \mp \; \delta\; e^{iA}\; N^{-A}_{-\delta} \;)  \;\right ] \;
\Omega^{(+)}_{j+1/2,m-1/2} \; +  \right.
$$
$$
\left. \left [\; (\;-M^{A}_{-\delta} \; \mp \; \delta\; e^{iA}\; M^{-A}_{-\delta}\;) \; +  \;
(\;N^{A}_{-\delta} \; \pm \; \delta\; e^{iA}\; N^{-A}_{-\delta}\;) \;\right ] \;
\Omega^{(-)}_{j+1/2,m-1/2} \;\;  \right \} \;\; +
$$
$$
T_{-1/2} \otimes \; D \;\; \left \{ \;\;
\left [\; (\;-K^{A}_{-\delta} \; \pm \; \delta\; e^{iA}\; K^{-A}_{-\delta}\;) \; +  \;
(\;-L^{A}_{-\delta} \; \pm \; \delta\; e^{iA}\; L^{-A}_{-\delta}\;)\; \right ] \;
\Omega^{(+)}_{j-1/2,m+1/2} \; +  \right.
$$
$$
\left. \left [\; (\;K^{A}_{-\delta} \; \pm \; \delta\; e^{iA}\; K^{-A}_{-\delta}\;) \; +  \;
(\;-L^{A}_{-\delta} \; \mp \; \delta\; e^{iA}\; L^{-A}_{-\delta}\;) \;\right ] \;
\Omega^{(-)}_{j-1/2,m+1/2} \;\; \right \} \;\; +
$$
$$
T_{-1/2} \otimes \; E \;\;  \left \{ \;\; \left [\;
(\;M^{A}_{\delta} \; \pm \; \delta\; e^{iA}\; M^{-A}_{\delta}\;) \; +  \;
(\;N^{A}_{\delta} \; \pm \; \delta\; e^{iA}\; N^{-A}_{\delta}\;)\; \right ] \;
\Omega^{(+)}_{j+1/2,m+1/2} \; +   \right.
$$
$$
\left. [\; (\;-M^{A}_{\delta} \; \pm \; \delta\; e^{iA}\; M^{-A}_{\delta}\;) \; +  \;
(\;N^{A}_{\delta} \; \mp \; \delta\; e^{iA}\; N^{-A}_{\delta}\;) \;] \;
\Omega^{(-)}_{j+1/2,m+1/2} \;\; \right \}
$$

\noindent where the~symbols $B, C, D, E $ denote respectively
$$
B \; = \; {\sqrt{j+m} \over 2j+1} \; {1 \over \sqrt{2}} \;
{\sqrt{4\pi} \over 2 \sqrt{j}(-1)^{m} } \; , \;\;
C\; = \; {\sqrt{j-m+1} \over 2j+1} \;{1 \over \sqrt{2}} \;
{\sqrt{4\pi} \over 2 \sqrt{j+1}(-1)^{m} } \; ,
$$
$$
D \; = \; {\sqrt{j-m} \over 2j+1} \; {1 \over \sqrt{2}} \;
{\sqrt{4\pi} \over 2 \sqrt{j}(-1)^{m+1} } \; ,   \;\;
E \; =\; {\sqrt{j+m+1} \over 2j+1} \; {1 \over \sqrt{2}} \;
{\sqrt{4\pi} \over  2 \sqrt{j+1} (-1)^{m+1} } \;\; .
$$

The~representation (A.12) will be significantly simplified if $A=0$; so
one can find
$$
A=0 \; : \qquad \Sigma^{(\pm)}_{\epsilon jm\delta}(x)  \; = \;
{e^{-i\epsilon t}\over r}\; \times
\eqno(A.13)
$$
$$
\left [ \; T_{+1/2} \otimes \; B \;\;
(\;\pm \; \delta\;  K_{\delta}\; +  \; L_{\delta} \; )  \;
\Omega^{(\pm \delta)}_{j-1/2,m-1/2} \; +   \right.
$$
$$
T_{+1/2} \otimes \; C \;\;
( \;\mp \delta \; M_{-\delta} \; +   \; N_{-\delta} \;) \;
\Omega^{(\mp \delta )}_{j+1/2,m-1/2} \; +
$$
$$
T_{-1/2} \otimes \; D \;\;
(\;\pm \delta \; K_{-\delta} \; - \;L_{-\delta} \; )  \;
\Omega^{(\mp \delta)}_{j-1/2,m+1/2} \; +
$$
$$
\left. T_{-1/2} \otimes \; E \;\;
(\;\pm \delta M_{\delta} \; +  \; N_{\delta} \;)  \;
\Omega^{(\pm \delta)}_{j+1/2,m+1/2} \; \right ] \; .
$$

In particular, the~formula (A.13) apparently exhibits the~Abelian
fermion-like sub-structure thta stems from the~Abelian-like
 $P$-inversion operation (6.4).

\end{document}